\documentclass[twocolumn]{autart}
\pdfoutput=1
\usepackage{amsmath}
\usepackage{amssymb}
\usepackage{algorithmic}
\usepackage{algorithm} 
\usepackage{color}
\usepackage{url} 
\usepackage{graphicx}          
\usepackage{dsfont}
\usepackage{multirow,array}
\usepackage[table]{xcolor}
\usepackage{arydshln}
\renewcommand{\thefootnote}{\fnsymbol{footnote}}
\usepackage[export]{adjustbox}
\usepackage[a4paper, text={170mm,257mm},left=16mm,top=16mm,centering]{geometry}
\usepackage[title]{appendix}
\usepackage{placeins}
\usepackage{mathtools}
\usepackage{subcaption}
\usepackage{physics}
\usepackage{bm}
\usepackage{soul}
\usepackage{wrapfig}
\usepackage[sectionbib]{chapterbib}

\definecolor{Gray}{gray}{0.9}

\newcommand{\generealweight}{W}

\newcommand{\R}{\mathbb{R}}

\newcommand{\bw}{W}

\newcommand{\net}{\mathcal{W}}
\newcommand{\hyp}{\mathcal{H}}
\newcommand{\data}{\mathcal{D}}

\def\rmH{{\mathbf{H}}}
\newtheorem{remark}{Remark}
\DeclareMathOperator{\diag}{diag}
\DeclareMathOperator*{\argmax}{argmax}
\DeclareMathOperator*{\argmini}{argmin}
\newcommand{\ones}{\mathds{1}}
\newcommand{\eye}{\mathbf{I}} 
\newtheorem{lemma}{Lemma}
\newtheorem{definition}{Definition}

\newtheorem{proof}{Proof}

\newcommand{\modify}[1]{{\color{black}  {#1}}}
\newcommand{\modifyfive}[1]{{\color{black}  {#1}}}
\newcommand{\modifytwo}[1]{{\color{black}  {#1}}}
\begin{document}
\begin{frontmatter}    
		
\title{Sparse Bayesian Deep Learning for \\ Dynamic System Identification\thanksref{footnoteinfo}}

\thanks[footnoteinfo]{This paper was not presented at any IFAC 
meeting.
}

\author[First]{Hongpeng Zhou}\ead{h.zhou-3@tudelft.nl}, 
\author[First]{Chahine Ibrahim} \ead{ibrahim-chahine@outlook.com},
\author[Second]{Wei Xing Zheng}\ead{W.Zheng@westernsydney.edu.au},
\author[Third,First]{Wei Pan}\ead{wei.pan@manchester.ac.uk}

\address[First]{Department of Cognitive Robotics, Delft University of Technology, Delft, 2628 CD, the Netherlands}
\address[Second]{School of Computer, Data and Mathematical Sciences, Western Sydney University, Sydney, NSW 2751, Australia}
\address[Third]{Department of Computer Science, The University of Manchester, Manchester, M13 9PL, United Kingdom}

\begin{abstract}
This paper proposes a sparse Bayesian treatment of deep neural networks (DNNs) for system identification. Although DNNs show impressive approximation ability in various fields, several challenges still exist for system identification problems. First, DNNs are known to be too complex that they can easily overfit the training data. Second, the selection of the input regressors for system identification is nontrivial. Third, uncertainty quantification of the model parameters and predictions are necessary. The proposed Bayesian approach offers a principled way to alleviate the above challenges by marginal likelihood/model evidence approximation and structured group sparsity-inducing priors construction. The identification algorithm is derived as an iterative regularised optimisation procedure that can be solved as efficiently as training typical DNNs. 
Remarkably, an efficient and recursive Hessian calculation method for each layer of DNNs is developed, turning the intractable training/optimisation process into a tractable one.
Furthermore, a practical calculation approach based on the Monte-Carlo integration method is derived to quantify the uncertainty of the parameters and predictions. 
The effectiveness of the proposed Bayesian approach is demonstrated on several linear and nonlinear system identification benchmarks by achieving good and competitive simulation accuracy. The code to reproduce the experimental results is open-sourced and available online.
\end{abstract}

\end{frontmatter}

\section{Introduction}
\vspace{-0.2cm}
System identification (SYSID) has a long history in natural and social sciences \cite{ljung1999system}.
Various approaches have been proposed for both linear/nonlinear systems and static/dynamical processes~\cite{chen2014system,chiuso2012,ced_other4,bw_other1}. 
\modify{Among these, neural networks (NNs) are prominent black-box models and recently regained research interest in the SYSID community
~\cite{ljung_lstm,beintema2021nonlinear,gedon2021deep,forgione2021continuous}, 
thanks to the boom of deep learning.
}

The deep neural network (DNN) models have their advantages and disadvantages. An early paper on feed-forward NNs proved the universal approximation capabilities of any measurable function, using one hidden layer on a compact set~\cite{NN_Identification}.
\modify{The training of DNN is mainly based on data which does not require much prior information about the system~\cite{lecun2015deep}.}
\modify{Several works also achieved competitive results by using feed-forward NNs~\cite{NN_Identification2} and recurrent neural networks (RNNs)~\cite{delgado1995dynamic,weber2021non} in the context of dynamical systems.}
However, it is not easy to design a proper NN structure. 
First of all, the trade-off between the model complexity and (simulation) prediction accuracy should be considered. An over-simplified model cannot reveal the underlying relation between input and output data. On the other hand, an over-complex model may overfit the training data, thus reducing its generalisation ability.
Besides, the inevitable (non-Gaussian and non-additive) noise and non-smooth characteristics of some nonlinear processes may also cause the overfitting problem. 
Furthermore, NNs can also be underspecified by the data and constitute a large space of hypotheses for high-performing models~\cite{bayesian_case}. 
Another challenging problem for SYSID is input regressor selection,
which is defined as follows: given input regressors $z(t+1) = [u(t+1), u(t), \cdots, u(t-l_u),y(t),y(t-1),\cdots, y(t-l_y)]^{\top} \in \mathbb{R}^{l_u+l_y+1}$ with $l_u$ and $l_y$ denoting respectively the input and output lag, the most relevant input regressor features, which can explain the intrinsic phenomenon of the system, are selected~\cite{castellano2000variable}.
An effective input regressor selection can improve the prediction performance, and generalisation ability of the identified model. 

For these challenges, the sparse Bayesian learning method offers a principled way to tackle them simultaneously: a) A more efficient exploration of the hypothesis space (corresponding to saddle points) of NN models is possible~\cite{bayesian_case,zhou2019bayesnas}; b) Over-fitting can be alleviated, and model redundancies can be eliminated through marginalisation and by choice of sparsity inducing prior distribution over parameters~\cite{mackay1992a}; 
c) Important input variables can be selected automatically by imposing structured sparsity on the NN;
d) Model parameters and prediction uncertainties can be quantified, which is particularly useful in decision making and safety-critical applications such as autonomous driving and structural health monitoring~\cite{huang2019}. 

Diverse Bayesian SYSID methods have been developed in the last decades. To name a few, a practical sparse Bayesian approach to state-space identification of nonlinear systems was proposed in~\cite{pan2016} in the context of biochemical networks. A Bayesian identification algorithm of nonlinear autoregressive exogenous (NARX) models using variational inference with a demonstration on the electroactive polymer was introduced in~\cite{Jacobs2018}. A framework for identifying the governing interactions and transition logics of subsystems in cyber-physical systems was presented in~\cite{yuan2019data} by using Bayesian inference and pre-defined basis functions. A variational expectation maximisation approach to SYSID when the data includes outliers was developed in~\cite{tianshi2020}.
Two approaches to SYSID using Bayesian networks were proposed in~\cite{chiuso2012}. The first one combines kernel-based stable spline and group Least Angle Regression while the other combines stable splines with the hyper-prior definition in a fully Bayesian model. However, this work did not discuss how to apply the Bayesian approach to the NN model. 
Another typical probabilistic nonparametric modelling method is the Gaussian process (GP), which can perform excellently for linear and nonlinear SYSID tasks, but suffers from the high computational burden for large datasets and cannot conduct input regressor selection efficiently.
Overall, specific to the use of NNs as a model form, little attention has been given to the identification of dynamic systems in a Bayesian framework.

Several approaches have been proposed to treat the NNs in a Bayesian manner, e.g., Laplace approximation, expectation propagation, variational inference, etc. Among these methods, the Laplace approximation is an approximated inference approach that can only represent local properties but is closer in efficiency to maximum a posteriori (MAP)~\cite{mackay1992a}. 
However, to update the posterior variance of parameters, the Laplace approximation method requires computing the inverse Hessian of log-likelihood, which is infeasible for large-scale NNs. 
To address this issue, a fast Hessian calculation technique was devised for convolutional NNs and successfully obtained an impressive image classification performance~\cite{zhou2019bayesnas}. 

In this paper, a companion technique for recurrent layers is also developed. 
\modify{Specifically, by unfolding a recurrent layer with its equivalent Fully Connected (FC) layers, the Hessian calculation of a recurrent layer can be treated as the Hessian calculation for the FC layers. Besides, since the Hessian is a diagonally dominant matrix~\cite{martens2015optimizing}, we develop a recursive and efficient method to compute the diagonal blocks of the Hessian matrix. Each block represents the Hessian diagonal entries of each layer and can be calculated recursively along with a backward propagation through time (BPTT) process. 
It should be noted that the Hessian is a necessity for the Laplace approximation method and can accelerate the optimisation process. In this paper, by incorporating the Hessian information to update the loss function, it can be observed that the proposed Bayesian approach can converge faster than the conventional optimisation method without capturing the Hessian information. 
Similar rapid convergence is also observed in the previous works related to the second-order optimisation methods~\cite{Hessian_fclayer,botev2020gauss,boyd2004convex,nocedal1980updating}.}

In addition, a sparse Bayesian approach is proposed to address several challenges for system identification based on deep neural networks, including overfitting the training data, the selection of input regressors, and the uncertainty quantification of model parameters. 
We will consider two typical DNNs, i.e., Multi-Layer Perceptron (MLP) and Long Short-Term Memory networks (LSTM).
The simulation error is adopted as the evaluation metric, which is a more challenging criterion compared with one-step-ahead prediction. The simulation error is equivalent to the $N$-step-ahead prediction error, with $N$ denoting a user-defined temporal horizon. 
In order to identify the system in a Bayesian framework, the group priors are introduced over network parameters to induce structured sparsity, and the Laplace approximation is used to approximate the intractable integral of the evidence. The main contributions of this paper have four folds: 

\begin{itemize}
	\item A practical iterative algorithm using
	Bayesian deep learning is proposed for SYSID. The first identification cycle of the algorithm is equivalent to the conventional sparse group lasso regularisation method.
	This algorithm can be used with both MLP and LSTM networks for linear and nonlinear processes.
	\item \modify{An efficient Hessian calculation method is proposed for each layer of DNNs, both for MLPs and RNNs. By calculating the block-diagonal entry of the Hessian, the proposed method can turn an intractable training/optimisation procedure into a tractable one.
	The sparsification process is also accelerated by recursively updating the Hessian information.}
	\item  
	The structured sparsity is incorporated in the Bayesian formulation of the identification problem to alleviate the overfitting issue and select the input regressor. 
	As a consequence, the number of hidden neurons in both MLP and LSTM networks can be significantly reduced. 
	\item The proposed algorithm achieves good and competitive simulation accuracy on five benchmark datasets. 
	The datasets of three linear processes are provided in the MATLAB System Identification Toolbox\footnote{\scriptsize\url{https://nl.mathworks.com/help/ident/examples.html}}, including the Hairdryer, Heat exchanger, and the Glass Tube manufacturing process. 
	The datasets of two nonlinear processes are provided on the Nonlinear System Identification Benchmarks website\footnote{\scriptsize\url{https://sites.google.com/view/nonlinear-benchmark/}}, including the Cascaded Tanks~\cite{ct_benchmark} and Coupled Electric Drives~\cite{ced_benchmark}.
\end{itemize}
The organisation of this paper is as follows.
Section~\ref{sec:preliminaries} formulates the identification problem using DNNs and introduces the Bayesian approach. 
Section~\ref{sec:algorithm_developement} presents the iterative procedure of the proposed sparse Bayesian learning algorithm and a recursive Hessian computation method. 
The illustration of structured sparsity regularisation, uncertainty quantification, and the proposed training algorithm are introduced in Section~\ref{sec:sysID}.
The identification results and detailed analysis are given in Section~\ref{sec:experiments}.
Section~\ref{sec:conclusion} concludes the paper.
A discussion on the limitations and future work are also included in Section~L of Appendix.

\vspace{-0.2cm}
\section{Preliminaries} \label{sec:preliminaries}
\vspace{-0.2cm}
\subsection{Problem formulation} \label{subsec:problem_formulation}
The chosen mathematical model structure is generated by training the network $\mathtt{Net}(\net,z)$, where $\net$ represents an array of the weights in the network and $z$ represents the input regressors of size $1 \times(l_y+l_u+1)$.
These are best defined by the prediction model:.
\begin{align}
    \hat{y}(t+1) &= \mathtt{Net}(\net, z(t+1),\epsilon) 
	\label{eq:nn_pred}
\end{align}
where $\epsilon$ represents the noise term. It should be noted that the $\epsilon$ can be in any distribution of exponential family. And the model parameter can be identified with a maximum likelihood method in the case of Gaussian noise (see Chapter 7.3 in~\cite{ljung1999system}).
The input regressor of the model is defined as a combination of lagged elements of the system input $u$ and outputs $y$. The input lag is denoted $l_u$ and output lag $l_y$, resulting in the expression $z(t+1) = [u(t+1), u(t), \cdots, u(t-l_u),y(t),y(t-1),\cdots, y(t-l_y)]^{\top}$.
\modifytwo{With such a network model, we aim to address two typical problems in SYSID. First, how to promote the sparsity of $\net$ to relieve the overfitting issue of DNNs? Second, how to select the input regressors automatically by identifying and removing the redundant features from $z(t+1)$?
}

The first DNN model considered is the LSTM network, a type of RNN. Benefiting from the advantages of processing sequential data and memorizing information, LSTM can also be applied straightforwardly for dynamic SYSID~\cite{delgado1995dynamic}. The BPTT method is used to train LSTM, where the network is unfolded in time and weights are updated based on an accumulation of gradients across time steps.
\begin{figure}[ht]
    \centering
    \includegraphics[width=\linewidth]{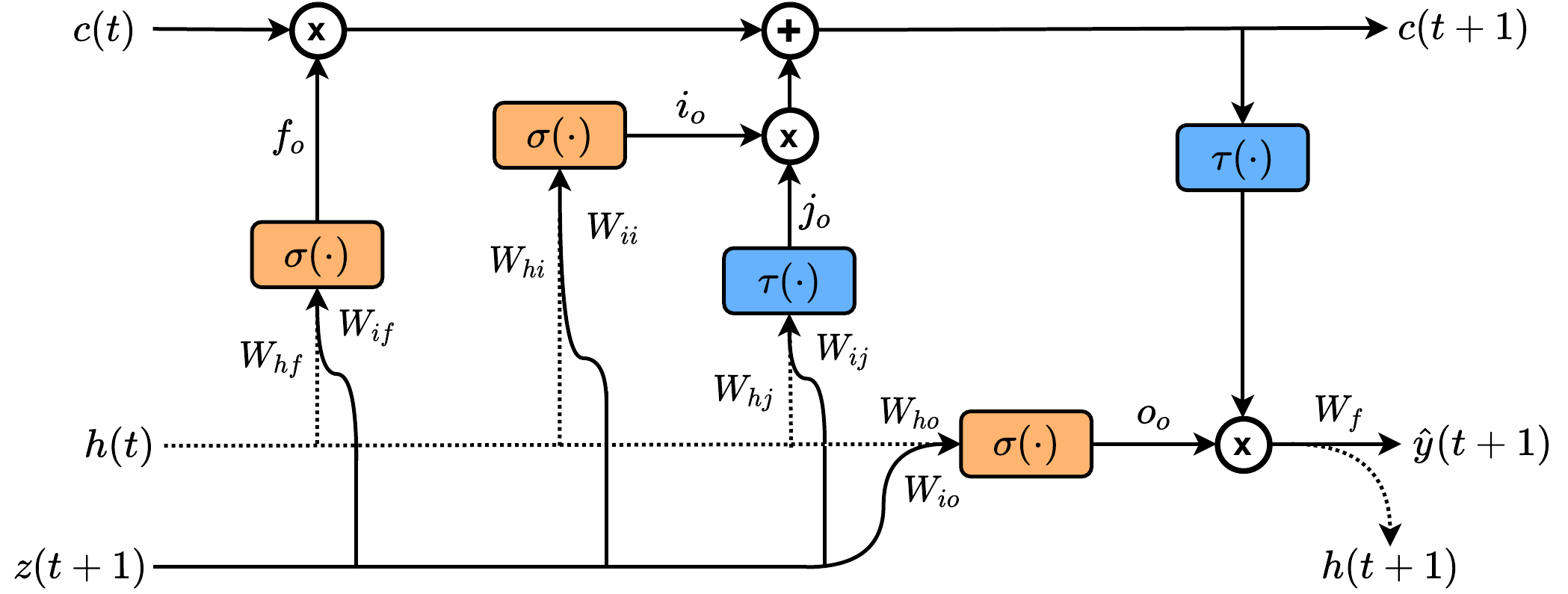}
    \caption{Single layer long short term memory network.}
    \label{fig:lstm}
\end{figure}

The second DNN model considered is the MLP, a type of feed-forward NN. Backpropagation with stochastic gradient descent algorithm and variations are often used to train a MLP network.
\begin{figure}[ht]
    \centering
    \includegraphics[width=\linewidth]{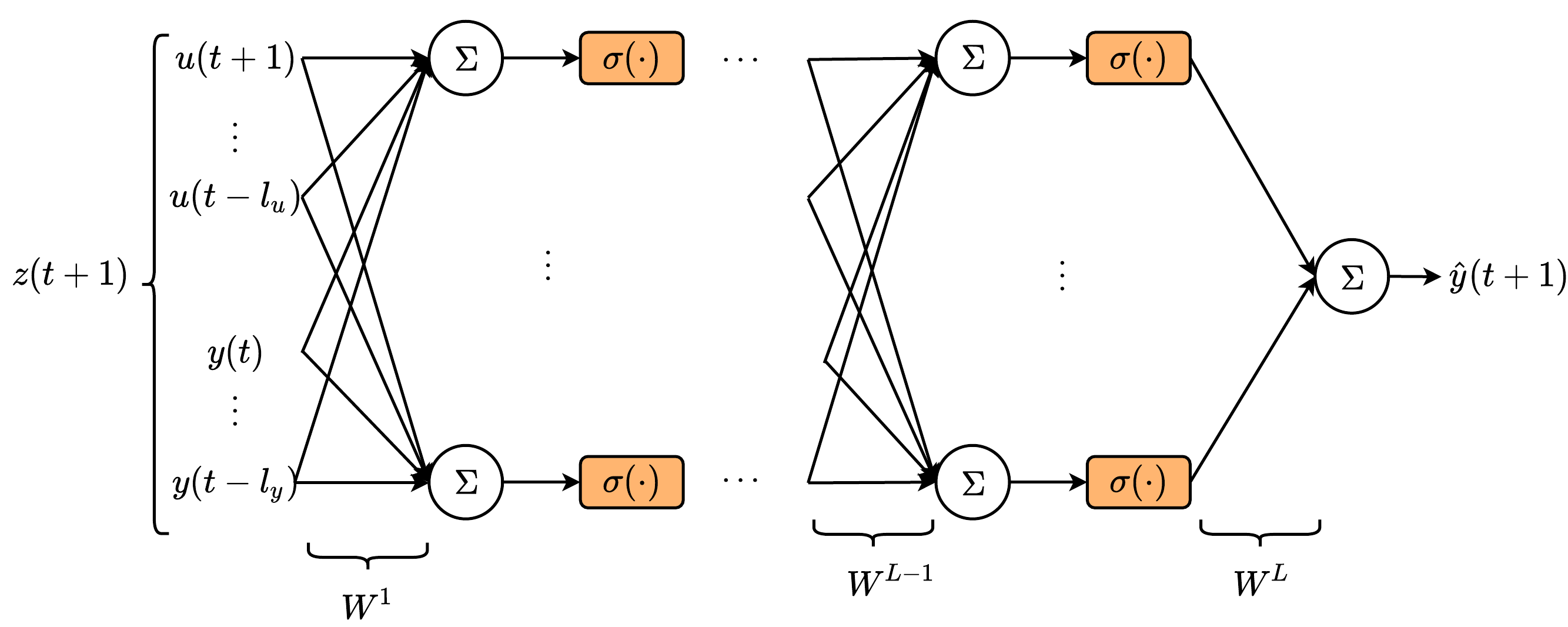}
    \caption{Multi-Layer Perceptron with $L$ layers}
    \label{fig:mlp}
\end{figure}  
\subsection{Learning in a Bayesian framework} \label{subsec:bayesian}

\modifytwo{Given a dataset $\data = (U, Y)$ where the input $U = [u(1), u(2), \ldots, u(T)]$ and output $Y = [y(1), y(2), \ldots, \\, y(T)]$ with $T$ referring to the number of samples, the posterior estimation for network weights $\net$ is given by Bayes' rule:}
\begin{equation}
    p(\net|\data,\hyp) = \frac{p(\data|\net,\hyp)p(\net,\hyp)}{p(\data|\hyp)} \label{eq:bayes_rule}
\end{equation}
$p(\data|\net,\hyp)$ designates the likelihood function, where $p(\net,\hyp)$ denotes the prior over the weights $\net$ and $p(\data|\hyp)$ is the evidence of the hypothesis $\hyp$ given $\data$. The hypothesis generally incorporates model and inference assumptions. For simplicity of notations, the hypothesis term is dropped in the rest of the paper. 
Assume that the likelihood function belongs to the exponential family: 
\vspace{-0.5cm}
\modifytwo{
\begin{align}
p(\data|\net,\theta) 
&= g(\theta) \exp\big\{\sum_{s=1}^S\eta_s(\net,\sigma)T_s(\sigma)+B(\net)\big\}
\nonumber \\
&= g(\theta) \exp\big\{- \mathbf{E}(\net,\theta)\big\} \label{eq:likelihood}
\end{align}
where $g(\cdot),T_s(\cdot), \eta_s(\cdot)$ and $B(\cdot)$ are known functions corresponding to a specific exponential family distribution,
$\theta$ is the parameter of the family, and
$\mathbf{E}(\net,\theta)$ denotes an energy function.}

\vspace{-0.2cm}
The prior probabilities $p(\net)$ takes a Gaussian relaxed variational form 
\begin{math}
    p(\net) \geq p(\net,\psi) 
    = \mathcal{N}(\net|0,\Psi)\ \phi(\psi) 
\end{math}
, where $\phi(\psi)$ represents the hyperprior probability of $\psi\triangleq [\psi_{11}^{1}, \cdots, \psi_{n_11}^{1}, \cdots, \psi_{n_1n_2}^{1}, \cdots, \psi_{11}^{L}, \cdots, \psi_{n_{L-1}n_L}^{L}] $
and $\Psi \triangleq \text{diag}(\psi)$.
With the principle of minimizing the misaligned probability mass, the hyper-parameter $\psi$ can be obtained by
\vspace{-0.3cm}
\begin{align}
    \hat{\psi} &= \argmini_{\psi\geq0} \int p(\data|\net,\theta) |p(\net)-p(\net,\psi)|\ d\net 
    \notag
    \\
    &= \argmax_{\psi\geq0} \int p(\data|\net,\theta) p(\net,\psi)\ d\net.
   \label{eq:max_evidence}
\end{align}
The resulting problem is known as a type II maximum likelihood. The integration is intractable and can be obtained by the Laplace approximation method, which is explained in detail in Section~\ref{subsec:laplace}. 

\vspace{-0.2cm}
\section{Sparse Bayesian Deep Learning} \label{sec:algorithm_developement}
\vspace{-0.2cm}
\subsection{Laplace approximation} \label{subsec:laplace}
The Laplace approximation method is adopted to compute the intractable integral in Eq.~\eqref{eq:max_evidence}.
$\mathbf{L}(\net,\theta)$ can be approximated by a second-order Taylor series expansion around a set of connection weights $\net^*$ with the operator $\Delta \net = \net-\net^{*}$, so we have
\begin{align}
     \mathbf{E} \approx \mathbf{L}(\net^{*},\theta)+\frac{1}{2}\Delta \net^T \mathbf{H} \Delta \net + \Delta \net^T  \mathbf{g}.
\end{align}
The resulting expression for the likelihood in a compact form is given by
\begin{align}
    p(\data|\net,\theta) &= \mathbf{A}\big(\net^{*},\theta\big)\exp\big\{-\frac{1}{2}\net^T \mathbf{H}\net - \net^T\mathbf{\hat{g}}\big\}\label{eq:laplace}\\
    \mathbf{\hat{g}}\big(\net^{*},\theta\big) &= \mathbf{g}\big(\net^{*},\theta\big) - \mathbf{H}\big(\net^{*},\theta\big)\net^{*} \nonumber
\end{align}
where $\mathbf{H}\big(\net^{*},\theta\big)$ and $\mathbf{g}\big(\net^{*},\theta\big)$ are respectively the Hessian and the gradient of the loss function $\mathbf{E}$ with respect to $\net$ at $\net^{*}$. Eq.~\eqref{eq:laplace} is obtained by grouping elements independent of the target variable $\net$ in $\mathbf{A}\big(\net^{*},\theta\big)$. The approximated likelihood is an exponent of a quadratic function corresponding to the Taylor series expansion of the energy loss. This form can be recast into a Gaussian function. 
In effect of the conjugacy of the prior and posterior, the posterior $p(\net|\data)$ is Gaussian given by:
 \modify{
\begin{align}
   p(\net|\data) &= \mathcal{N}(\net|\mu_\net,\Sigma_\net) 
    \label{eq:posterior_1}
    \\ 
    \mu_\net = \Sigma_\net \mathbf{\hat{g}},& \quad \Sigma_\net = \big[\mathbf{H} + \Psi^{-1}\big]^{-1}
    \label{eq:posterior}
\end{align}}
\hspace{-0.1cm}A more detailed derivation of the Laplace approximation is given in Appendix~A~\cite{zhou2022sparse}.
\vspace{-0.1cm}
\subsection{Evidence maximisation}
The evidence in Eq.~\eqref{eq:max_evidence} attempts to find the volume of the product $p(\data|\net,\theta) p(\net,\psi)$, which is Gaussian and proportional to the posterior. Thus, one can approximate the evidence as the volume around the most probable value (here posterior $\mu_\net$). 
\vspace{-0.3cm}
\begin{align}
    \hat{\psi} &=  \argmax_{\psi\geq0} \int p(\data|\net,\theta)p(\net|\psi)p(\psi) d\net \label{eq:evidence}\\
    &\approx \argmax_{\psi\geq0} \underbrace{p(\data|\mu_\net,\theta)}_\text{Best Fit Likelihood} \underbrace{p(\mu_\net|\psi)|\Sigma_\net|^\frac{1}{2}}_\text{Occam Factor}. \label{eq:evidence_result}
\end{align}
In David Mackay's words, the evidence is approximated by the product of the data likelihood given the most probable weights and the Occam factor~\cite{mackay1992a}. It can also be interpreted as a Riemann approximation of the evidence, where the best-fit likelihood represents the peak of the evidence. And the Occam's factor is the Gaussian curvature around the peak.

By realizing that the posterior mean $\mu_\net$ maximises  $p(\data|\net,\theta)p(\net|\psi)$, Eq.~\eqref{eq:evidence_result} can be rewritten into a joint maximisation in $\net$ and $\psi$. By applying the $-2\log(\cdot)$ operation, the evidence maximisation in Eq.~\eqref{eq:max_evidence} can be recast into a joint minimisation of an objective function $\mathcal{L}(\net,\psi,\theta)$ given by:
\vspace{-0.75cm}
{\small{
\begin{align}
    \mathcal{L}(\net,\psi,\theta) =& \net^T \mathbf{H}\net + 2\net^T\mathbf{\hat{g}} +\net^T\Psi^{-1} \net
    + \log|\Psi| \nonumber \\ &+\log|\mathbf{H}+\Psi^{-1}| -T\log(2\pi\theta) \label{eq:objective_function}
\end{align}
}}For a more thorough mathematical derivation that leads to Eq.~\eqref{eq:objective_function} and insight into the Laplace approximation, please refer to Appendix~A and~B~\cite{zhou2022sparse}.
\subsection{Convex-concave procedure}
The objective function in Eq.~\eqref{eq:objective_function} can be seen as a sum of a convex $u$ and concave $v$ functions in $\psi$ with:
\begin{align}
    u(\net,\psi) &= \net^T \mathbf{H}\net + 2\net^T\mathbf{\hat{g}} + \net^T \Psi^{-1} \net  \label{eq:u}\\
    v(\psi) &= \log|\Psi|+\log|\mathbf{H}+\Psi^{-1}|. \label{eq:v}
\end{align}
$ \net^T \Psi^{-1} \net$ is positive definite, since $\psi>0$. Thus, $u$ is convex in $\Psi$. $v$ can be reformulated as a log-determinant of an affine function of $\Psi$. By using the Schur complement determinant identity: 
\begin{align}
    |\Psi||\mathbf{H}+\Psi^{-1}| = \Bigg|\begin{matrix} \mathbf{H} & \\ & -\Psi\end{matrix}\Bigg|  = |\mathbf{H}||\mathbf{H}^{-1}+\Psi| \label{eq:det_schur_comp}
\end{align}
and taking the log of Eq.\eqref{eq:det_schur_comp},
\begin{align}
     \log|\Psi|+\log|\mathbf{H}+\Psi^{-1}| = \log|\mathbf{H}| + \log|\mathbf{H}^{-1}+\Psi| 
    \nonumber
\end{align}
one finds an equivalent expression of $v$ that is concave in $\Psi$.
The minimisation problem can therefore be reformulated as a convex-concave procedure (CCCP)~\cite{chen2014system}. $\net$ and $\psi$ are obtained by the iterative minimisation of Eq.~\eqref{eq:cccp1}-\eqref{eq:cccp2}.
\begin{align}
    \net(k+1) &= \argmini_{\net} u\big(\net,\psi(k)\big)  \label{eq:cccp1}\\
    \psi(k+1) &= \argmini_{\psi\geq0} u\big(\net(k+1),\psi\big) + \alpha(k) \cdot \psi 
    \label{eq:cccp2}
\end{align}
where $\alpha(k)= \nabla_\psi v\big(\psi(k)\big)^T$ is the gradient of $v$ evaluated at the current iterate $\psi(k)$. Using the chain rule, its analytical form is given by:
\begin{align}
    \alpha(k)=& \nabla_\psi \Big(\log|\Psi|+\log|\mathbf{H}+\Psi^{-1}|\Big)\Big|_{\psi=\psi(k)} \nonumber \\
    =& -\diag \Big(\Psi^{-1}(k)\Big) \odot \diag \Big(\big(\mathbf{H}+\Psi^{-1}(k)\big)^{-1}\Big) \nonumber \\ 
    &\odot \diag \Big(\Psi^{-1}(k)\Big) +\diag  \Big(\Psi^{-1}(k)\Big) 
    \label{eq:derive_alpha_1}
\end{align}
$\odot$ is the point-wise Hadamard product. Since $\Psi$ is a diagonal matrix, Eq.~\eqref{eq:cccp2} can be expressed per connection independently. 
With $\Sigma_{W^{l}_{ab}}(k)$ being the connection weight posterior variance, the analytical form for $\alpha$ is:
\begin{align}
    \Sigma_{\net}(k) &= \big(\mathbf{H}(k)+ \Psi(k)^{-1}\big)^{-1} \label{eq:posterior update}\\
    \alpha^{l}_{ab}(k) &= -\frac{\Sigma_{W^{l}_{ab}}(k)}{{\psi^{l}_{ab}}(k)^{2}} + \frac{1}{{\psi^{l}_{ab}}(k)}. \label{eq:alpha_update}
\end{align}
The optimisation step in Eq.~\eqref{eq:cccp2} for $\psi^l_{ab}$ becomes
\begin{align}
        \psi^l_{ab}(k+1) &= \argmini_{\psi\geq0} \frac{W^l_{ab}(k+1)^2}{\psi}+ \alpha^l_{ab}(k) \cdot \psi. \label{eq:psi_min}
\end{align}
\vspace{-0.1cm}By noting that
\begin{align}
    \frac{{W^l_{ab}}^2}{\psi}+ \alpha^l_{ab} \cdot \psi &\geq 2\Big|\sqrt{\alpha^l_{ab} \cdot} W^l_{ab}\Big|
\end{align}
the analytical solution is given by $$\psi^l_{ab}(k+1) = \frac{|W^l_{ab}(k+1)|}{\omega^l_{ab}(k)}$$ where $\omega^l_{ab}(k) = \sqrt{\alpha^l_{ab}(k)}$.

For the second part, finding $\net$ can be done with stochastic gradient descent on Eq.~\eqref{eq:cccp1}, which can be reformulated as
 the minimisation of a regularised loss function as follows:
\begin{align}
    \net(k+1) & = \argmini_{\net} \mathbf{L} = \argmini_{\net} \net^T \mathbf{H}\net + 2\net^T\mathbf{\hat{g}} \nonumber \\
              &  + \sum\limits_{l=1}^{L}\sum\limits_{a=1}^{n_{l-1}}\sum\limits_{b=1}^{n_l}||\omega^{l}_{ab} \cdot W^{l}_{ab}||_{l_1} \\
              & \approx \argmini_{W} \mathbf{E}(\cdot) + \lambda \sum\limits_{l=1}^{L} \rho(\omega^{l},W^{l}).
              \label{eq:loss}
\end{align}
$\mathbf{E}(\cdot)$ designates the energy loss function defined in Eq.~\eqref{eq:likelihood} and $\rho(\cdot)$ is the regularisation term.

\section{Hessian Computation} \label{sec:hessian}
\subsection{\modify{Definitions and properties of the Hessian}}
\label{subseec:hessian_introduction}

\modify{
For a DNN model, the Hessian of a weight matrix $\bw \in \mathbb{R}^{m\times n}$ is a square matrix of the second-order of partial derivatives of the loss function and can be formulated as:
\begin{equation}
    \label{eq:Hessian_definition}  
    {\rmH}_{\bw} = \begin{bmatrix}
    \pdv[2]{\mathbf{L}}{\vec \bw_1} &
\pdv{\mathbf{L}}{\vec \bw_1}{\vec \bw_2}   & \cdots & \pdv{\mathbf{L}}{\vec \bw_1}{\vec \bw_{mn}} \\
\pdv{\mathbf{L}}{\vec \bw_2}{\vec \bw_1} &\pdv[2]{\mathbf{L}}{\vec \bw_2}  & \cdots & \pdv{\mathbf{L}}{\vec \bw_2}{\vec \bw_n} \\
\vdots & \vdots & \ddots & \vdots \\
\pdv{\mathbf{L}}{\vec \bw_{mn}}{\vec \bw_1} &\pdv{\mathbf{L}}{\vec \bw_{mn}}{\vec \bw_2}  & \cdots & \pdv[2]{\mathbf{L}}{\vec \bw_{mn,mn}}
\end{bmatrix}
\end{equation}
So the $(i,j)$ element of ${\rmH}_{\bw}$ is:
\begin{equation}
    \label{eq:genreal_Hessian_element_chap7}  
    [{\rmH}_{\bw}]_{ij} = \pdv{\mathbf{L}}{\vec \bw_i}{\vec \bw_j}
\end{equation}
where $\vec \bw \in \mathbb {R}^{mn}$ is the vectorisation of the multi-dimensional weight matrix $\bw \in \mathbb {R}^{m\times n}$. 
As the dimension of the Hessian is the square of the number of unknown parameters (${\rmH}_{\bw} \in \mathbb{R}^{mn\times mn}$), it would be convenient to conduct the Hessian calculation by treating the matrix as a vector (the vectorisation operator is defined in Definition~1 of Appendix~D~\cite{zhou2022sparse}).}

\modify{
The Hessian information can benefit the training of DNNs from two aspects. 
First, it can accelerate the optimisation process. 
Several previous works on second-order optimisation methods (e.g., the Quasi-Newton methods~\cite{boyd2004convex,nocedal1980updating}) have presented that by incorporating the Hessian information in the optimisation process, the rapid convergence can be obtained without a lot of tuning work~\cite{Hessian_fclayer,botev2020gauss}. 
Besides,~\cite{martens2010deep} demonstrated that the Hessian information, also known as curvature matrix, could address the typical pathological curvature problem, where the first-order optimisation method often falls into the ``canyon" with large varying curvature because of their lack of ability to capture the curvature information~\cite{dauphin2014identifying,martens2015optimizing}. 
Second, the Hessian of the weight matrix is a required component for the Laplace approximation method. 
The Hessian is not only used to calculate the posterior distribution of weight parameters as in Eq.~\eqref{eq:posterior} but also used to update the loss function in each cycle (see Eq.~\eqref{eq:derive_alpha_1}-Eq.\eqref{eq:loss}.)}

\modify{
However, as the dimension of the Hessian is the square of the number of parameters, the calculation and storage of the Hessian for large-scale neural networks are infeasible considering their millions of parameters or more~\cite{botev2020gauss}. 
To address this problem, an efficient Hessian calculation method for a FC layer was in presented~\cite{Hessian_fclayer,botev2020gauss}. The proposed method therein can compute the diagonal blocks of the Hessian, where each block represents the diagonal entries of the Hessian in each layer and can be calculated recursively along with the back-propagation process using Kronecker products.
}

\modify{
Inspired by this method~\cite{Hessian_fclayer,botev2020gauss} and the diagonal dominant feature of the Hessian~\cite{martens2015optimizing}, we develop two efficient and recursive block-diagonal
calculation methods for the Hessian computation of FC layer and recurrent layer in this section. 
}


\subsection{\modify{Compute the Hessian of fully-connected layer}}
\label{subsec:hessian_fc_layer}
\modify{
Given a MLP as shown in Fig.~\ref{fig:mlp}, the output of the hidden layer $l$ can be calculated as:
\begin{align}
h^l = {\bw^l}a^{l-1} + b^l  , \ \ \
a^l = \sigma(h^l)
\label{eq:fc_layer_defintion}
\end{align}
where $b^l$ is the bias, $\sigma(\cdot)$ is the nonlinear activation function. The superscript $l$ denotes the layer index. $a^l$ and $h^l$ represent the activation value and the pre-activation value, respectively.}
\modify{
With these definitions, the proposed Hessian calculation method for a FC layer is summarised in Lemma~\ref{lemma:fc_hessian}. }
\begin{lemma}
\label{lemma:fc_hessian}
\modify{
For a fully-connected layer, given the activation function $\sigma(\cdot)$, the activation value $a^l,a^{l-1}$ and the pre-activation value $h^l$, the Hessian of the weight matrix $\generealweight^l$ is calculated recursively as follows:
\begin{equation}
    \label{eq:fc_Hessian_approximation_chap7}
    {{\rmH}}^l = \diag((a^l)^2 \otimes {H}^l)
\end{equation}
where $\otimes$ stands for Kronecker product. ${H}^l$ is the pre-activation Hessian and is updated as:
\begin{align}
\label{eq:fc_pre_Hessian_approximation_chap7}
        {H}^l = {(B^l)}^2 \odot \left(\left(({\bw^{l+1})^{\top}}\right)^2 {H}^{l+1}\right)+ {D}^l
\end{align}
in which ${B}^l$ and ${D}^l$ are defined as:
\begin{align}
\label{eq:fc_pre_Hessian_approximation_diagnoal} 
        {B}^l = \sigma'(h^l), \ \ \ {D}^l=\sigma''(h^l) \odot \frac{{\partial}L}{{\partial}a^l} 
\end{align}
where $\odot$ represents the element-wise multiplication.}

\modify{The above procedures can be calculated along with a backward propagation process. }
\end{lemma}
\modify{
\begin{remark}
It should be noted that Lemma~\ref{lemma:fc_hessian} is a modification of the Hessian calculation method proposed in~\cite{Hessian_fclayer}. 
The proposed approach can be computed more efficiently. Specifically, 
if the Hessian of a FC layer is computed as Eq.~\eqref{eq:fc_Hessian_approximation_chap7}-\eqref{eq:fc_pre_Hessian_approximation_diagnoal}, then the multiply accumulate operation (MACs) for the pre-activation Hessian $H$ and Hessian $\rmH$ could be reduced from $n(2m^2+2n^2+4mn+3m-1)$ to $n(2+4m)$ with $W \in \R^{m \times n}$ (e.g., if $n=100, m=100$, then the original method requires $107.97\times 10^{6}$ MACs compared with only $0.04\times 10^{6}$ MACs for the approximate method.).
Lemma~\ref{lemma:fc_hessian} is also the inspiration of the proposed Hessian calculation method for a recurrent layer. 
We will revisit Lemma~\ref{lemma:fc_hessian} many times in the following. 
\end{remark}
}

\subsection{\modify{Compute the Hessian of recurrent layer}}
\label{subsec:hessian_recurrent}
\modify{
The challenge of the Hessian calculation for a recurrent layer comes from the recurrent operation, where the weight matrices in a RNN cell will be revisited iteratively through time~\cite{martens2018kronecker}. This behaviour is different to the FC layer, where the weight matrices only join once through the operation in a forward propagation process.
Since a LSTM cell is a special form of the RNN,
for the convenience of explanation, we use a simplified RNN structure to illustrate the Hessian calculation process. 
As shown in Fig~\ref{fig:lstm_bptt}, we denote
$z(t)$, $h(t)$ and $y(t)$ as the input, hidden state and output of the time step $t$, respectively. The behaviour of this RNN layer can be described by
\begin{align}
    h(t) &= \sigma\left(\bar h(t)\right) = \sigma(W_i z(t) + W_h h(t - 1)) 
    \label{eq:rnn1}\\
    y(t) &= g(W_o h(t))
    \label{eq:rnn2}
\end{align}
where $W_i$, $W_h$ and $W_o$ represent the weight matrix of the input layer, hidden layer and output layer, respectively, and $\sigma$ is the activation function.
\begin{figure}[ht]
    \centering
    \includegraphics[scale = 0.46]{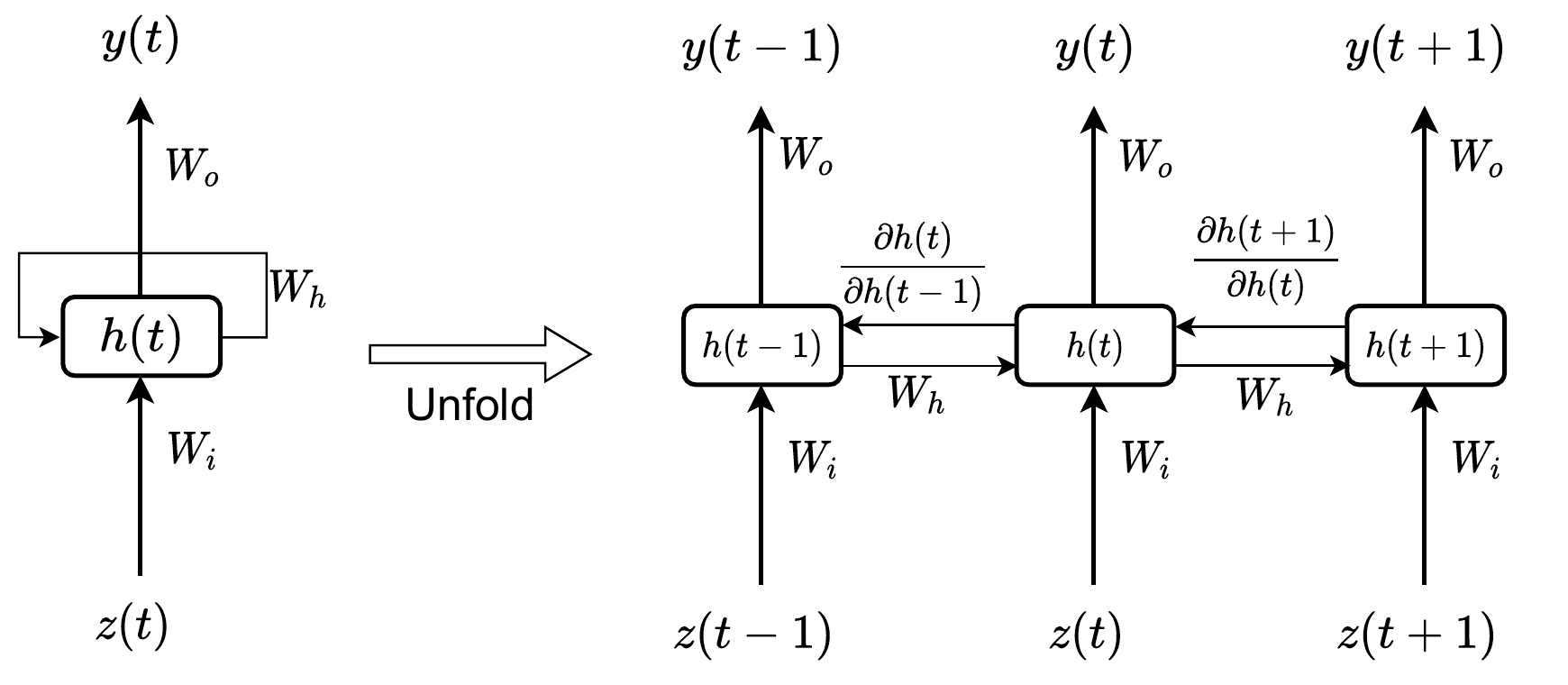}
    \caption{\modify{An unrolled RNN layer.}}
    \label{fig:lstm_bptt}
\end{figure}
It can be found that an unrolled RNN layer can be unfolded as several FC layer. Therefore the Hessian calculation for a recurrent layer can be regarded as the calculation of its equivalent FC layer. 
Inspired by Lemma~\ref{lemma:fc_hessian}, we propose a recursive and efficient method to compute the Hessian of a recurrent layer as follows.   
}

\begin{lemma}
\label{lemma:rnn_hessian}
\modify{
For a recurrent layer, given $\sigma$ representing the activation function,  $\tau$ representing backward propagation time horizon, $T$ representing the number of data samples, $z(t)$, $h(t)$ and $y(t)$ representing the input, hidden state and output at the time step $t$, $W_i$, $W_h$ and $W_o$ representing the weight matrix of the input layer, hidden layer and output layer, the Hessian of $W_i$, $W_h$, $W_o$ within the RNN layer is calculated as follows:
}

\modify{
1) The Hessian for $W_o$ is:
\begin{equation}
\label{eq:rnn_hessian_output_weight_chap7}
    \mathbf{H}_o = \frac{1}{T}\sum\limits^{T}_{t=1} \mathbf{H}^{\top}_o , \ \ \
     \mathbf{H}^{\top}_o = h(t)^2 \otimes H^{\top}_o
\end{equation}
where $H^{\top}_o$ is the pre-activation Hessian. 
}

\modify{
2) The Hessian for $W_h$ is:
\begin{equation} 
\label{eq:rnn_hessian_hidden_weight_1}
    \mathbf{H}_h = \mathbb{E}\left(\sum\limits_{t=1}^{T}\sum\limits_{j=\max(1,t - \tau + 1)}^{t} \mathbf{H}_h^{t,j}\right)
\end{equation}
\begin{equation}
\label{eq:rnn_hessian_hidden_weight_2}
    \mathbf{H}_h^{t,j} = h(j-1)^2 \otimes H_h^{t,j}
\end{equation}
where $\mathbf{H}_h^{t,j}$ and $H_h^{t,j}$ represent the Hessian and the pre-activation Hessian, respectively. In particular, 
\begin{math}
    H_h^{t,j} = B_h^2 \odot \left(\left(W_h^\top\right)^2 H^{t, j + 1}_h\right) + D_h
\end{math}, 
where
\begin{math}
    B_h = \sigma '(\bar h(j)), D_h = \sigma ''(\bar h(j)) \odot \frac{\partial{L}}{\partial{h(j)}}
\end{math}.
}

\modify{
3) The Hessian for $W_i$ is:
\begin{equation}
\label{eq:rnn_hessian_input_weight_1}
    \mathbf{H}_i = \mathbb{E}\left(\sum\limits_{t=1}^{T}\sum\limits_{k=\max(1,t - \tau + 1)}^{t} \mathbf{H}_i^{t,k}\right)
\end{equation}
\begin{equation}
\label{eq:rnn_hessian_input_weight_2}
    \mathbf{H}_i^{t,k} = \left(z(k)\right)^2 \otimes H_i^{t,k}
\end{equation}
where \begin{math}
    H_i^{t,k} = \prod^{t}_{j=k+1} B_i^2 \odot \left(\left((W_i)^\top\right)^2 H^{j-1, j}_i\right)
\end{math} with
\begin{math}
    B_i = \sigma '(\bar h(j))
\end{math}.
}

\modify{
The above procedures can be calculated along with a BPTT process.}
\end{lemma}
\modify{
It should be noted that Lemma~\ref{lemma:fc_hessian} and Lemma~\ref{lemma:rnn_hessian} elaborate the detailed procedures to calculate the Hessian with respect to a single data sample (i.e., $T=1$). If the number of data points is more than $1$ (i.e., $T>1$), the Hessian is calculated by averaging the Hessian of an individual data sample. The detailed proof of Lemma~\ref{lemma:fc_hessian} and Lemma~\ref{lemma:rnn_hessian} are given in Section~D.1 and Section~D.2 of Appendix~D~\cite{zhou2022sparse}.
}

\vspace{-0.2cm}
\section{Regularised Identification Algorithm} \label{sec:sysID}
\vspace{-0.2cm}
\subsection{Input regressor selection and structured sparsity regularisation}

As illustrated in Section~\ref{subsec:problem_formulation}, the input regressor is $z(t+1) = [u(t+1), u(t), \cdots, u(t-l_u),y(t),y(t-1),\cdots, y(t-l_y)]^{\top} \in \mathbb{R}^{l_u+l_y+1}$. The feature selection means identifying and removing the redundant features from $z(t+1)$.
The proposed method can select the input regressors automatically by imposing structured sparsity regularisation on the DNN.

\begin{figure}[ht]
    \centering
    \includegraphics[width=\linewidth]{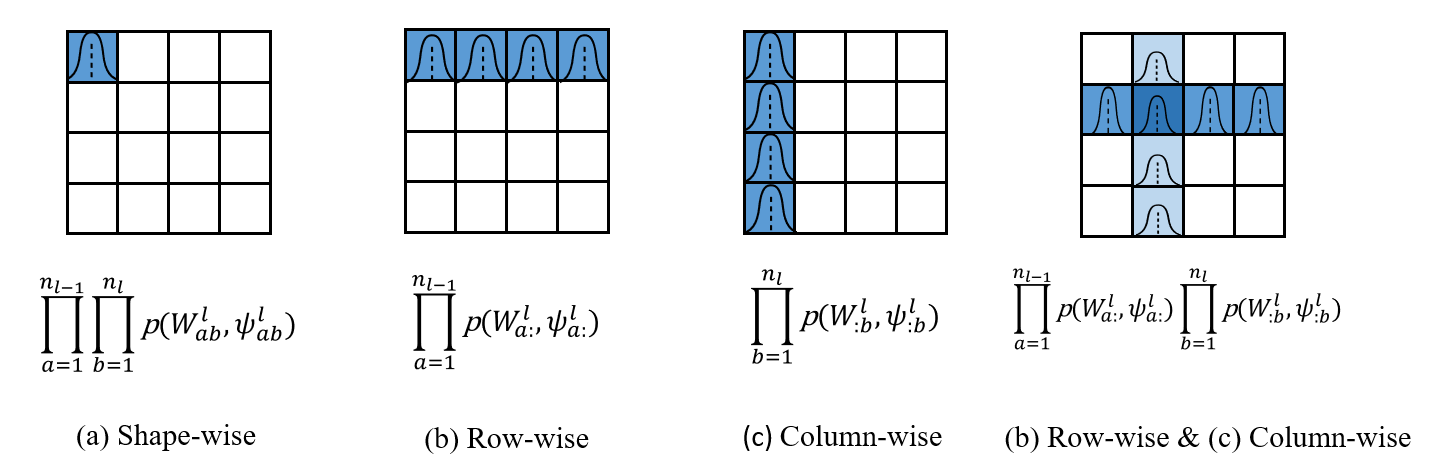}
    \caption{Priors for structured sparsity of weight matrices. \modify{$l$ is the layer index. a and b denote the row and column index of a 2-D weight matrix, respectively.}}
    \hspace{-0.2cm}
    \vspace{-0.6cm}
    \label{fig:sparsity}
\end{figure}

Specifically, the iterative procedure derived in Section~\ref{sec:algorithm_developement} includes an assumption on the independence and non-stationarity of connection weights, resulting in a shape-wise regularisation as shown in Fig.~\ref{fig:sparsity}(a). This drives the individual connection weight to $0$. 
In some applications, one may want to enforce more structured sparsity by pre-defining groups and re-expressing the regularisation term as a function of these groups~\cite{zhou2019bayesnas}. This paper uses a structured regularisation of rows and columns (Fig.~\ref{fig:sparsity}(b-d)). The benefits of such an approach, specific to this paper, are obtaining compact sparse models and the suppression of input nodes in $z$ that are deemed less pertinent without loss of accuracy. 
The reduction in the dimensionality of the input vector $z$ represents the selection of input regressors.

To extend this approach to the Bayesian framework, one has to revisit the prior formulation. The prior of a weight matrix is formulated based on the designated group of weight matrices (row or column or both). These groups are considered independent, but the connection weights of a specific group share the same prior Gaussian relaxation (see Fig.~\ref{fig:sparsity}(b-d)). This results in a slightly different iterative update rule for the identification algorithm. 

For each of the cases shown in Fig.~\ref{fig:sparsity}, the update rules for $\psi$, $\omega$ and the regularisation function $\rho$ are given in Appendix~C Table~C.1~\cite{zhou2022sparse}. There also stands more insight into how the adopted group priors slightly change the regularisation update rules on group Lasso regularisers~\cite{simon2013sparse}.

\vspace{-0.2cm}
\subsection{Algorithm}
\vspace{-0.2cm}
A pseudocode for the iterative procedure is given by Algorithm~\ref{algo:algorithm}. 
\begin{algorithm}[h]
    \caption{Identification Algorithm}
    \label{algo:algorithm}
    \begin{algorithmic}
        \REQUIRE
        \begin{itemize}
            \item Collect input-output data $u(t)$ and $y(t)$ for $t = 1,2,\cdots,T$.
            \item Arrange input regressors according to the chosen lags $l_u,l_y$. 
            \item Set regularisation parameter $\lambda$ (empirically tuned) and DNN pruning thresholds $\kappa_\psi, \kappa_W$ ($\approx 10^{-3}$). 
            \item Set the number of repeated experiments $M$, identification cycles $C_{\max}$ and the number of epochs in each cycle $E_{\max}$.
            \item Initialise hyper-parameters $\Psi(0)=\eye$ and $\omega(0)=\ones$.
        \end{itemize}

        \ENSURE 
        Return the set of connection weights $\net$
        \FOR{$m=1$ to $M$}
        \FOR{$c=1$ to $C_{\max}$}
        \FOR{$e=1$ to $E_{\max}$}
            \STATE $1$ Stochastic Gradient Descent with loss function ($\rho$ is defined in Table~C.1):
        		\vspace{-0.6cm}
        		{\small{
        		\begin{align}
        		\net(k+1) = \min_{\net} \mathbf{E}(\cdot) +  \lambda\sum\limits_{i=1}^{N} \pmb{\rho}(\omega^{l},W^{l})
        		\label{eq:loss_function}
        		\end{align}}}
        		\vspace{-0.4cm}
			\ENDFOR
        	\STATE $2$ Update $\alpha$ according to Eqs.~\eqref{eq:posterior update}-\eqref{eq:alpha_update}
        	\STATE $3$ Update $\psi$ and $\omega$ according to Table~C.1 
        	\STATE $4$ Dynamic pruning:\\
        \IF{$\psi^{l}_{ab}(k)<\kappa_{\psi}$ \textbf{or} $|W^{l}_{ab}(k)|<\kappa_{W}$}
        \STATE \textbf{prune} $W^{l}_{ab}(k)$
        \ENDIF
        \ENDFOR
        \STATE Simulate on the validation data and choose the model with the smallest root mean square error (RMSE).
        \ENDFOR
    \end{algorithmic}
    \label{algorithm}
\end{algorithm}
\begin{remark}\rm
We now give some clarifications on the definition of cycle and epoch in Algorithm~\ref{algo:algorithm}.
One identification “cycle” has $E_{\max}$ epochs.
One “epoch” refers to that the entire dataset is processed forward and backward by the NN for one time.
In the first identification cycle, the regularisation is conventional ($\omega(0)=\ones$). That is, the first obtained model is a sparse model corresponding to the conventional sparse group lasso regularisation method (as shown in~\eqref{eq:loss_function}), and sparser models are expected to result from the subsequent identification cycles.
\end{remark}
\begin{remark}\rm
The proposed algorithm shares the local convergence properties (local minima, saddle point) of the adopted stochastic gradient descent method~\cite{zhou2021local}. This is because the Laplace approximation is a local approximation method and includes an assumption on the unimodality of the posterior. However, the pruning and regularisation techniques introduced are heuristics that help speed up the algorithm and improve convergence and optimality. Nonetheless, the identification experiments are run multiple times with different initialisations. The identified model with the best simulation accuracy is chosen. 
\end{remark}
\vspace{-0.3cm}
\subsection{Making predictions with uncertainties} 
In the Bayesian procedure, predictions are made using the posterior predictive distribution, which is given by: 
\vspace{-0.5cm}
\begin{align}
    p(\hat{y}|z,\data) &= \int p(\hat{y}|\net,z)\ p(\net|\data)d\net. \label{eq:post_pred_dist} 
\end{align}
The first term of the integral is the likelihood of the prediction conditional on the network parameters. The second term is the inferred posterior distribution over the weights $\net$, which can be calculated as Eq~\eqref{eq:posterior_1}.
The expected value of the prediction is:
\begin{equation}
\begin{aligned}
    \mathbb{E}\big[\hat{y}\big] &= \int \hat{y}\ p(\hat{y}|z,\data)d\hat{y} \\
    &= \int \big( \int \hat{y}\ p(\hat{y}|\net,z)d\hat{y} \big) \ p(\net|\data)d\net \\
    &= \int\mathtt{Net}(\net,z)\ p(\net|\data)d\net
\end{aligned}
\end{equation}
Using the inferred posterior distribution over the weights, one can approximate this integral by the Monte-Carlo sampling method. An unbiased estimate of the prediction is given by the average predictions using $\net$ sampled by the posterior $M$ times as below: 
\vspace{-0.3cm}
\begin{align}
    \mathbf{\mu_{\hat{y}}} \approx \frac{1}{M} \sum_{m=1}^M \mathtt{Net}(\net(m),z). \label{eq:mean_mc}
\end{align}
In an analogous way, to estimate the variance in the posterior predictive distribution, the expected value $\mathbb{E}\big[\hat{y}^T\hat{y}\big]$ is analytically derived as follows:
\vspace{-0.3cm}
\modify{
\begin{equation}
\begin{aligned}
    \mathbb{E}\big[\hat{y}^T\hat{y}\big] &= \int \hat{y}^T\hat{y}\ p(\hat{y}|z,\data)d\hat{y} \\ 
    &= \int\big( \int \hat{y}^T\hat{y}\ p(\hat{y}|\net,z)d\hat{y}\big)\ p(\net|\data)d\net \\ 
    &= \int\big( \zeta  +\mathtt{Net}(\net,z)^2\big)\ p(\net|\data)d\net. \label{eq:expectation_variance3}
\end{aligned}
\end{equation}
}
\modify{where $\zeta$ represents the aleatoric uncertainty.} An unbiased estimate of the variance is given by Monte-Carlo integration methods~\cite{gal_thesis}, with M samples from the inferred posterior distribution of $\net$ as below: 
\modify{
\begin{align}
    \mathbf{\Sigma_{\hat{y}}} &\approx \zeta +\frac{1}{M} \sum_{m=1}^M \mathtt{Net}(\net(m),z)^2-\mathbf{\mu_{\hat{y}}}^T\mathbf{\mu_{\hat{y}}}. \label{eq:std_mc}
\end{align}
}
\hspace{-0.2cm}This variance (Eq.~\eqref{eq:std_mc}) represents the model uncertainty in the prediction. It is approximated by the sum of an aleatoric uncertainty and epistemic uncertainty. The aleatoric uncertainty is generally known to be irreducible corresponding to the noise covariance of the measurement and is generally incorporated in the likelihood form~\cite{gal_thesis}.
\modify{
For example, if the likelihood is given as Gaussian distribution, then $\zeta$ should represent the noise variance.
}
The epistemic uncertainty corresponds to the model's uncertainty in a prediction that is often called reducible uncertainty~\cite{gal_thesis} and grows when moving away from the training data~\cite{bayesian_case}.

\section{Experiments} \label{sec:experiments}
An overview of the simulation accuracy of our experiments compared with other methods can be found in Tables~E.1--E.2 in Appendix~E~\cite{zhou2022sparse}.
The code to reproduce the experimental results is open-sourced and available online\footnote{\scriptsize\url{https://github.com/hongpengzhou/Deep-Bayesian-System-Identification}}.

\subsection{Dataset and experiment setup}
This section is to summarise the identification experiments of three linear processes and two nonlinear processes using the proposed algorithm. For linear systems, the identification procedure is repeated $M=20$ times with $C_{\max}=6$ identification cycles. For nonlinear systems, the identification is also repeated $M=20$ times but with $C_{\max}=10$ identification cycles each. Table \ref{tab:models_sysID} provides a summary of the model structure used for identification as well as the mean, standard deviation, and minimum validation $\mathtt{RMSE}$ of the $M$ best-identified models and the percentage of sparse parameters in the best-identified model.
In Appendix~F--Appendix~J~\cite{zhou2022sparse}, the benchmarks are described more thoroughly with sparsity plots, simulation plots, and posterior predictive mean and uncertainty plots corresponding to the best-identified model.
\begin{table*}[ht]
	\caption{Models are trained to identify linear and nonlinear processes with validation information}
	\label{tab:models_sysID}
	\begin{center}
		 \resizebox{1.00\textwidth}{!}{
		\begin{tabular}{|c|c|c||c|c|c|c|}
			\hline
			\textbf{Process-Model}  & \textbf{Layers-Units}  & \textbf{Lags} & $\mathtt{RMSE}_{\mathtt{val}}$ ($\mu \pm \sigma$)  & $\mathtt{RMSE}_{\mathtt{val}}$ ($\mathtt{min}$)  & \textbf{Sparsity} & \textbf{Appendix}   \\ \hline \hline
			Hairdryer-MLP           & 1 - 50                 & 5     & 0.074 $\pm$ 0.0005  & 0.073 &  88.1\% & Appendix F  \\ \hline
  			Hairdryer-LSTM          & 1 - 10                 & 5     & 0.093 $\pm$ 0.0166  & 0.081 &  93.5\% & Appendix F  \\ \hline
			Heat Exchanger-MLP      & 1 - 50                 & 150   & 0.086 $\pm$ 0.0002  & 0.086 &  99.3\% & Appendix G  \\ \hline
			Heat Exchanger-LSTM     & 1 - 10                 & 150   & 0.114 $\pm$ 0.0299  & 0.088 &  96.4\% & Appendix G  \\ \hline
			GT Manufacturing-MLP    & 1 - 50                 & 5     & 0.660 $\pm$ 0.0013  & 0.657 &  97.8\% & Appendix H   \\ \hline
			GT Manufacturing-LSTM   & 1 - 10                 & 5     & 0.671 $\pm$ 0.0019  & 0.669 &  99.0\% & Appendix H   \\ \hline \hline
			Cascaded Tanks-MLP      & 3 - 10                 & 20    & 0.428 $\pm$ 0.1032  & 0.257 &  84.5\% & Appendix I   \\ \hline
  			Cascaded Tanks-LSTM     & 1 - 50                 & 20    & 0.500 $\pm$ 0.1012  & 0.362 &  60.3\% & Appendix I   \\ \hline
			CED-MLP                 & 2 - 50                 & 10    
			& \begin{tabular}[c]{@{}c@{}} 0.187 $\pm$ 0.0285 \\ 0.134 $\pm$ 0.0192 \end{tabular} 
			& \begin{tabular}[c]{@{}c@{}} 0.149 \\ 0.120 \end{tabular}                             &  78.4\%  & Appendix J \\ \hline
			CED-LSTM                & 1 - 10                 & 10 
			& \begin{tabular}[c]{@{}c@{}} 0.155 $\pm$ 0.0257 \\ 0.126 $\pm$ 0.0201 \end{tabular}
			& \begin{tabular}[c]{@{}c@{}} 0.121 \\ 0.097 \end{tabular}                             &  72.8\%  & Appendix J \\ \hline
		\end{tabular}
	}
	\end{center}
\end{table*}

Three linear processes are identified, the Hairdryer, 
Heat exchanger and Glass Tube (GT) manufacturing process. 
The datasets of these processes are provided by Matlab in the corresponding tutorials (\url{https://nl.mathworks.com/help/ident/examples.html}) on linear SYSID. The chosen best validated models are compared to the methods used in the corresponding tutorials. Additional model structures used for the identification of the Hairdryer are taken from Chapter 17.3 of~\cite{ljung_sysID} and run in Matlab. The comparisons are in Appendix~E Table~E.1~\cite{zhou2022sparse}.

Two nonlinear processes, the Cascaded Tanks~\cite{ct_benchmark}, Coupled Electric Drives~\cite{ced_benchmark} are also identified. 
Information and datasets of these benchmarks are compiled on the web page of the Workshop on Nonlinear System Identification Benchmarks (\url{https://sites.google.com/view/nonlinear-benchmark/}). The cascaded tank system is a fluid level control system consisting of two tanks with free outlets fed by a water pump~\cite{ct_benchmark}. The fluid levels of these two tanks are adjusted by the input signal that controls the water pump. 
The coupled electric drive is a system that drives a pulley by controlling a flexible belt. Two electric motors provide the driving force, and the spring is used to fix the pulley. 
A more detailed description of the system and datasets of these benchmarks are compiled on the web page of the Nonlinear System Identification Benchmarks.
The models with the best validation performance are compared with the best models obtained using conventional NN methods for multiple experiments ($M=20$) and the previous works in the literature for every benchmark in Appendix~E Table~E.2~\cite{zhou2022sparse}.

\subsection{Analysis of experimental results} \label{sec:analysis}
In this subsection, the results will be discussed and analysed concerning the claims made on sparsity, uncertainty quantification, and simulation accuracy.

\begin{figure}[ht]
    \centering
    \includegraphics[width=\linewidth]{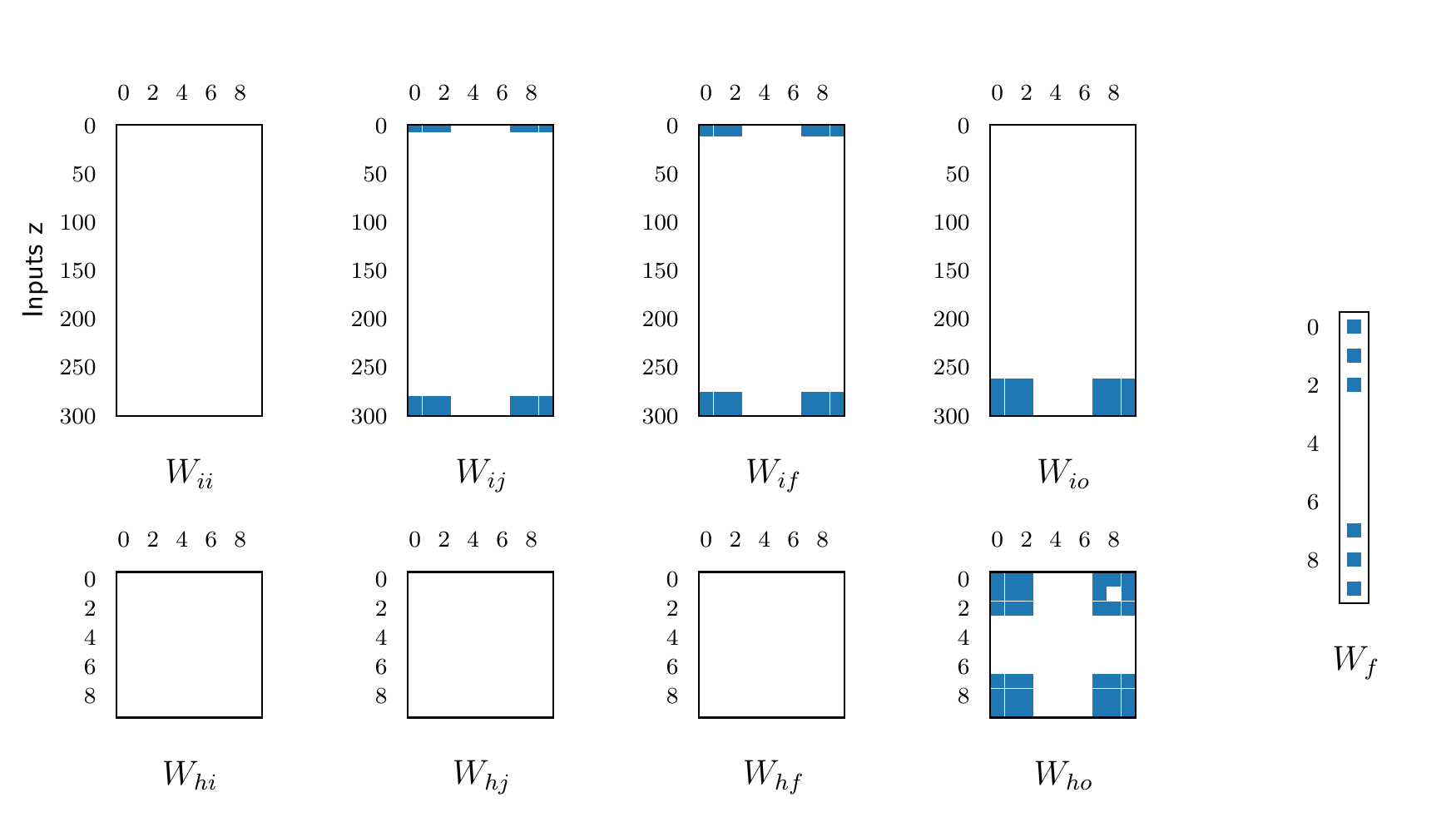}
    \caption{Sparsity plot of the identified LSTM for Heat Exchanger. (\textit{Blue represents non-pruned connection weights})}
    \label{fig:hex_lstm_sparse}
\end{figure}

\textbf{Sparsity:}
In most cases, the obtained networks are sparse models with structured sparsity. For example, Fig~\ref{fig:hex_lstm_sparse} is a sparsity plot of the Heat Exchanger identified LSTM model, \modifyfive{where half of the weight matrices related to hidden states are removed from the input gate ($W_{hi},W_{hj}$) and forget gate ($W_{hf}$)}.

According to Table~\ref{tab:models_sysID}, sparsity is more prominent in the identified linear systems than in nonlinear systems. This demonstrates that the nonlinearity that the data exhibits requires a higher complexity than in the linear case.

Starting with the linear systems, one can note that structured sparsity induces a recognised transport delay in the Heat Exchanger MLP and LSTM models, which characterises this system. Furthermore, the LSTM models for linear systems have complete operators pruned. This means that the cell state can be well regulated with fewer parameters than imposed by the initialised model structure in the Heat Exchanger case. Similar behaviour is seen across linear benchmarks.

Structured sparsity is also observed in the identified networks for nonlinear systems (Table~\ref{tab:models_sysID}). 
In addition to that, similar to LSTM models identified for linear systems, a lot of parameters involving the hidden states are pruned. A possible explanation for this behaviour is that the hidden states of LSTM units attempt to retain short-term information from the time series that is also available as lagged elements in the input regressor. The simulation result further shows that the input regressor with lagged elements can achieve better simulation performance for a LSTM model (see Appendix~E Table~E.1-E.2~\cite{zhou2022sparse}). 
Another observation related to the structured sparsity regularisation is the effect of input regressor selection. 
As shown in Fig.~H.2a in Appendix~H~\cite{zhou2022sparse}, the number of input regressors is reduced from $40$ to $2$ after applying the sparse Bayesian algorithm with row-wise and column-wise prior as shown in Appendix~C Table~C.1. 
The redundant input regressors are also identified for other benchmarks and removed from the NN, thereby reducing the model complexity.

We also find that DNNs (MLP models) with more hidden layers are necessities to approximate complex systems. For example, the optimal MLP model of the nonlinear cascaded tanks system includes three hidden layers. In contrast, the optimal MLP model of the linear hairdryer system only has one hidden layer. 
The MLP model with only one hidden layer and $10$ hidden neurons is also applied for the cascaded tank system. However, the obtained simulation error is around $0.663$, which is worse than the MLP model with three hidden layers ($0.257$ as in Table~\ref{tab:models_sysID}). 
It should also be noted that although the number of hidden layers of the MLP models is not reduced in these experiments, the number of hidden neurons is reduced, which provides a more suitable network structure for different systems. 
For example, as shown in Fig.~G.2a, the number of hidden neurons in the MLP model of the Heat Exchanger model is reduced from $50$ to $7$.

\textbf{Predictive distributions:} 
The posterior predictive distributions for each model result from the forward propagation of the parameters' posterior uncertainty obtained with the estimation data. Hence, if the validation data holds information that the model does not learn from the estimation data, the posterior predictive distribution could spread a bigger range of predictions~\cite{bayesian_case}.

\begin{figure}[ht]
    \centering
    \includegraphics[width=0.8\linewidth]{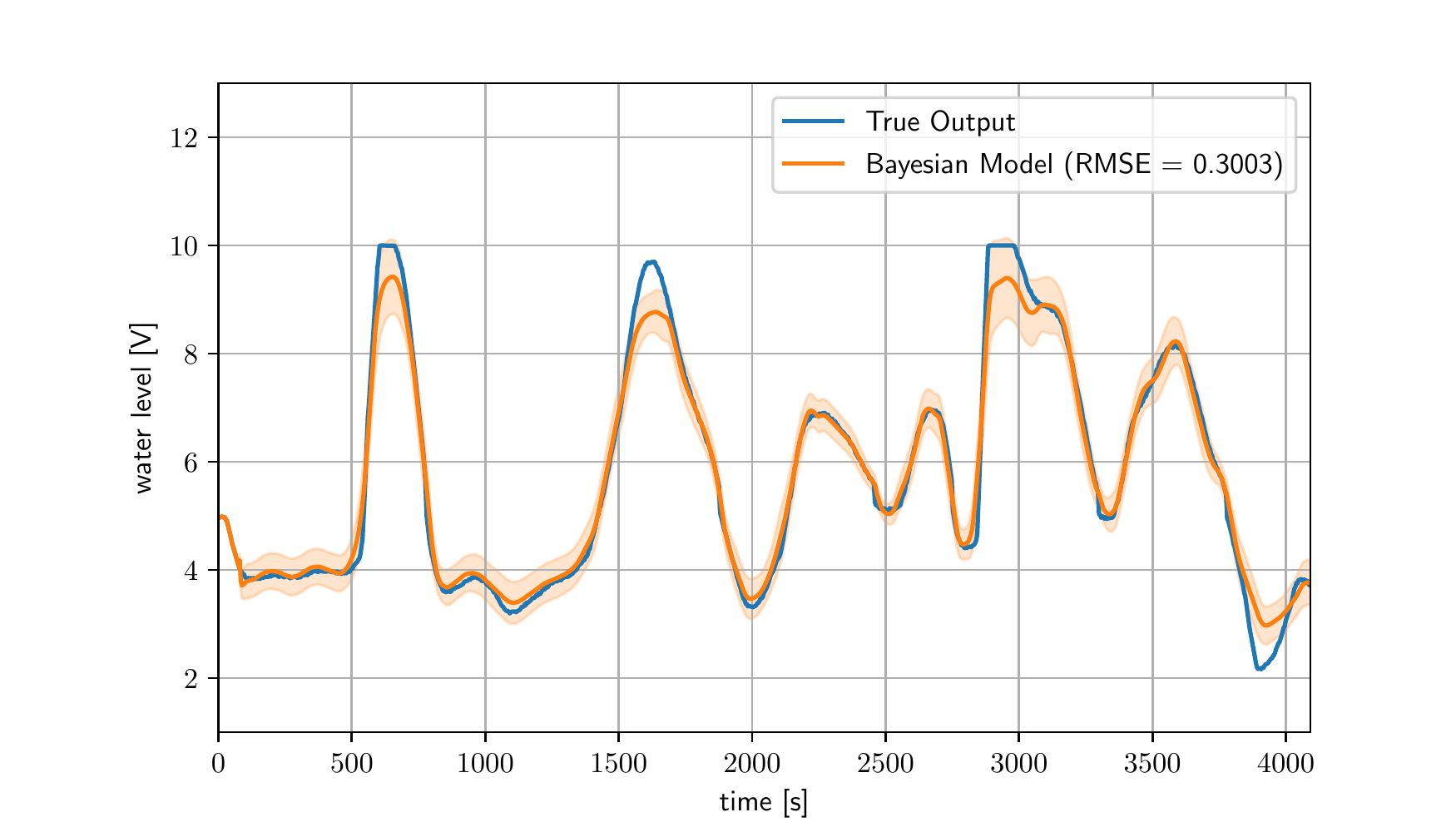}
    \caption{Posterior predictive $\mu_y \pm 2\sigma_y$ of the identified LSTM on Cascaded Tanks benchmark.}
    \label{fig:ct_post_pred}
\end{figure}

In some cases, the identified models show an unevenly distributed predictive uncertainty related to nonlinearities or disturbances characteristics of the process and regions where the model can be improved. Fig.~\ref{fig:ct_post_pred} shows that the identified model for Cascaded Tanks makes less robust predictions when overflow occurs. The Heat Exchanger shows evenly distributed predictions with uncertainty possibly coming from the ambient temperature disturbance. Furthermore, the model type also affects the predictive distribution. Examples include the LSTM models identified for the Glass Tube Manufacturing Process and Cascaded Tanks. In these benchmarks, the identified MLP model provides more robust predictions than the identified LSTM model. 

\textbf{Free run simulation performance:} 
The free run simulation is a good measure of the model's approximation ability to represent a dynamic process by propagating a model's prediction error while forecasting. 
In this paper, we select the simulation error as the evaluation metric.
It is important to note that, for the studied linear processes, a non-regularised LSTM performs worse when compared to other identification methods. This supports the previous concerns made on using LSTM for the identification of linear systems. The Bayesian MLP model outperforms the Bayesian LSTM model in most presented applications except for the Coupled Electric Drive.

Table~\ref{tab:models_sysID} displays the mean and standard deviation of the validation simulation errors and the minimum corresponding to the best-chosen model. The minimum is seen to fall close to the range of one standard deviation from the mean. In addition, the variance of validation errors for linear systems is overall less than for nonlinear systems. A possible explanation is that the added complexity in identifying nonlinear processes and the usage of more complex nonlinear structures (LSTM in this case), increases the likelihood of convergence towards saddle points. This is mainly because the Laplace method adopted is a local approximation of the evidence, which is a limitation of the proposed method and justifies running the identification experiment $M$ times.

The Bayesian approach to the identification of each benchmark constitutes an improvement over the conventional MLP and LSTM methods in simulation errors and pushes these methods to perform competitively with other literature (see Table~E.1--E.2). Besides, we also make a comparison with the well-known Gaussian process (GP) in machine learning by exploring different kernels (i.e., squared exponential kernel, rational quadratic kernel).
However, the GP method cannot perform input regressor selection efficiently, i.e., all regressors flow into the black box model without any priority. 

\section{Conclusion} \label{sec:conclusion}
In this paper, we combined sparse Bayesian learning and deep learning for SYSID. An iterative procedure for dynamic SYSID has been derived and evaluated with datasets of three linear and two nonlinear dynamic processes. 
The Bayesian approach in this paper has used the Laplace approximation to approximate the model evidence/marginal likelihood. The structured sparsity regularisation has been implemented on NNs by enforcing group-sparsity inducing priors. 
\modify{
An efficient Hessian calculation method for the recurrent layer has been presented by calculating the block-diagonal value of the Hessian.
}
The identified models for the dynamic systems are sparse models that have contributed to input regressor selection and performed competitively with other used SYSID methods in a free run simulation setting. 
In addition, uncertainties in the inferred predictions and connection weights have been quantified using Monte-Carlo integration methods.

\section{Acknowledgement}
We thank Jens Kober and Martijn Wisse from Delft University of Technology for helpful discussions. This work is supported by China Scholarship Council No.201706120017 (H.Zhou) and Huawei (W.Pan).

\bibliographystyle{plain}

\begin{thebibliography}{10}

\bibitem{botev2020gauss}
A.Botev.
\newblock {\em The {Gauss-Newton} matrix for Deep Learning models and its
  applications}.
\newblock PhD thesis, UCL (University College London), 2020.

\bibitem{Hessian_fclayer}
H.Ritter A.Botev and D.Barber.
\newblock Practical {Gauss-Newton} optimisation for deep learning.
\newblock In {\em Proceedings of the 34th International Conference on Machine
  Learning - Volume 70}, ICML'17, pages 557--565. JMLR.org, 2017.

\bibitem{chiuso2012}
A.Chiuso and G.Pillonetto.
\newblock A {B}ayesian approach to sparse dynamic network identification.
\newblock {\em Automatica}, 48(8):1553--1565, 2012.

\bibitem{delgado1995dynamic}
A.Delgado, C.Kambhampati, and K.Warwick.
\newblock Dynamic recurrent neural network for system identification and
  control.
\newblock {\em IEE Proceedings-Control Theory and Applications},
  142(4):307--314, 1995.

\bibitem{bayesian_case}
A.G.Wilson.
\newblock The case for {B}ayesian deep learning.
\newblock {\em arXiv preprint arXiv:2001.10995}, 2020.

\bibitem{SVENSSON2017189}
A.Svensson and T.B.Schön.
\newblock A flexible state–space model for learning nonlinear dynamical
  systems.
\newblock {\em Automatica}, 80:189--199, 2017.

\bibitem{van2012system}
P.van den Hof.
\newblock System identification-data-driven modelling of dynamic systems.
\newblock {\em Lecture notes, Eindhoven University of Technology}, 2012.

\bibitem{gedon2021deep}
D.Gedon, N.Wahlstr{\"o}m, T.B.Sch{\"o}n, and L.Ljung.
\newblock Deep state space models for nonlinear system identification.
\newblock {\em IFAC-PapersOnLine}, 54(7):481--486, 2021.

\bibitem{mackay1992a}
D.MacKay.
\newblock {B}ayesian interpolation.
\newblock {\em Neural Computation}, 4(3):415--447, 1992.

\bibitem{weber2021non}
D.Weber and C.G{\"u}hmann.
\newblock Non-autoregressive vs autoregressive neural networks for system
  identification.
\newblock {\em arXiv preprint arXiv:2105.02027}, 2021.

\bibitem{lecun2015deep}
Y.~LeCun et~al.
\newblock Deep learning.
\newblock {\em Nature}, 521(7553):436--444, 2015.

\bibitem{beintema2021nonlinear}
G.Beintema, R.Toth, and M.Schoukens.
\newblock Nonlinear state-space identification using deep encoder networks.
\newblock In {\em Learning for Dynamics and Control}, pages 241--250. PMLR,
  2021.

\bibitem{castellano2000variable}
G.Castellano and A.M.Fanelli.
\newblock Variable selection using neural-network models.
\newblock {\em Neurocomputing}, 31(1-4):1--13, 2000.

\bibitem{ced_other4}
H.V.H.Ayala, L.F. da~Cruz, and R.Z.Freire et~al.
\newblock Cascaded free search differential evolution applied to nonlinear
  system identification based on correlation functions and neural networks.
\newblock In {\em Proceedings of the 2014 IEEE Symposium on Computational
  Intelligence in Control and Automation (CICA)}, pages 1--7, Dec 2014.

\bibitem{zhou2019bayesnas}
H.Zhou, M.Yang, and J.Wang et~al.
\newblock {B}ayesnas: A {B}ayesian approach for neural architecture search.
\newblock In {\em Proceedings of the 36th International Conference on Machine
  Learning}, pages 7603--7613. PMLR, 2019.

\bibitem{matlab_dry}
The~MathWorks Incorporation.
\newblock Estimating simple models from real laboratory process data.
\newblock \url{https://nl.mathworks.com/help/ident/examples.html}.
\newblock Accessed: 2020-11-25.

\bibitem{matlab_hex}
The~MathWorks Incorporation.
\newblock Estimating transfer function models for a heat exchanger.
\newblock \url{https://nl.mathworks.com/help/ident/examples.html}.
\newblock Accessed: 2020-11-25.

\bibitem{matlab_gt}
The~MathWorks Incorporation.
\newblock Glass tube manufacturing process.
\newblock \url{https://nl.mathworks.com/help/ident/examples.html}.
\newblock Accessed: 2020-11-25.

\bibitem{lstm_matlab_toolbox_2021}
The~MathWorks Incorporation.
\newblock System identification toolbox.
\newblock
  \url{https://nl.mathworks.com/help/ident/ug/use-lstm-for-linear-system-identification.html#responsive_offcanvas}.
\newblock Accessed: 2020-11-25.

\bibitem{belz2017}
J.Belz, T.M{\"u}nker, and T.O.Heinz et~al.
\newblock Automatic modeling with local model networks for benchmark processes.
\newblock {\em IFAC-PapersOnLine}, 50(1):470--475, 2017.

\bibitem{martens2010deep}
J.Martens et~al.
\newblock Deep learning via {Hessian}-free optimization.
\newblock In {\em ICML}, volume~27, pages 735--742, 2010.

\bibitem{martens2018kronecker}
J.Martens, J.Ba, and M.Johnson.
\newblock Kronecker-factored curvature approximations for recurrent neural
  networks.
\newblock In {\em International Conference on Learning Representations}, 2018.

\bibitem{martens2015optimizing}
J.Martens and R.Grosse.
\newblock Optimizing neural networks with kronecker-factored approximate
  curvature.
\newblock In {\em International conference on machine learning}, pages
  2408--2417. PMLR, 2015.

\bibitem{nocedal1980updating}
J.Nocedal.
\newblock Updating quasi-{Newton} matrices with limited storage.
\newblock {\em Mathematics of computation}, 35(151):773--782, 1980.

\bibitem{snoek2015scalable}
J.Snoek, O.Rippel, and K.Swersky et~al.
\newblock Scalable {B}ayesian optimization using deep neural networks.
\newblock In {\em Proceedings of the 2015 International Conference on Machine
  Learning}, pages 2171--2180. PMLR, 2015.

\bibitem{willard2020integrating}
J.Willard, X.Jia, and S.Xu et~al.
\newblock Integrating physics-based modeling with machine learning: A survey.
\newblock {\em arXiv preprint arXiv:2003.04919}, 2020.

\bibitem{NN_Identification}
M.Stinchcombe K.Hornik and H.White.
\newblock Multilayer feedforward networks are universal approximators.
\newblock {\em Neural networks}, 2(5):359--366, 1989.

\bibitem{ljung1999system}
L.Ljung.
\newblock System identification.
\newblock {\em Wiley encyclopedia of electrical and electronics engineering},
  pages 1--19, 1999.

\bibitem{ljung_sysID}
L.Ljung.
\newblock {\em System Identification: (2nd Ed.): Theory for the User 2nd Ed.}
\newblock Prentice Hall PTR, USA, 1999.

\bibitem{ljung_lstm}
L.Ljung, C.Andersson, and K.Tiels et~al.
\newblock Deep learning and system identification.
\newblock {\em IFAC-PapersOnLine}, 53(2):1175--1181, 2020.

\bibitem{bw_other1}
A.Janot M.Brunot and F.Carrillo.
\newblock Continuous-time nonlinear systems identification with output error
  method based on derivative-free optimisation.
\newblock {\em IFAC-PapersOnLine}, 50(1):464--469, 2017.

\bibitem{forgione2021continuous}
M.Forgione and D.Piga.
\newblock Continuous-time system identification with neural networks: model
  structures and fitting criteria.
\newblock {\em European Journal of Control}, 59:69--81, 2021.

\bibitem{NN_Identification2}
M.Leshno, V.Y.Lin, and A.Pinkus et~al.
\newblock Multilayer feedforward networks with a nonpolynomial activation
  function can approximate any function.
\newblock {\em Neural Networks}, 6(6):861--867, 1993.

\bibitem{tianshi2020}
M.Lindfors and T.Chen.
\newblock Regularized lti system identification in the presence of outliers: A
  variational em approach.
\newblock {\em Automatica}, 121:109152, 2020.

\bibitem{ced_other1}
M.Scarpiniti, D.Comminiello, and R.Parisi et~al.
\newblock Novel cascade spline architectures for the identification of
  nonlinear systems.
\newblock {\em IEEE Transactions on Circuits and Systems I: Regular Papers},
  62(7):1825--1835, July 2015.

\bibitem{schoukens2016modeling}
M.Schoukens and F.G.Scheiwe.
\newblock Modeling nonlinear systems using a {Volterra} feedback model.
\newblock In {\em Proceedings of the 2016 Workshop on Nonlinear System
  Identification Benchmarks}, 2016.

\bibitem{ct_benchmark}
M.Schoukens, P.Mattson, and T.Wigren et~al.
\newblock Cascaded tanks benchmark combining soft and hard nonlinearities.
\newblock In {\em Workshop on nonlinear system identification benchmarks},
  pages 20--23, 2016.

\bibitem{zhou2021local}
R.Ge M.Zhou and C.Jin.
\newblock A local convergence theory for mildly over-parameterized two-layer
  neural network.
\newblock {\em arXiv preprint arXiv:2102.02410}, 2021.

\bibitem{simon2013sparse}
N.Simon, J.Friedman, and T.Hastie et~al.
\newblock A sparse-group lasso.
\newblock {\em Journal of Computational and Graphical Statistics},
  22(2):231--245, 2013.

\bibitem{paszke2019pytorch}
Adam Paszke, Sam Gross, Francisco Massa, Adam Lerer, James Bradbury, Gregory
  Chanan, Trevor Killeen, Zeming Lin, Natalia Gimelshein, Luca Antiga, et~al.
\newblock Pytorch: An imperative style, high-performance deep learning library.
\newblock {\em Advances in neural information processing systems}, 32, 2019.

\bibitem{MATTSSON201840}
D.Zachariah P.Mattsson and P.Stoica.
\newblock Identification of cascade water tanks using a pwarx model.
\newblock {\em Mechanical systems and signal processing}, 106:40--48, 2018.

\bibitem{kim2020Auto}
R.Karagoz and K.Batselier.
\newblock Nonlinear system identification with regularized tensor network
  b-splines.
\newblock {\em Automatica}, 122:109300, 2020.

\bibitem{ct_relan}
R.Rishi, K.Tiels, and A.Marconato et~al.
\newblock An unstructured flexible nonlinear model for the cascaded water-tanks
  benchmark.
\newblock {\em IFAC-PapersOnLine}, 50(1):452--457, 2017.
\newblock 20th IFAC World Congress.

\bibitem{sb_ced_other7}
F.~{Sabahi} and M.~R. {Akbarzadeh-T}.
\newblock Extended fuzzy logic: Sets and systems.
\newblock {\em IEEE Transactions on Fuzzy Systems}, 24(3):530--543, June 2016.

\bibitem{boyd2004convex}
S.Boyd and L.Vandenberghe.
\newblock {\em Convex optimization}.
\newblock Cambridge university press, 2004.

\bibitem{ced_bw_schoukens}
S.C.Nechita, R.Toth, and D.Khandelwal et~al.
\newblock Toolbox for discovering dynamic system relations via tag guided
  genetic programming, 2020.

\bibitem{sjoberg1994}
J.~Sjöberg, H.~Hjalmarsson, and L.~Ljung.
\newblock Neural networks in system identification.
\newblock {\em IFAC Proceedings Volumes}, 27(8):359 -- 382, 1994.

\bibitem{lawrence1998size}
C.L.Giles S.Lawrence and A.C.Tsoi.
\newblock What size neural network gives optimal generalization? convergence
  properties of backpropagation.
\newblock Technical report, 1998.

\bibitem{brunton2016discovering}
J.L.Proctor S.L.Brunton and J.N.Kutz.
\newblock Discovering governing equations from data by sparse identification of
  nonlinear dynamical systems.
\newblock {\em Proceedings of the National Academy of Sciences},
  113(15):3932--3937, 2016.

\bibitem{chen2014system}
T.Chen, M.S.Andersen, and L.Ljung et~al.
\newblock System identification via sparse multiple kernel-based regularization
  using sequential convex optimization techniques.
\newblock {\em IEEE Transactions on Automatic Control}, 59(11):2933--2945,
  2014.

\bibitem{gao2020recalling}
T.Gao, X.Gong, and K.Zhang et~al.
\newblock A recalling-enhanced recurrent neural network: Conjugate gradient
  learning algorithm and its convergence analysis.
\newblock {\em Information Sciences}, 519:273--288, 2020.

\bibitem{ced_benchmark}
T.Wigren and M.Schoukens.
\newblock {\em Coupled electric drives data set and reference models}.
\newblock Department of Information Technology, Uppsala Universitet, 2017.

\bibitem{gt_benchmark}
G.Bastin V.Wertz and M.Haest.
\newblock Identification of a glass tube drawing bench.
\newblock {\em IFAC Proceedings Volumes}, 20(5):333--338, 1987.

\bibitem{pan2016}
W.Pan and et~al Y.Yuan, J.S.Gon{\c{c}}alves.
\newblock A sparse {B}ayesian approach to the identification of nonlinear
  state-space systems.
\newblock {\em IEEE Transactions on Automatic Control}, 61(1):182--187, 2016.

\bibitem{pan2016online}
F.Menolascina W.Pan and G.B.Stan.
\newblock Online model selection for synthetic gene networks.
\newblock In {\em Proceedings of the 55th IEEE Conference on Decision and
  Control (CDC)}, pages 776--782. IEEE, 2016.

\bibitem{Jacobs2018}
W.R.Jacobs and et~al T.Baldacchino, T.Dodd.
\newblock Sparse {B}ayesian nonlinear system identification using variational
  inference.
\newblock {\em IEEE Transactions on Automatic Control}, 63(12):4172--4187,
  2018.

\bibitem{ma2019bayesian}
X.Ma, A.R.Triki, and M.Berman et~al.
\newblock A {B}ayesian optimization framework for neural network compression.
\newblock In {\em Proceedings of the IEEE/CVF International Conference on
  Computer Vision}, pages 10274--10283, 2019.

\bibitem{dauphin2014identifying}
Y.Dauphin, R.Pascanu, C.Gulcehre, K.Cho, S.Ganguli, and Y.Bengio.
\newblock Identifying and attacking the saddle point problem in
  high-dimensional non-convex optimization.
\newblock {\em arXiv preprint arXiv:1406.2572}, 2014.

\bibitem{gal_thesis}
Y.Gal.
\newblock {\em Uncertainty in Deep Learning}.
\newblock PhD thesis, University of Cambridge, 2016.

\bibitem{huang2019}
Y.Huang, C.Shao, and B.Wu et~al.
\newblock State-of-the-art review on {B}ayesian inference in structural system
  identification and damage assessment.
\newblock {\em Advances in Structural Engineering}, 22(6):1329--1351, 2019.

\bibitem{yuan2019data}
Y.Yuan, X.Tang, and W.Zhou et~al.
\newblock Data driven discovery of cyber physical systems.
\newblock {\em Nature Communications}, 10(1):1--9, 2019.

\bibitem{zhou2022sparse}
Hongpeng Zhou, Chahine Ibrahim, Wei~Xing Zheng, and Wei Pan.
\newblock Sparse bayesian deep learning for dynamic system identification.
\newblock {\em arXiv preprint arXiv:2107.12910}, 2022.

\end{thebibliography}

\newpage
\appendix
\begin{appendices}
\vspace{-0.2cm}
\section{The Laplace Approximation} \label{ap:laplace}
In this section, a more detailed mathematical description of the Laplace approximation adopted is made.
The likelihood is given by an exponential family distribution as~\eqref{eq:likelihood} in Section \ref{subsec:laplace}.
As a typical exponential family distribution, Gaussian distribution will be adopted in the following to show how to derive the iterative procedures.
The formulation of the likelihood with Gaussian distribution is rewritten as:
\begin{align*}
    p(\data|\net,\sigma^2) &= \prod_{t=1}^T \mathcal{N}(y(t)|\mathtt{Net}(z(t),\net),\sigma^2) \notag\\
    &=  (2\pi\sigma^2)^{-\frac{T}{2}} \exp\Big\{- \mathbf{L}(\net,\sigma^2)\Big\}. 
\end{align*}
$\mathbf{L}(\net,\sigma^2)$ is denoted as the energy function, or the loss of the network given the data $\data$. It is given by
\begin{math}
    \mathbf{L}(\net,\sigma^2) = \frac{1}{2\sigma^2}\sum_{t=1}^T(y(t)-\text{Net}(z(t),\net))^2 
\end{math}.
The expression $\mathtt{Net(\cdot)}$ is the resulting network nonlinear map. To compute the intractable integral for the evidence, the energy function can be expanded by using a second-order Taylor series expansion around $\net^*$ as follows:
\begin{equation}
\label{eq:taylor}
\begin{aligned}
    \mathbf{E}(\net,\sigma^2) &\approx \mathbf{L}(\net^*,\sigma^2) + (\net-\net^*)^T  \mathbf{g}(\net^*,\sigma^2) 
    \\ &
    + \frac{1}{2}(\net-\net^*)^T \mathbf{H}(\net^*,\sigma^2) (\net-\net^*) 
\end{aligned} \notag
\end{equation}
where $\mathbf{g}= \nabla \mathbf{E}(\net,\sigma^2)|_{\net^*}$ and  $\mathbf{H} = \nabla\nabla \mathbf{E}(\net,\sigma^2)|_{\net^*} $. 
To ease notation, we use $\mathbf{g}$ and $\mathbf{H}$ to denote $\mathbf{E}(\net^*,\sigma^2)$ and $\mathbf{H}(\net^*,\sigma^2)$, respectively.
The quadratic expression is also adopted among Trust-Region methods, where a region is defined around the current iterate connection weights $\net$ and the expansion is considered as a reasonable local representation of the loss function. With this expansion, the likelihood function becomes:
\vspace{-0.5cm}
\begin{equation}
\label{eq:compact_form}
\begin{aligned}
p(\data|\net,\sigma^2) &\approx \mathbf{A} \cdot  \exp\Big\{-\Big( \frac{1}{2}\net^T \mathbf{H} \net 
+ \net^T \mathbf{\hat{g}}\Big)\Big\} 
\end{aligned}
\end{equation}
with
$
    \mathbf{\hat{g}} = \mathbf{g} - \mathbf{H}\net^{*},
    \mathbf{A} = (2\pi\sigma^2)^{-\frac{T}{2}} \cdot\exp\Big\{-\Big( \frac{1}{2}{\net^*}^T \mathbf{H} \net^* - {\net^*}^T \mathbf{g} +\mathbf{L}(\net^*,\sigma^2)\Big) \Big\} 
$.
A Gaussian form can be easily recuperated from \eqref{eq:compact_form} by completing the square in the exponent. Before that, we define the following quantities:$\mathbf{B} = \exp\Big\{\frac{1}{2}\mathbf{\hat{g}}^T \mathbf{H} \mathbf{\hat{g}} \Big\}$ and $\mathbf{C} = (2\pi)^{\frac{\kappa}{2}}|\mathbf{H}|^\frac{1}{2}$
then~\eqref{eq:compact_form} can be reformulated as:
\vspace{-0.5cm}
\begin{equation}
\begin{aligned}
     p(\data|\net,\sigma^2) \approx& \mathbf{A} \cdot  \exp\Big\{-\Big( \frac{1}{2}\net^T \mathbf{H} \net + \net^T \mathbf{\hat{g}}\Big)\Big\} \nonumber \\
     &\mathbf{\cdot} \exp\Big\{\frac{1}{2}\mathbf{\hat{g}}^T \mathbf{H} \mathbf{\hat{g}} - \frac{1}{2}\mathbf{\hat{g}}^T \mathbf{H} \mathbf{\hat{g}} \Big\}\\
     =&  \mathbf{A} \cdot \mathbf{B} 
     \cdot \exp\Big\{-\Big( \frac{1}{2}\net^T \mathbf{H} \net + \net^T \mathbf{\hat{g}}+ \\ &\frac{1}{2}\mathbf{\hat{g}}^T \mathbf{H} \mathbf{\hat{g}}\Big)\Big\}\\
     =& \mathbf{A} \cdot \mathbf{B} \cdot \mathbf{C} \cdot \mathcal{N}(\net|\hat{\net},\mathbf{H}^{-1})
\end{aligned}
\end{equation}
where $\hat{\net}=-\mathbf{H}^{-1} \mathbf{\hat{g}}$.
Given a Gaussian likelihood and a Gaussian prior defined in Section \ref{subsec:bayesian}, by effect of the conjugacy rule, the posterior is also Gaussian $\mathcal{N}(\mu_\net,\Sigma_\net)$ with
\begin{math}
        \mu_\net =  \big[\mathbf{H} + \Psi^{-1}\big]^{-1} \mathbf{\hat{g}} 
\end{math}
and
\begin{math}
        \Sigma_\net = \big[\mathbf{H} + \Psi^{-1}\big]^{-1}. 
\end{math}

\vspace{-0.2cm}
\section{Evidence Maximisation} \label{ap:evidence_maximisation}
\vspace{-0.2cm}
This section provides a mathematical proof of the derived objective function. Starting from the maximisation in Eq.~\eqref{eq:max_evidence}, the likelihood and prior are replaced by their expressions in the preliminary Section \ref{subsec:bayesian}. Then we have
\vspace{-0.3cm}
\begin{equation}
\begin{aligned}
     &\int p(\data|\net,\sigma^2)p(\net|\psi)p(\psi)\ d\net \label{eq:evidence2}\\
     =& \int \mathbf{A} \cdot  \exp\Big\{-\Big( \frac{1}{2}\net^T \mathbf{H} \net + \net^T \mathbf{\hat{g}}\Big)\Big\} \\
     &\cdot  \mathcal{N}(\net|0,\Psi) \cdot \phi(\psi)\ d\net\\
     =&  \frac{\mathbf{A}}{(2\pi)^{K/2} |\Psi|^{\frac{1}{2}}} \cdot \int \exp\Big\{-\Big( \frac{1}{2}\net^T \mathbf{H} \net + \net^T \mathbf{\hat{g}}\Big)\Big\} \\
     &\cdot \exp\Big\{-\Big( \frac{1}{2}\net^T \mathbf{\Psi^{-1}} \net\Big)\Big\}  \ d\net \cdot \prod_{l=1}^{L}\prod_{a=1}^{n^{{l-1}}} \prod_{b=1}^{n^{{l}}} \phi(\psi_{ab}^l) \\
     =& \frac{\mathbf{A}}{(2\pi)^{K/2} |\Psi|^{\frac{1}{2}}} \cdot \int \exp\Big\{-\mathcal{E}(\net,\sigma^2)\Big\}\ d\net 
     \\ & \cdot \prod_{l=1}^{L}\prod_{a=1}^{n^{{l-1}}} \prod_{b=1}^{n^{{l}}}\phi(\psi_{ab}^l) \notag
\end{aligned}
\end{equation}
where
\begin{math}
    \mathcal{E}(\net,\sigma^2) = \frac{1}{2}\net^T \mathbf{H} \net + \net^T \mathbf{\hat{g}} +  \frac{1}{2}\net^T \mathbf{\Psi^{-1}} \net
\end{math}.
The integral in Eq.~\eqref{eq:evidence2} is the integral of the product $p(\data|\net,\sigma^2)p(\net|\psi)$, which is proportional to the posterior $p(\net|\data,\psi)$. In most applications, the posterior peaks with respect to the prior, and the evidence can be approximated by the posterior volume. This approximation is analogous to the usage of the Laplace approximation of the posterior in David MacKay's Bayesian framework \cite{mackay1992a}. That is, 
\vspace{-0.3cm}
\begin{equation}
\begin{aligned}
     &\int p(\data|\net,\sigma^2)p(\net|\psi)\ d\net 
     \label{eq:evidence3}
     \\ \approx& p(\data|\mu_\net,\sigma^2)p(\mu_\net|\psi) \cdot |\Sigma_\net|^{\frac{1}{2}} \cdot (2\pi)^{\kappa/2} \\
    \iff & \int \exp\Big\{-\mathcal{E}(\net,\sigma^2)\Big\}\ d\net
    \\ \approx & \exp\Big\{-\mathcal{E}(\mu_\net,\sigma^2)\Big\} \cdot |\Sigma_\net|^{\frac{1}{2}} \cdot (2\pi)^{\kappa/2}
\end{aligned}
\end{equation}
where
\vspace{-0.3cm}
\begin{equation}
\label{eq:net_min}
\begin{aligned}
    \mathcal{E}(\mu_\net,\sigma^2) &= \frac{1}{2}\mu_\net^T \mathbf{H} \mu_\net + \mu_\net^T \mathbf{\hat{g}} +  \frac{1}{2}\mu_\net^T \mathbf{\Psi^{-1}} \mu_\net \\
    &=\min_\net \frac{1}{2}\net^T \mathbf{H} \net + \net^T \mathbf{\hat{g}} +  \frac{1}{2}\net^T \mathbf{\Psi^{-1}} \net. 
\end{aligned}
\end{equation}
Hence, the maximisation of the evidence becomes the maximisation below:
\begin{equation}
\label{eq:max_evidence2}
\begin{aligned}
    \psi =& \argmax_{\psi>0} \frac{\mathbf{A}}{(2\pi)^{\kappa/2} |\Psi|^{\frac{1}{2}}} \cdot \exp\big\{-\mathcal{E}(\mu_\net,\sigma^2)\big\} 
    \\
    &\cdot |\Sigma_\net|^{\frac{1}{2}}  \cdot \prod_{l=1}^{L}\prod_{a=1}^{n^{{l-1}}} \prod_{b=1}^{n^{{l}}}\phi(\psi_{ab}^l). \notag
\end{aligned}
\end{equation}
By applying a $-2\log(\cdot)$ operation and using Eq.~\eqref{eq:net_min}, one obtains
\begin{equation}
\begin{aligned}
    \psi =& \argmini_{\psi>0} -2\log\Bigg[ \frac{\mathbf{A}\cdot\exp\big\{-\mathcal{E}(\mu_\net,\sigma^2)\big\}}{(2\pi)^{\kappa/2} 
    |\Psi|^{\frac{1}{2}}}  \\
	&\cdot |\Sigma_\net|^{\frac{1}{2}}  \cdot \prod_{l=1}^{L}\prod_{a=1}^{n^{{l-1}}} \prod_{b=1}^{n^{{l}}}\phi(\psi_{ab}^l)\Bigg] \\
    =& \argmini_{\psi>0} -2\log(\mathbf{A})  + \mathcal{E}(\mu_\net,\sigma^2) + \log|\Psi| \\
    &- \log|\Sigma_\net| -2 \sum_{l=1}^{L}\sum_{a=1}^{n^{{l-1}}} \sum_{b=1}^{n^{{l}}}\log(\phi(\psi_{ab}^l))\\
    \net,\psi =& \argmini_{\net,\psi>0} \frac{1}{2}\net^T \mathbf{H} \net + \net^T \mathbf{\hat{g}} +  \frac{1}{2}\net^T \mathbf{\Psi^{-1}} \net \\
    &
    +\log|\Psi| + \log|\mathbf{H} + \Psi^{-1}| 
    -2\log(\mathbf{A}) \\ & 
    -2 \sum_{l=1}^{L}\sum_{a=1}^{n^{{l-1}}} \sum_{b=1}^{n^{{l}}}\log(\phi(\psi_{ab}^l)). \nonumber
\end{aligned}
\end{equation}
Since the hyperprior $\phi(\psi)$ is a non-informative hyper-prior, the final objective function is given by:
\begin{equation}
\label{eq:obj_function2}
\begin{aligned}
    \mathcal{L}(\net,\psi,\sigma^2) =& \net^T \mathbf{H}\net + 2\net^T\mathbf{\hat{g}} +\net^T\Psi^{-1} \net \\
    &
    + \log|\Psi|+\log|\mathbf{H}+\Psi^{-1}| -T\log(2\pi\sigma^2). \notag
\end{aligned}
\end{equation}

\section{Regularisation Update Rules} \label{ap:structured_sparsity}
To enforce a group regularisation on network parameters, the prior formulation is revisited, which does not alter the derivation of the loss function. The difference is with the optimisation step for $\psi$. Parameters in the same row share the prior uncertainty parameter ${\psi}^l_{a:}$ and in the same column the prior uncertainty ${\psi}^l_{b:}$. 
The optimisation step in Eq.~\eqref{eq:psi_min} for $\psi^l_{:b}$, the prior width shared among the connection weights in the same column, becomes
\vspace{-0.3cm}
\begin{equation}
\begin{aligned}
  \psi^l_{:b}(k+1) = \argmini_{\psi\geq0} & \sum\limits_{b=1}^{n^l}\frac{W^l_{:b}(k+1)^T W^l_{:b}(k+1)}{\psi} \\
  & + \alpha^l_{:b}(k) \cdot \psi 
\end{aligned}
\end{equation}
where $\alpha^l_{:b} = \sum\limits_{a=1}^{n_{l-1}}{\alpha^l_{ab}(k)}.$ By noting that
\begin{align}
    \sum\limits_{b=1}^{n^l}\frac{{W^l_{:b}}^T W^l_{:b}}{\psi}+ \alpha^l_{:b} \cdot \psi &\geq 2\Big|\Big|\sqrt{\alpha^l_{:b} \cdot} W^l_{:b}\Big|\Big|_{l_2}
\end{align}
the analytical solution is given by $\psi^l_{:b}(k+1) = \frac{||W^l_{:b}(k+1)||_{2}}{\omega^l_{:b}(k)}$$ where $$\omega^l_{:b}(k) = \sqrt{\alpha^l_{:b}(k)} = \sqrt{\sum\limits_{a=1}^{n_{l-1}}\alpha^l_{ab}(k)}$.
The row-wise regularisation can be analogously derived. Note that the update rules for $\alpha^l_{ab}$ remains similar to Eq.~\eqref{eq:posterior update} and \eqref{eq:alpha_update}. However, when using both row-wise and column-wise group regularisation, the posterior is updated according to a combined prior expressed with a prior width given by :
\begin{align}
	{\psi}^l_{ab}(k) = \frac{1}{(\frac{1}{{\psi}^l_{a:}(k)}+\frac{1}{{\psi}^l_{:b}(k)})}.
\end{align}
Table \ref{tab:mlp} summarises the update rules according to the category of regularisation techniques adopted.
\begin{table*}[ht]
	\caption{Hyper-parameters update rule based on regularisation technique.}
	\label{tab:mlp}
	\begin{center}
	\resizebox{2.00\columnwidth}{!}{\begin{tabular}{|c|c|c|c|c|}
		\hline
		Category  & Prior Formulation     & $\rho(\omega^l,W^l)$  & $\omega^l$    & ${\psi}^l$  \\ \hline
		(a) Shape-wise 
		& $\prod\limits_{a=1}^{n_{l-1}}\prod\limits_{b=1}^{n_l}p(W^l_{ab}, \psi^l_{ab})$               
		& $\sum\limits_{a=1}^{n_{l-1}}\sum\limits_{b=1}^{n_l}\|\omega_{ab}^l(k) \cdot W^l_{ab}(k)||_{l_1}$   
		& ${\omega}^l_{ab}(k) = \sqrt{{\alpha}^l_{ab}(k)}$
        & ${\psi}^l_{ab}(k) = \frac{|W^l_{ab}(k)|}{\omega^l_{ab}(k-1)}$   \\ \hline 
		(b) Row-wise  & $\prod\limits_{a=1}^{n_{l-1}}p(W^l_{a:},\psi^l_{a:}) $
		& $\sum\limits_{a=1}^{n_{l-1}}||\omega^l_{a:}(k)\cdot W^l_{a:}(k)||_{l_2}$
		& ${\omega}^l_{a:}(k) = \sqrt{\sum\limits_{b=1}^{n_l}{\alpha^l_{ab}(k)}}$
		& ${\psi}^l_{a:}(k) = \frac{||W^l_{a:}(k)||_{2}}{\omega^l_{a:}(k-1)}$
		\\ \hline
		(c) Column-wise &    $\prod\limits_{b=1}^{n_{l}}p(W^l_{:b},\psi^l_{:b}) $ 
		& $\sum\limits_{b=1}^{n_l}||\omega^l_{:b}(k)\cdot W^l_{:b}(k)||_{l_2}$    
		& ${\omega}^l_{:b}(k) = \sqrt{\sum\limits_{a=1}^{n_{l-1}}{\alpha^l_{ab}(k)}}$
		& ${\psi}^l_{:b}(k) = \frac{||W^l_{:b}(k)||_{2}}{\omega^l_{:b}(k-1)}$
        \\ \hline
        \begin{tabular}[c]{@{}c@{}}
		(b) Row-wise  \\ + \\ (c) Column-wise 
		\end{tabular}
		& \begin{tabular}[c]{@{}c@{}}
			$\prod\limits_{a=1}^{n_{l-1}} p(W^l_{a:},\psi^l_{a:})$ \\
			$\times  \prod\limits_{b=1}^{n_{l}} p(W^l_{:b},\psi^l_{:b})$
        \end{tabular}
		& \begin{tabular}[c]{@{}c@{}}
			$\sum\limits_{a=1}^{n_{l-1}}\|\omega_{a:}^l(k) \cdot W^l_{a:}(k)||_{l_2}$ \\
			$+\sum\limits_{b=1}^{n_l}\|\omega_{:b}^l(k) \cdot W^l_{:b}(k)||_{l_2}$
        \end{tabular}
		& \begin{tabular}[c]{@{}c@{}}
		    ${\omega}^l_{a:}(k) = \sqrt{\sum\limits_{b=1}^{n_l}{\alpha^l_{ab}(k)}}$\\
		    ${\omega}^l_{:b}(k) = \sqrt{\sum\limits_{a=1}^{n_{l-1}}{\alpha^l_{ab}(k)}}$ \\
		\end{tabular} 
		& \begin{tabular}[c]{@{}c@{}c@{}}
		    ${\psi}^l_{a:}(k) = \frac{||W^l_{a:}(k)||_{2}}{\omega^l_{a:}(k-1)}$\\
			${\psi}^l_{:b}(k) = \frac{||W^l_{:b}(k)||_{2}}{\omega^l_{b:}(k-1)}$\\ 
			${\psi}^l_{ab}(k) = 1/(\frac{1}{{\psi}^l_{a:}(k)}+\frac{1}{{\psi}^l_{:b}(k)})$\\
		\end{tabular}                                                     
        \\ \hline
	\end{tabular}
}
\end{center}
\end{table*}

\section{Calculation of the inverse of the Hessian}
\label{appsec:hessian_calculation}
\modify{
Before we prove the Lemma~\ref{lemma:fc_hessian} and Lemma~\ref{lemma:rnn_hessian}, we first give the definition about the vectorisation operator for the $2D$ matrix in Eq.~\eqref{eq:vectorisation_2d}.
}

\modify{
\begin{definition}
\label{define:vectorisation_operator_2d}
The vectorisation operator for a $2D$ matrix $A \in \mathbb {R}^{m\times n}$ is defined as stacking the elements of $A$ into a vector $\vec A \in \mathbb {R}^{mn}$ by assembling the columns of $A$ sequentially. Formally, the vectorisation process $\mathbb{V}_{2D}: \mathbb {R}^{m\times n} \to \mathbb {R}^{mn}$ is:
\begin{equation}
    \label{eq:vectorisation_2d}  
    \mathbb{V}_{2D}(A) = \begin{bmatrix}
    A_{:,1} \\ A_{:,2} \\ \cdots \\ A_{:,n}
\end{bmatrix} 
\end{equation}
where \begin{math}
    A_{:,i} = \begin{bmatrix}
    A_{1,i}  A_{2,i}  \cdots  A_{m,i} \end{bmatrix}^\top
\end{math}
The operator $\mathbb{V}_{2D}$ represents the identity map if the input $A$ is a vector itself. 
\end{definition}
}

\subsection{\modify{Derivation of the Hessian calculation method for Fully-connected layer}}
\label{appsubsec:MLP_hessian_calculation}
\modify{
The proof Lemma~\ref{lemma:fc_hessian} is based on the Hessian calculation method proposed in~\cite{Hessian_fclayer}.
}
\begin{proof}\rm
\modify{
Since the Hessian is diagonal dominant, we mainly consider how to obtain the diagonal value of the Hessian matrix. Suppose that $\generealweight^l$ is a $2D$ matrix with $\generealweight^l \in \mathbb {R}^{m\times n}$. Then the diagonal value of ${\rmH}^l$ can be computed as the Jacobian matrix of the gradient of $\generealweight$ with respect to the vectorisation of $\generealweight^l$:
\begin{equation}
\begin{aligned}
 &\pdv[2]{\mathbf{L}}{\mathbb{V}_{2D}( \generealweight^l)}
 = \frac{\partial}{\partial{\mathbb{V}_{2D}( \generealweight^l)}}	\left(\frac{\partial{\mathbf{L}}}{\partial{\mathbb{V}_{2D}( \generealweight^l)}}\right) \\
  =& \frac{\partial}{\partial{ \generealweight^l}}\left(\frac{\partial{\mathbf{L}}}{\partial{h^l}}\frac{\partial{h^l}}{ \generealweight^l}\right) = \frac{\partial}{\partial{ \generealweight^l}}\left(\left(1\otimes\frac{\partial{\mathbf{L}}}{\partial{h^l}}\right)\left(\frac{\partial{h^l}}{ \generealweight^l}\otimes\mathbf{I} \right)\right) \\
  =&\frac{\partial}{\partial{ \generealweight^l}}\left( ({\mathbf{a}}^{l-1})^{\top} \otimes \frac{\partial \mathcal{L}}{\partial \mathbf{h}^{l}}\right) = ({\mathbf{a}}^{l-1})^{\top} \otimes \frac{\partial^{2} \mathcal{L}}{\partial \generealweight^{l} \partial h^{l}} \\
  =& ({\mathbf{a}}^{l-1})^{\top} \otimes\left(\frac{\partial h_{l}^{\top}}{\partial \generealweight_{l}} \frac{\partial^{2} \mathcal{L}}{\partial (h^{l})^{2}}\right) 
  \\
  =&  
  ({\mathbf{a}}^{l-1})^{\top} \otimes
  \left(\left(({a}^{l-1})^{\top} \otimes \mathbf{I}\right)^{\top} \frac{\partial^{2} \mathcal{L}}{\partial (h^{l})^{2}}\right)
  \\
  =&
  ({a}^{l-1})^{\top} \otimes\left({a}^{l-1} \otimes \frac{\partial^{2} \mathcal{L}}{\partial (h^{l})^{2}}\right) 
  \\
  =& ({a}^{l-1})^{\top} \otimes{a}^{l-1} \otimes \frac{\partial^{2} \mathcal{L}}{\partial (h^{l})^{2}} \\
  =& \left( {a}^{l-1}({a}^{l-1})^{\top}\right) \otimes \frac{\partial^{2} \mathcal{L}}{\partial (h^{l})^{2}}
  \label{eq:fc_Hessian_derive} 
\end{aligned}
\end{equation}
where $\frac{\partial^{2}\mathcal{L}}{\partial (h^{l})^{2}}$ is defined as the pre-activation Hessian $H^l$, which can be calculated as follows:
\begin{equation}
\begin{aligned}
&H^l=\frac{\partial^{2} \mathcal{L}}{(\partial h^{l})^{2}}=\frac{\partial}{\partial h^{l}}\left(\frac{\partial \mathcal{L}}{\partial h^{l}}\right)\\
=&\frac{\partial}{\partial h^{l}}\left(\frac{\partial \mathcal{L}}{\partial h^{l+1}} \frac{\partial h^{l+1}}{\partial a^{l}} \frac{\partial a^{l}}{\partial h^{l}}\right) \\
=& \frac{\partial^{2} \mathcal{L}}{\partial h^{l} \partial h^{l+1}} \bw^{l+1} \frac{\partial a^{l}}{\partial h^{l}}+\frac{\partial \mathcal{L}}{\partial h^{l+1}} \frac{\partial h^{l+1}}{\partial a^{l}} \frac{\partial}{\partial h^{l}}\left(\frac{\partial a^{l}}{\partial h^{l}}\right) \\
=&\frac{\partial (a^{l})^{\top}}{\partial h^{l}} (\bw^{l+1})^{\top} H^{l+1} \bw^{l+1} \frac{\partial a^{l}}{\partial h^{l}}+\sum_{k} \frac{\partial  \mathcal{L}}{\partial a^{l}_{k}} \frac{\partial^{2}a^{l}_{k}}{\partial (h^{l})^{2}}
\label{eq:fc_pre-Hessian_derive_1} 
\end{aligned}
\end{equation}
Define the diagonal matrices $B^l$ and $D^l$ as:
\begin{align}
    B^l &= \diag(\frac{\partial (a^{l})^{\top}}{\partial h^{l}}) =\diag(\sigma'(h^l)), \\
    D^l &= \diag(\sum_{k} \frac{\partial  \mathcal{L}}{\partial a^{l}_{k}} \frac{\partial^{2}a^{l}_{k}}{\partial (h^{l})^{2}})=\diag(\sigma''(h^l) \circ \frac{{\partial}L}{{\partial}a^l})
    \label{eq:fc_pre_Hessian_diagnoal_derive} 
\end{align}
Further,~\eqref{eq:fc_pre-Hessian_derive_1} can be reformulated as:
\begin{equation}
\begin{aligned}
&H^l=B^l ({\generealweight^{l+1}})^{\top} H^{l+1} \generealweight^{l+1} B^l + D^l
\label{eq:fc_pre-Hessian_derive_2} 
\end{aligned}
\end{equation}
where $\diag ()$ in~\eqref{eq:fc_pre_Hessian_diagnoal_derive} refers to generating a diagonal matrix whose diagonal value is extracted from a vector.
The initialised $H^l$ is the second-order derivative of the loss function with respect to the output of the neural network. 
}

\modify{
The above derivation process is a recap of the Hessian calculation methods proposed in~\cite{Hessian_fclayer}.
In this paper, in order to reduce computational complexity, we make a further simplification by extracting the diagonal values of the pre-activation Hessian $H$ in~\eqref{eq:fc_pre-Hessian_derive_1} and Hessian $\rmH$~\eqref{eq:fc_Hessian_derive} for recursive computation. Thus, the matrix multiplication could be reduced to vector multiplication. The Hessian calculation process can be approximated as:
\begin{equation}
    \label{eq:fc_Hessian_approximation_derive}
    \hat{{\rmH}}^l = \diag\left((a^{l-1}\right)^2 \otimes \hat{H}^l)
\end{equation}
\begin{align}
\label{eq:fc_pre_Hessian_approximation_derive}
        \hat{H}^l = {(\hat{B}^l)}^2 \circ \left(\left(({\bw^{l+1})^{\top}}\right)^2 \hat{H}^{l+1}\right)+ \hat{D}^l, \\ \notag
        \hat{B}^l = \sigma'(h^l), \ \ \ \hat{D}^l=\sigma''(h^l) \circ \frac{{\partial}L}{{\partial}a^l}
\end{align}
}
\end{proof}

\subsection{\modify{Derivation of the Hessian calculation method for recurrent layer}}
\label{appsubsec:RNN_hessian_calculation}
\modify{
The proof of Lemma~\ref{lemma:rnn_hessian} is as follows.
}

\begin{proof}\rm
\modify{
As explained before, a RNN layer normally consists of three matrices (i.e., $W_i, W_o, W_h$), which will be revisited many times for a complete recurrent operation. Therefore, the Hessian calculation for a RNN layer can be divided into three parts to calculate the Hessian of $W_i, W_o, W_h$, respectively. 
The procedures are summarised as follows: 
}
\begin{enumerate}
    \item \modify{
    Extend the RNN layer to its equivalent FC layers through sequence $t$ (see Fig.~\ref{fig:lstm_bptt}).
    }
    \item \modify{
    Calculate the Hessian for $W_o$:
    }
    \begin{figure}[ht]
        \centering
        \includegraphics[scale = 0.44]{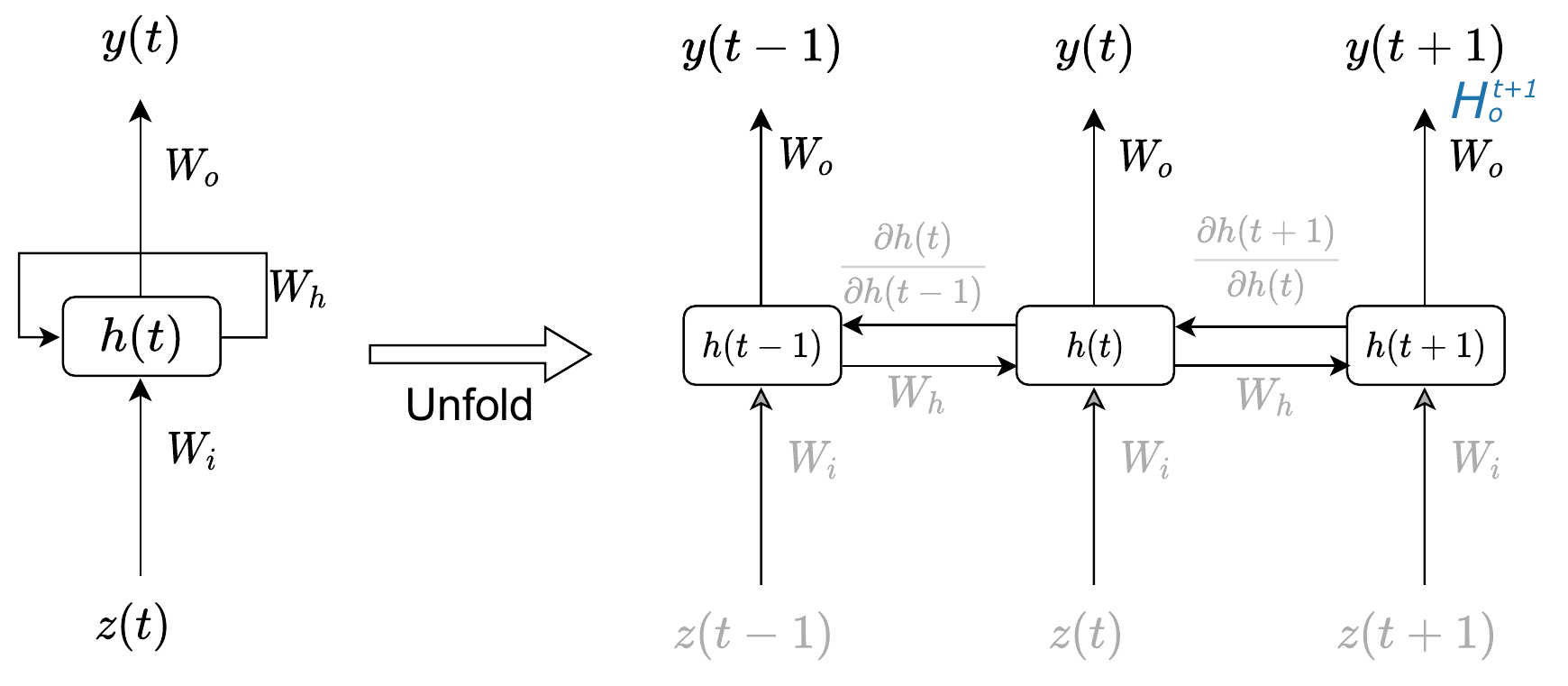}
        \caption{\modify{The equivalent FC layers about output weight.}}
        \label{fig:lstm_bptt_output_weight}
    \end{figure}
    
    \modify{
    If a RNN layer is unfolded through the time, then the $W_o$ can be regarded as the weight matrix for a FC layer, whose input is h(t) and output is $y(t)$. As explained in~Eq.\eqref{eq:rnn2}, the output $y(t)$ is computed by applying the activation function $g(\cdot)$ on the matrix multiplication between $h(t)$ and $W_o$. Such matrix multiplication is implemented $T$ times for a complete recurrent operation (as shown in the unblurred part in Fig.\ref{fig:lstm_bptt_output_weight}). According to the Hessian calculation method of the FC layers as in Lemma~\ref{lemma:fc_hessian}, the Hessian for $W_o$ is:
    \begin{equation}
    \mathbf{H}_o = \frac{1}{T}\sum\limits^{T}_{t=1} \mathbf{H}^{\top}_o, \ \ \
    \mathbf{H}^{\top}_o = h(t)^2 \otimes H^{\top}_o
    \end{equation}
    where $H^{\top}_o$ is the initialised pre-activation Hessian of $W_o$. 
    It should be noted that $H^{\top}_o$ will be updated along the BPTT process and be used as the initialised pre-activation Hessian for $W_h$ and $W_i$. 
    \begin{align}
    \label{eq:rnn_Hessian_approximation_1}
            H^{t,t}_h = H^{t,t}_i = ({{B}})^2 \circ \left(\left(\bw_o^{\top}\right)^2 {H}^{t}_o\right)+ {D}
    \end{align}
    with ${B}$ and ${D}$ being defined as:
    \begin{align}
    \label{eq:rnn_Hessian_approximation_2} 
            {B} = g'(M_i), \ \ \ 
            {D}= g''(M_i) \circ \frac{{\partial}L}{{\partial}y(t)} 
    \end{align}
    }
    
    \item \modify{
    Calculate the Hessian for $W_h$:}
        \begin{figure}[ht]
        \centering
        \includegraphics[scale = 0.44]{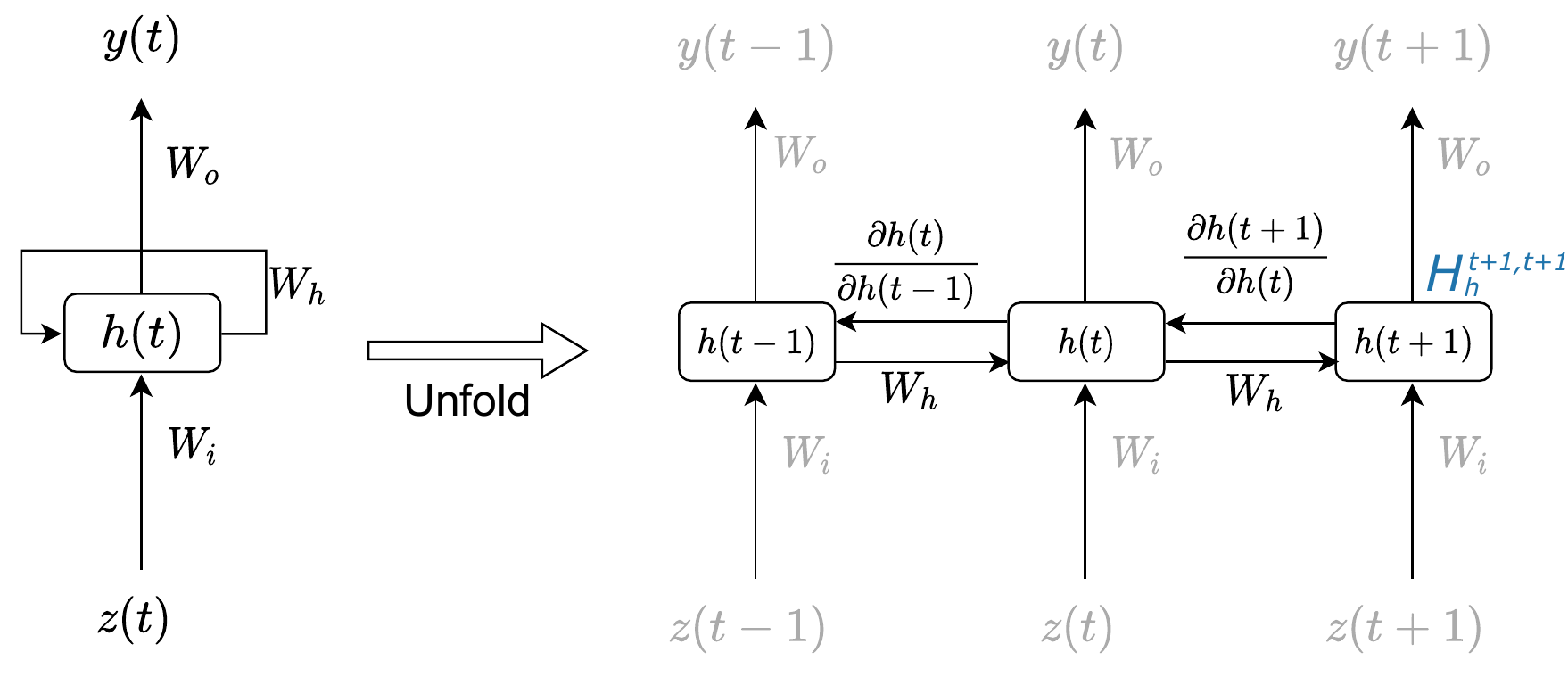}
        \caption{\modify{The equivalent FC layers about hidden weight.}}
        \label{fig:lstm_bptt_hidden_weight}
    \end{figure}
    
    \modify{
    As shown in Fig.~\ref{fig:lstm_bptt_hidden_weight}, the $W_h$ can be regarded as the weight matrix for a FC layer, whose input is $h(t-1)$ and output is $h(t)$.
    As explained in~\eqref{eq:rnn1}, one component of $h(t)$ is the matrix multiplication between $h(t-1)$ and $W_h$. Such matrix multiplication is implemented $\min(t, \tau)$ times for the time step $t$, where $\tau$ is the backward propagation time horizon, (as shown in the unblurred part in Fig.\ref{fig:lstm_bptt_hidden_weight}). Therefore, if we only calculate the Hessian of $W_h$ referring to a single data sample at time $t$, then the Hessian can be calculated by averaging $\min(t, \tau)$ individual Hessians as follows:
    \begin{equation} 
        \mathbf{H}_h = \mathbb{E}\left(\sum\limits_{j=\max(1,t - \tau + 1)}^{t} \mathbf{H}_h^{t,j}\right)
    \end{equation}  
    If we consider the complete time steps $T$, then the Hessian for $W_h$ is:
\begin{equation} 
    \mathbf{H}_h = \mathbb{E}\left(\sum\limits_{t=1}^{T}\sum\limits_{j=\max(1,t - \tau + 1)}^{t} \mathbf{H}_h^{t,j}\right)
\end{equation}
where
\begin{align}
    \mathbf{H}_h^{t,j} &= h(t-1)^2 \otimes H_h^{t,j}, \\
    H_h^{t,j} &= {(B_h)}^2 \circ \left(\left(W_h^\top\right)^2 H^{t, j + 1}_h\right) + D_h
\end{align}
with
\begin{math}
    B_h = \sigma '(\bar h(t)), D_h = \sigma ''(\bar h(t)) \circ \frac{\partial{L}}{\partial{h(t)}}
\end{math}.
$H_h^{t,j}$ is the pre-activation Hessian whose initialised value is given by~\eqref{eq:rnn_Hessian_approximation_1}. 
}

    \item \modify{
    Calculate the Hessian for $W_i$:
    }
        \begin{figure}[ht]
        \centering
        \includegraphics[scale = 0.44]{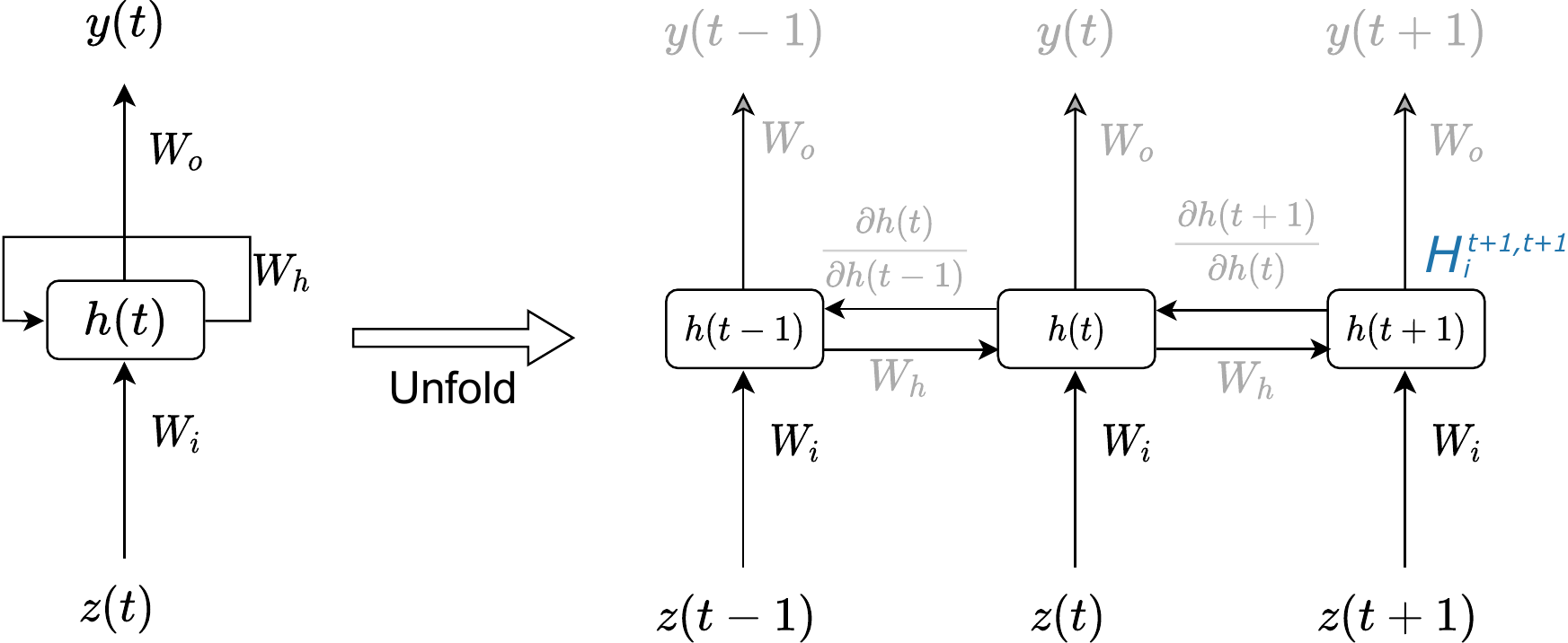}
        \caption{\modify{The equivalent FC layers about input weight.}}
        \label{fig:lstm_bptt_input_weight}
    \end{figure}
  
    \modify{
    As shown in Fig.~\ref{fig:lstm_bptt_input_weight}, the $W_i$ can be regarded as the weight matrix for a FC layer, whose input is $z(t)$ and output is $h(t)$.
    Similar to the calculation method for $W_h$, the Hessian of $W_i$ can be calculated by averaging $\min(t, \tau) \times T$ individual Hessians as follows:
 \begin{equation}
    \mathbf{H}_i = \mathbb{E}\left(\sum\limits_{t=1}^{T}\sum\limits_{k=\max(1,t - \tau + 1)}^{t} \mathbf{H}_i^{t,k}\right)
\end{equation}
where the individual Hessian is updated as:
\begin{align}
    \mathbf{H}_i^{t,k} &= (z(k))^2 \otimes H_i^{t,k}, \\
    H_i^{t,k} &= \prod^{t}_{j=k+1} {(B_i)}^2 \circ \left(\left((W_i)^\top\right)^2 H^{j-1, j}_i\right)
\end{align}
with 
\begin{math}
    B_i = \sigma '(\bar h(j))
\end{math}.   
It should be noted that the initialised pre-activation Hessian $H_i^{t,t}$ is calculated by~\eqref{eq:rnn_Hessian_approximation_1}.
}
\end{enumerate}
\end{proof}

\section{Free Run Simulation Results} \label{ap:simulation}
The identification experiments are implemented using the PyTorch library~\cite{paszke2019pytorch}. The MLP and LSTM models are randomly initialised, trained based on a one-step-ahead prediction approach, and validated in a free run simulation setting. The stochastic gradient descent method adopted is the ADAM optimiser. The learning rate is scheduled using Cosine Annealing for each identification experiment. 

For the evaluation metric, it should be noted that prediction and simulation error are two typical evaluation metrics for SYSID. 
Given the current and past system input and output measurements, the prediction means predicting the system response to the future $k$ steps, where $k$ denotes the prediction horizon. 
Simulation is to predict the system response based only on the input data and initial conditions. 
Therefore, the simulation error is a more challenging evaluation metric used in this paper. 
The figure of merit used is given by the root mean square error (RMSE) of the simulation experiment:
$
     RMSE(\hat{y},y) = \sqrt{\frac{1}{T}\sum\nolimits_{i=1}^{T} ({y_i - \hat{y}_i})^2}
$.
\modify{Besides, the model sparsity refers to the number of zero-valued parameters divided by the total number of parameters.}

\begin{figure*}[h]
	\centering
	\begin{subfigure}[h]{0.3\textwidth}
		\centering
		\includegraphics[scale=0.3]{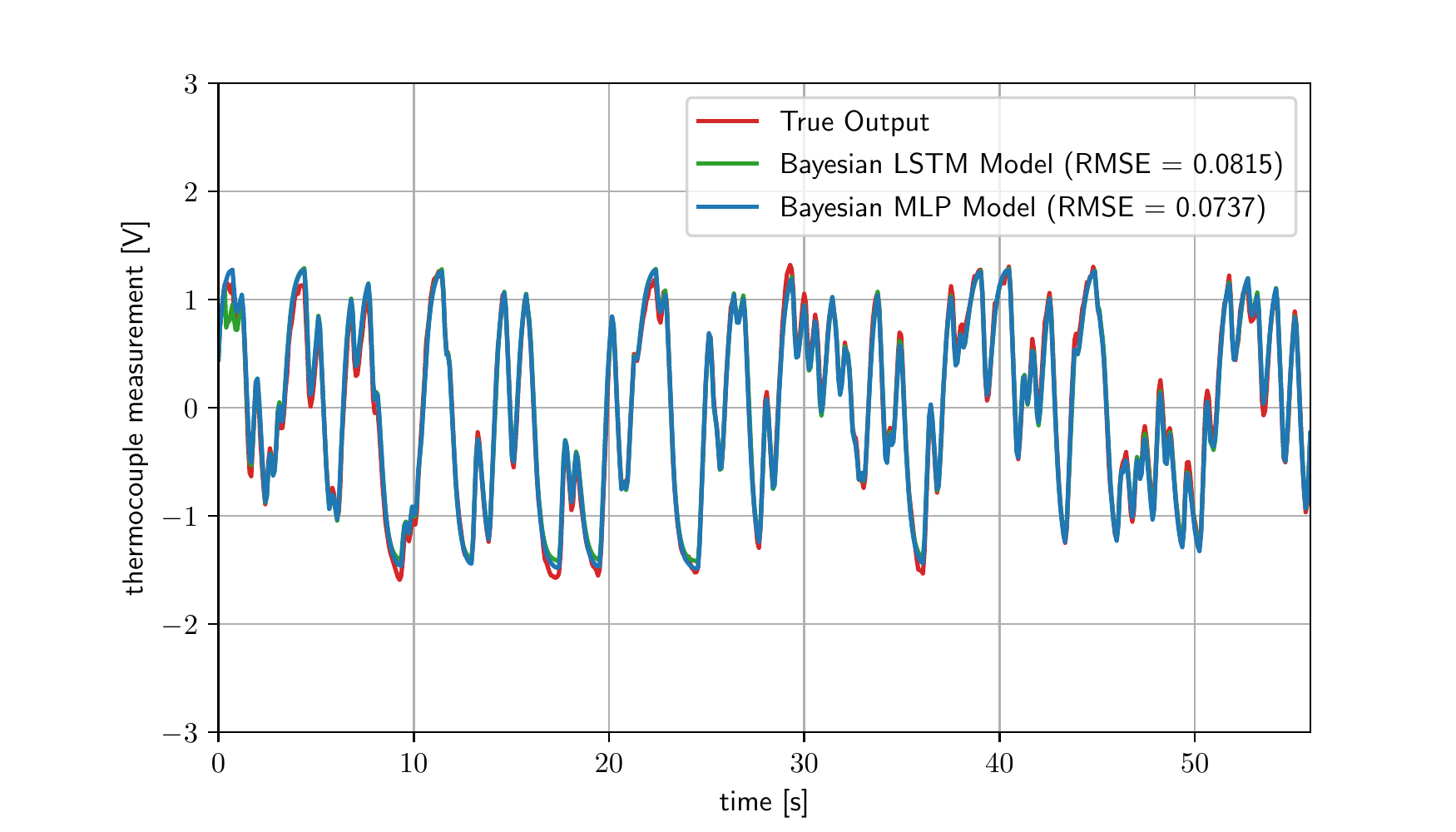}
		\caption{Hairdryer simulation comparison}
		\label{fig:dry_sim}
	\end{subfigure}
	\begin{subfigure}[h]{0.3\textwidth}
		\centering
		\includegraphics[scale=0.3]{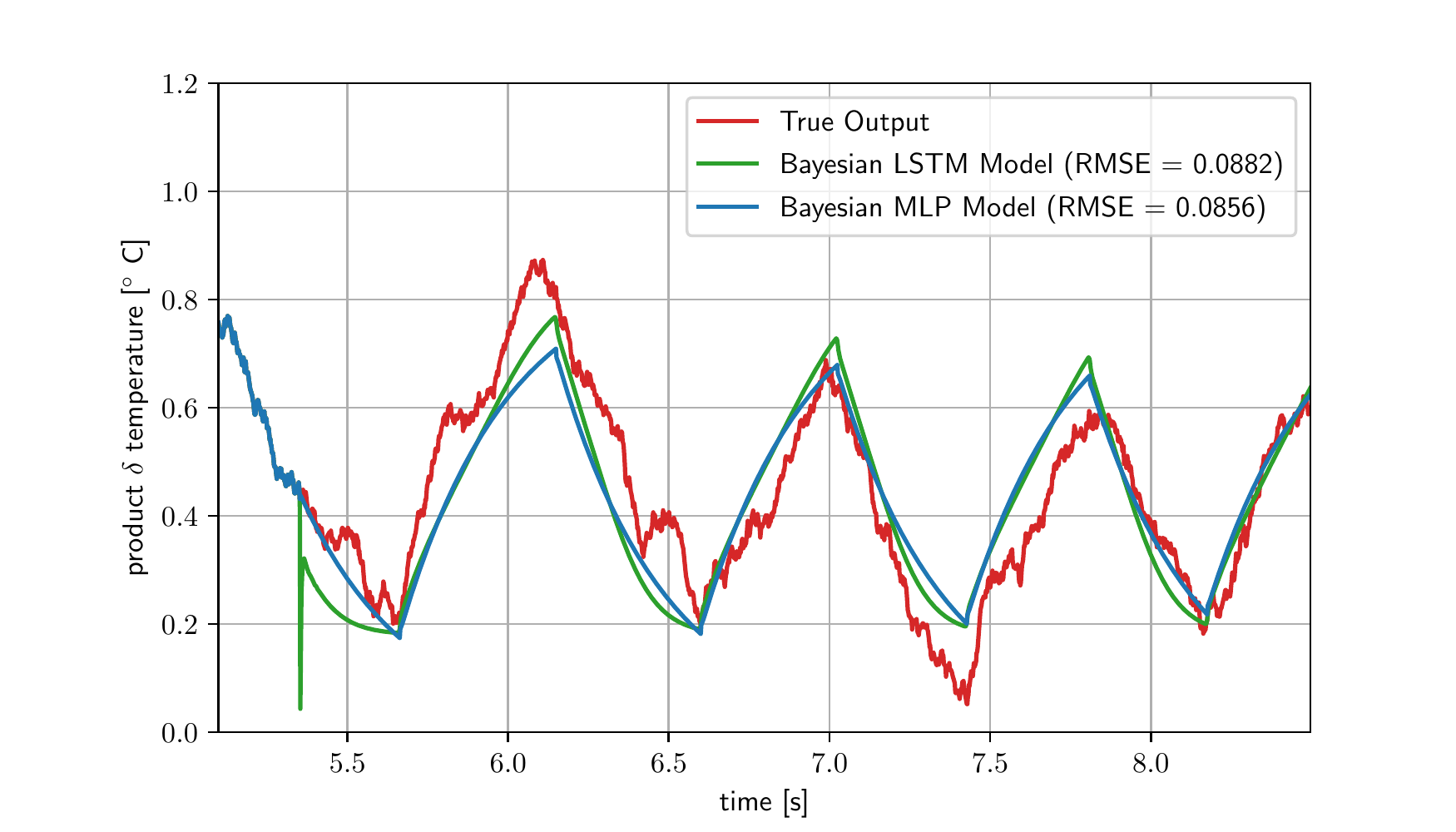}
		\caption{Heat exchanger simulation comparison}
		\label{fig:hex_sim}
	\end{subfigure}   
	\centering
	\begin{subfigure}[h]{0.3\textwidth}
		\centering
		\includegraphics[scale=0.3]{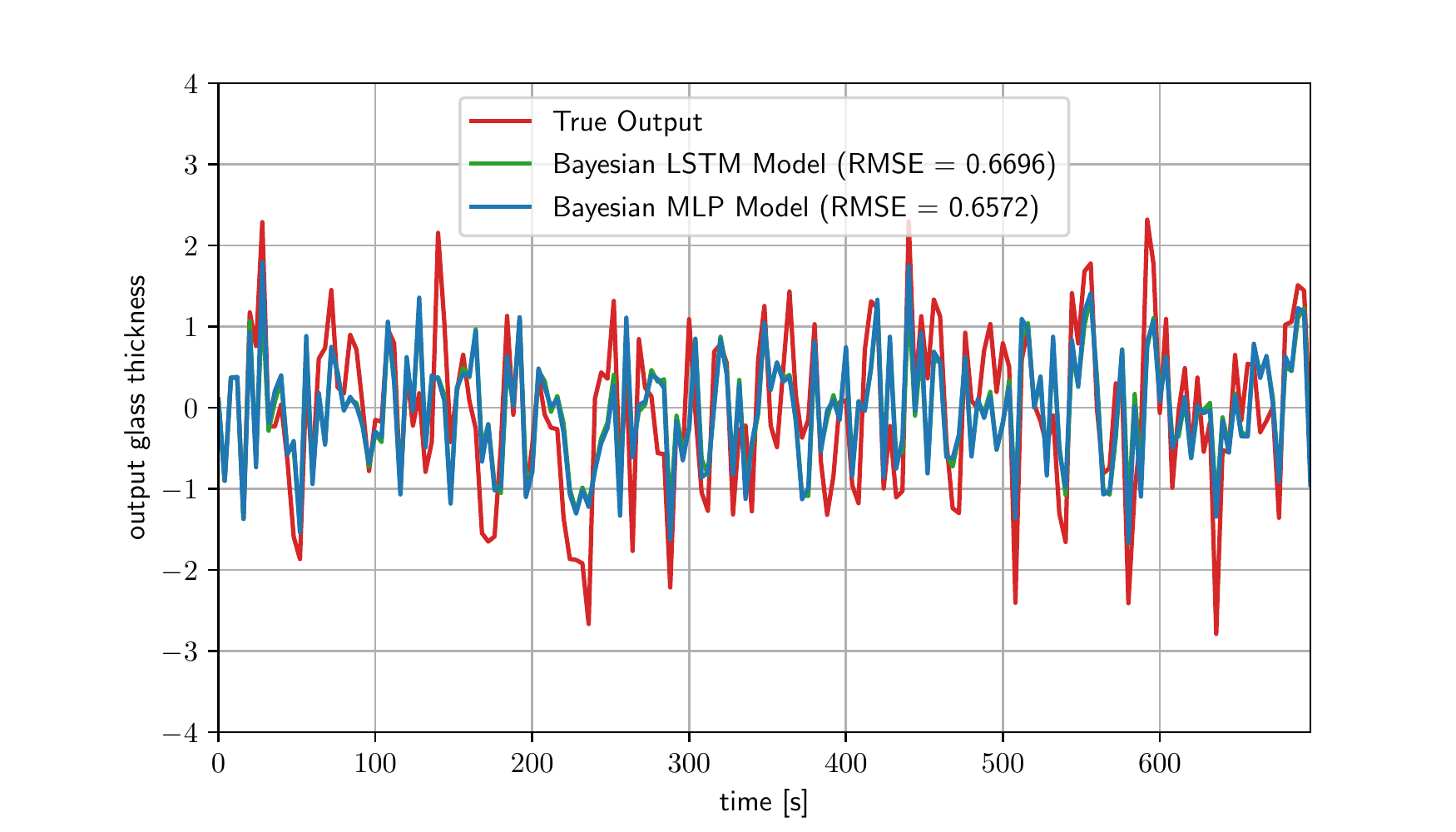}
		\caption{Glass tube manufacturing simulation comparison}
		\label{fig:gt_sim}
	\end{subfigure}
	\\
	\begin{subfigure}[h]{0.3\textwidth}
		\centering
		\includegraphics[scale=0.3]{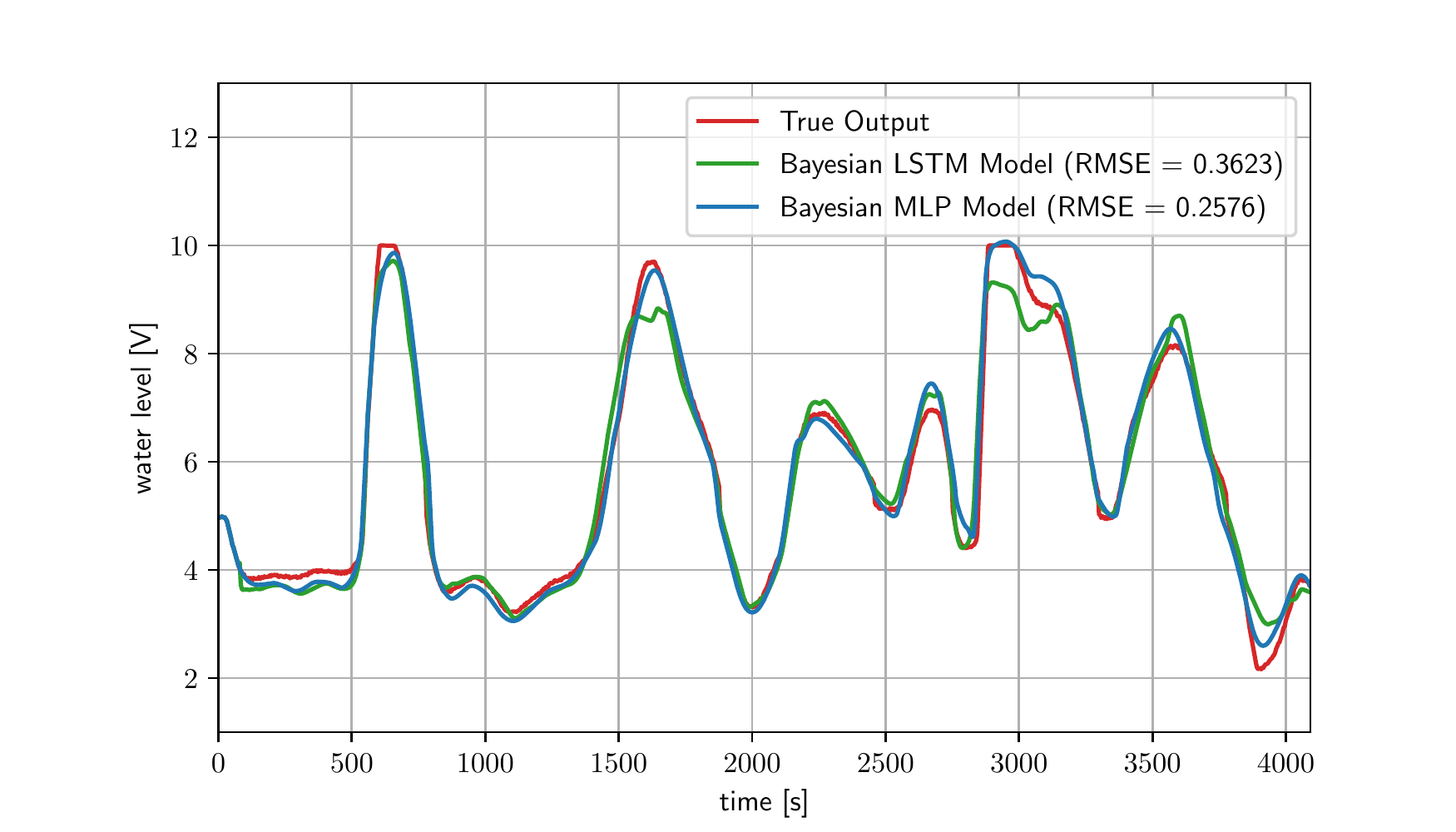}
		\caption{Cascaded tanks simulation comparison}
		\label{fig:ct_sim}
	\end{subfigure}   
	\centering
	\begin{subfigure}[h]{0.3\textwidth}
		\centering
		\includegraphics[scale=0.3]{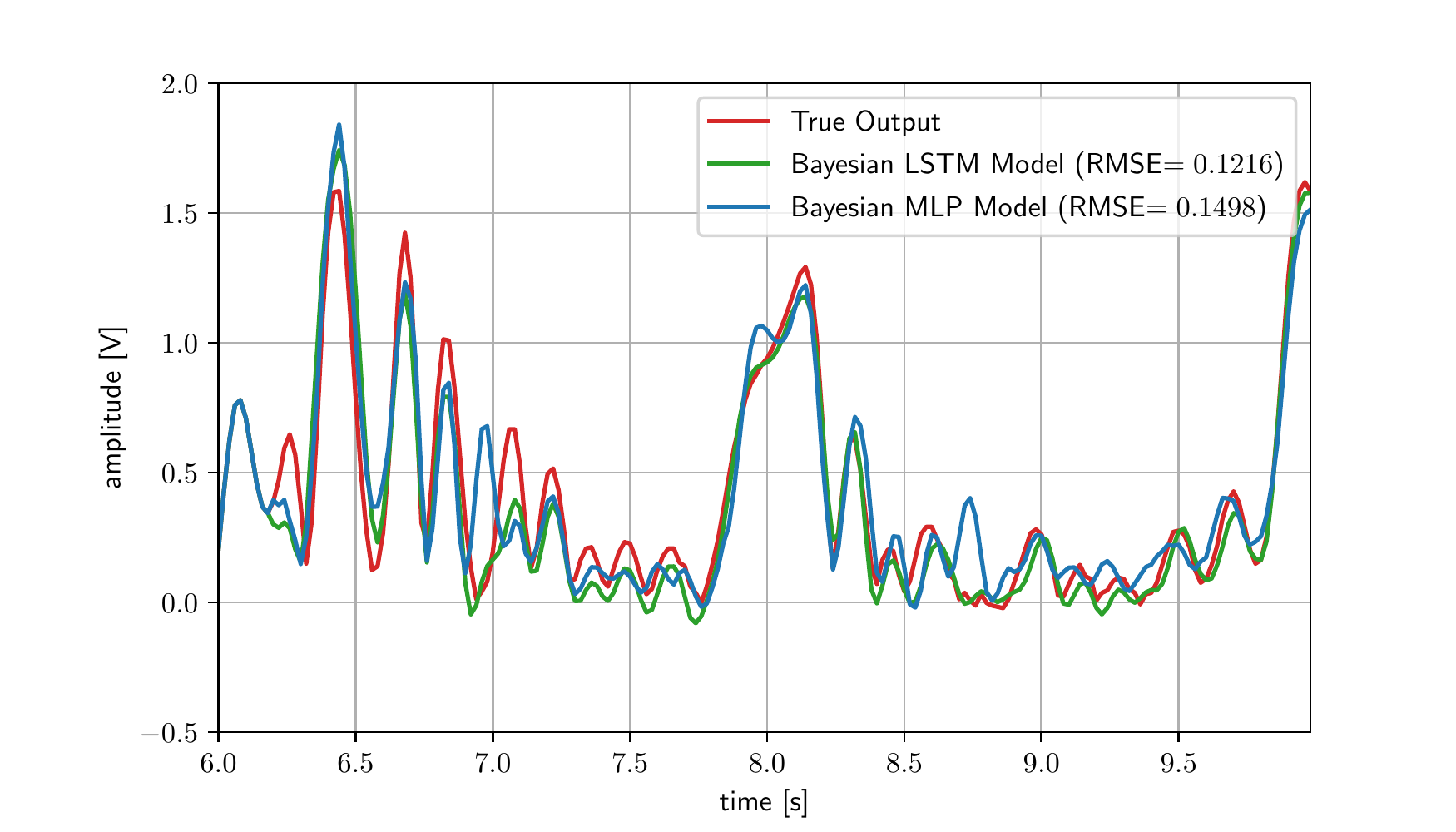}
		\caption{Coupled electric drives simulation comparison for the first validation dataset}
		\label{fig:ced_sim1}
	\end{subfigure}
	\begin{subfigure}[h]{0.3\textwidth}
		\centering
		\includegraphics[scale=0.3]{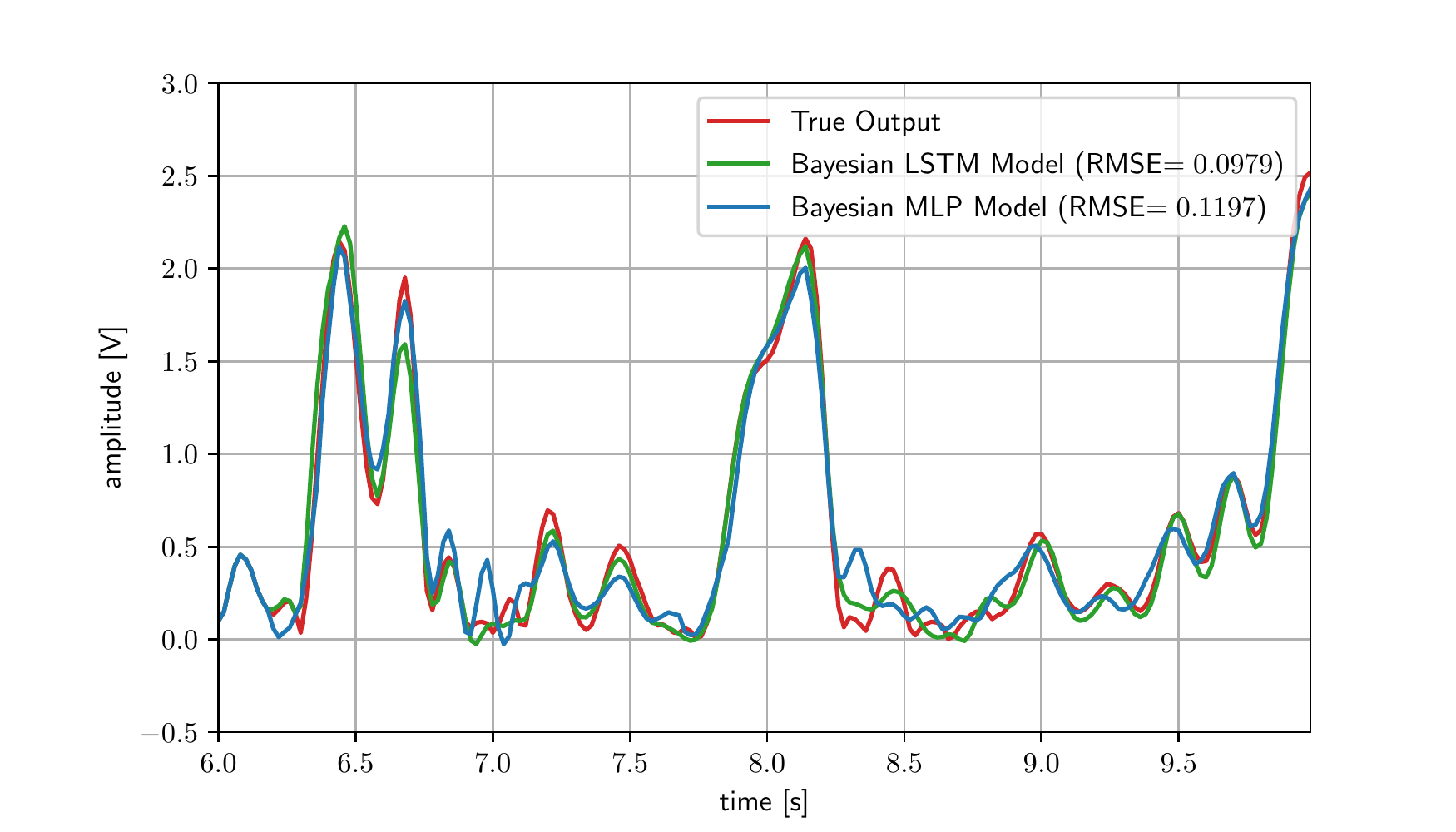}
		\caption{Coupled electric drives simulation comparison for the second validation dataset}
		\label{fig:ced_sim2}
	\end{subfigure}
	\caption{The comparison on simulation result for different datasets}
	\label{fig:simulation_comparison_result}
\end{figure*}

As shown in Fig.~\ref{fig:simulation_comparison_result} and Table~\ref{tab:comparision_linear_simulation} and~\ref{tab:comparision_non_linear_simulation}, this appendix includes the plots of the simulated experiments using the models identified and a comparison with previous models used in literature. 
The details of experiment settings on each benchmark are given in following sections (see Section~\ref{ap:dry}--Section~\ref{ap:ced}).
\begin{table}[h]
	\centering
	\caption{Comparison of RMSE on identified linear systems with other works
	}
	\begin{tabular}{|lc|}
		\hline
		\textbf{Hairdryer} & \textbf{RMSE [V]} \\ \hline
		Transfer Function Estimation \cite{matlab_dry} & 0.108  \\
		Subspace Identification \cite{ljung_sysID} & 0.105 \\
		ARMAX Model \cite{ljung_sysID} & 0.104 \\ 
		ARX Model \cite{ljung_sysID} & 0.103 \\
		GP$^a$ with rational quadratic kernel & 0.066 \\
		GP$^a$ with squared exponential kernel& 0.066 \\
		\cdashline{1-2}
		LSTM without lags & 0.219\\
		LSTM without regularisation & 0.205 \\ 
		\textbf{Bayesian LSTM} & \textbf{0.081} \\ \cdashline{1-2}
		MLP without regularisation & 0.076 \\ 
		\textbf{Bayesian MLP} & \textbf{0.073} \\ \hline
		\textbf{Heat Exchanger} & \textbf{RMSE [$^\circ C$]}  \\ \hline
		Transfer Function Estimation \cite{matlab_hex} & 0.140 \\
		Process and Disturbance Model \cite{matlab_hex} & 0.089 \\
		Process Model \cite{matlab_hex} & 0.088 \\
		GP$^a$ with rational quadratic kernel & 0.187 \\
		GP$^a$ with squared exponential kernel& 0.187 \\
		\cdashline{1-2}
		LSTM without lags & 0.185 \\
		LSTM without regularisation & 0.158 \\ 
		\textbf{Bayesian LSTM} & \textbf{0.088} \\ \cdashline{1-2}
		MLP without regularisation & 0.092 \\ 
		\textbf{Bayesian MLP} & \textbf{0.086} \\ \hline 
		\textbf{Glass Tube Manufacturing} & \textbf{RMSE [$\cdot$]}  \\ \hline
		Subspace Identification \cite{matlab_gt} & 0.688 \\
		ARX Model \cite{matlab_gt} & 0.676 \\
		GP$^a$ with rational quadratic kernel & 0.654 \\
		GP$^a$ with squared exponential kernel  & 0.656 \\
		\cdashline{1-2}
		LSTM without lags & 1.099 \\
		LSTM without regularisation & 1.056 \\ 
		\textbf{Bayesian LSTM} & \textbf{0.669} \\ \cdashline{1-2}
		MLP without regularisation & 0.663 \\ 
		\textbf{Bayesian MLP} & \textbf{0.657} \\ \hline
	\end{tabular}
	\label{tab:comparision_linear_simulation}
\end{table}
\begin{table}[h]
	\centering
	\caption{Comparison of RMSE on identified nonlinear systems with other works}
	\begin{tabular}{|lcc|}
		\hline
		\textbf{Cascaded Tanks} & \multicolumn{2}{c|}{\textbf{RMSE [V]}} \\ \hline
		LMN$^b$ with NFIR \cite{belz2017} & \multicolumn{2}{c|}{0.669}  \\
		Flexible State Space Model \cite{SVENSSON2017189} & \multicolumn{2}{c|}{0.450} \\ 
		Voltera Feedback Model \cite{schoukens2016modeling} & \multicolumn{2}{c|}{0.397} \\
		OEM$^c$ with NOMAD \cite{bw_other1} & \multicolumn{2}{c|}{0.376} \\
		Piecewise ARX Models \cite{MATTSSON201840} & \multicolumn{2}{c|}{0.350} \\ 
		NLSS$^d$ \cite{ct_relan} & \multicolumn{2}{c|}{0.343} \\
		Tensor network B-splines \cite{kim2020Auto} & \multicolumn{2}{c|}{0.302} \\ 
		GP$^a$ with rational quadratic kernel  & \multicolumn{2}{c|}{0.344}  \\
		GP$^a$ with squared exponential kernel  & \multicolumn{2}{c|}{0.344} \\
		\cdashline{1-3}
		LSTM without lags & \multicolumn{2}{c|}{0.954} \\
		LSTM without regularisation  & \multicolumn{2}{c|}{0.494} \\ 
		\textbf{Bayesian LSTM}   & \multicolumn{2}{c|}{\textbf{0.362}} \\ \cdashline{1-3}
		MLP without regularisation  & \multicolumn{2}{c|}{0.432} \\ 
		\textbf{Bayesian MLP}   & \multicolumn{2}{c|}{\textbf{0.257}} \\ \hline 
		
		\textbf{Coupled Electric Drives} & \multicolumn{2}{c|}{\textbf{\textbf{RMSE [ticks/s]}}} \\  
		& Drive 1 & Drive 2 \\ \hline
		Extended Fuzzy Logic \cite{sb_ced_other7} & 0.150 & 0.092  \\ 
		Cascaded Splines \cite{ced_other1} &  0.216 & 0.110 \\
		TAG3P$^d$ \cite{ced_bw_schoukens} & - & 0.128 \\
		RBFNN - FSDE$^f$  \cite{ced_other4} & 0.130 & 0.185 \\ 
		GP$^a$ with rational quadratic kernel  & 0.150 & 0.167 \\
		GP$^a$ with squared exponential kernel  & 0.153 & 0.132 \\
		\cdashline{1-3}
		LSTM without lags & 0.394 & 0.252 \\
		LSTM without regularisation & 0.149 & 0.131\\ 
		\textbf{Bayesian LSTM} &  \textbf{0.121} & \textbf{0.097} \\ \cdashline{1-3}
		MLP without regularisation & 0.206 & 0.111 \\ 
		\textbf{Bayesian MLP} & \textbf{0.149} & \textbf{0.120}\\ \hline 
	\end{tabular}
	\label{tab:comparision_non_linear_simulation}
	{\raggedright \\$^a$ \textit{Gaussian process model.\\ $^b$ Tree based Local Model Networks with external dynamics represented by NARX or NFIR.\\ $^c$ Output Error parametric Model estimation based on derivative free method.\\ $^d$ nonlinear State Space model.\\$^e$ Tree Adjoining Grammars \\$^f$ Free Search Differential Evolution is used to determine the regressors.} \par}
\end{table}
\vspace{-0.2cm}
\section{Hairdryer} \label{ap:dry}
In common industrial settings with heating, temperature control is a highly desired objective given the high transport lags and process delay. The ``hairdryer" is a small scale laboratory apparatus that designates the PT326 process trainer \cite{matlab_hex}. A mass of air is heated with thermal resistors and flows in a tube. The temperature at the outlet is measured by a thermocouple in volts. The objective is to identify the dynamic relationship between the input voltage to the thermal resistors and the thermocouple voltage at the outlet.
The dataset specific to this device is given by MATLAB in a tutorial on linear system identification. The sampling time is 0.08 seconds and the dataset contains 1000 data points. The dataset is detrended, bringing data to a zero mean. The first 300 data points are used for identification and the remaining 700 are used for validation.

\begin{figure}[h]
	\centering
	\begin{subfigure}[h]{0.5\textwidth}
       \centering
        \includegraphics[scale=0.4]{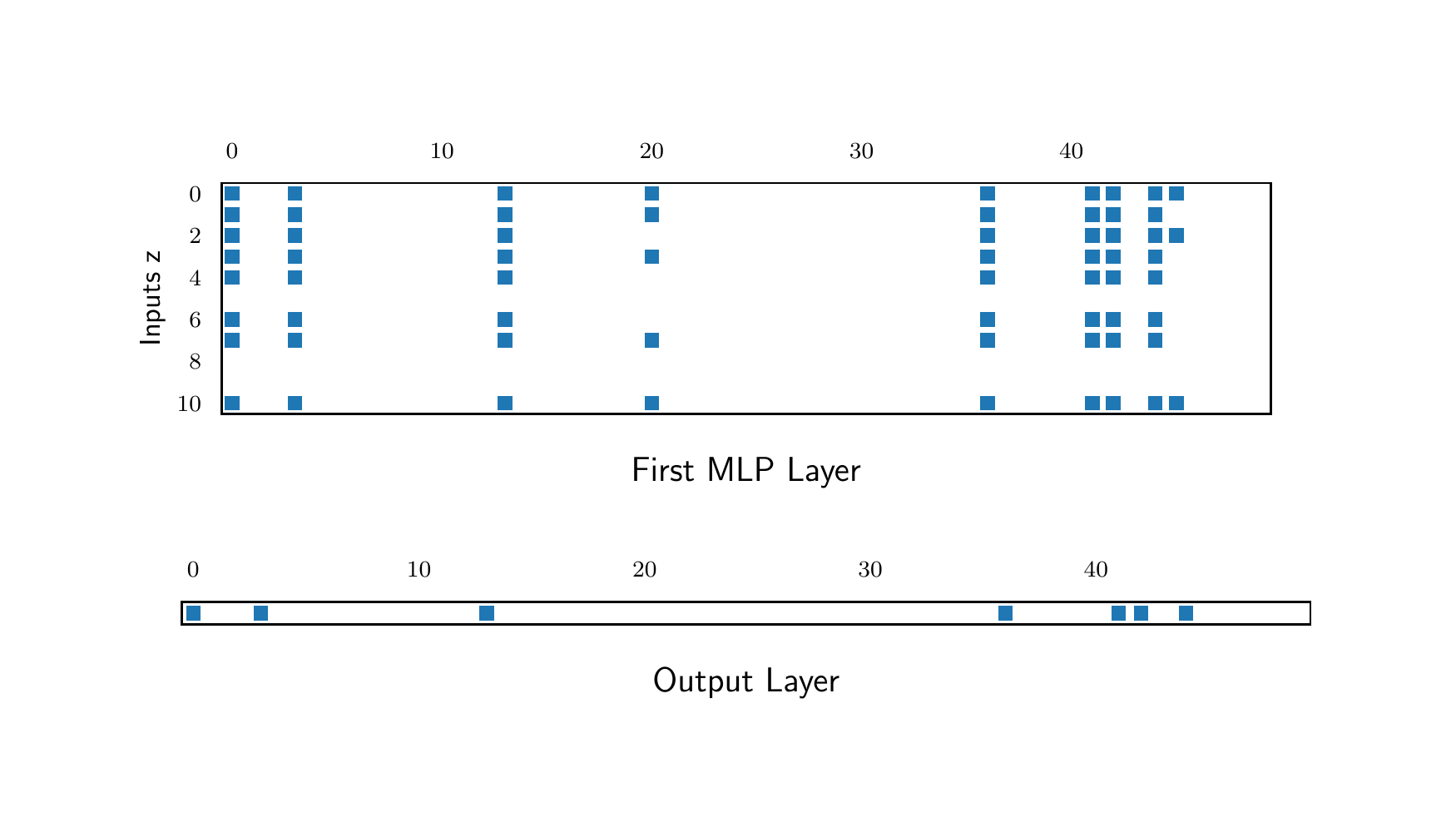}
        \vspace{-0.5cm}
        \caption{MLP model}
        \label{fig:dry_mlp_model}
	\end{subfigure}
	\begin{subfigure}[h]{0.5\textwidth}
       \centering
        \includegraphics[scale=0.4]{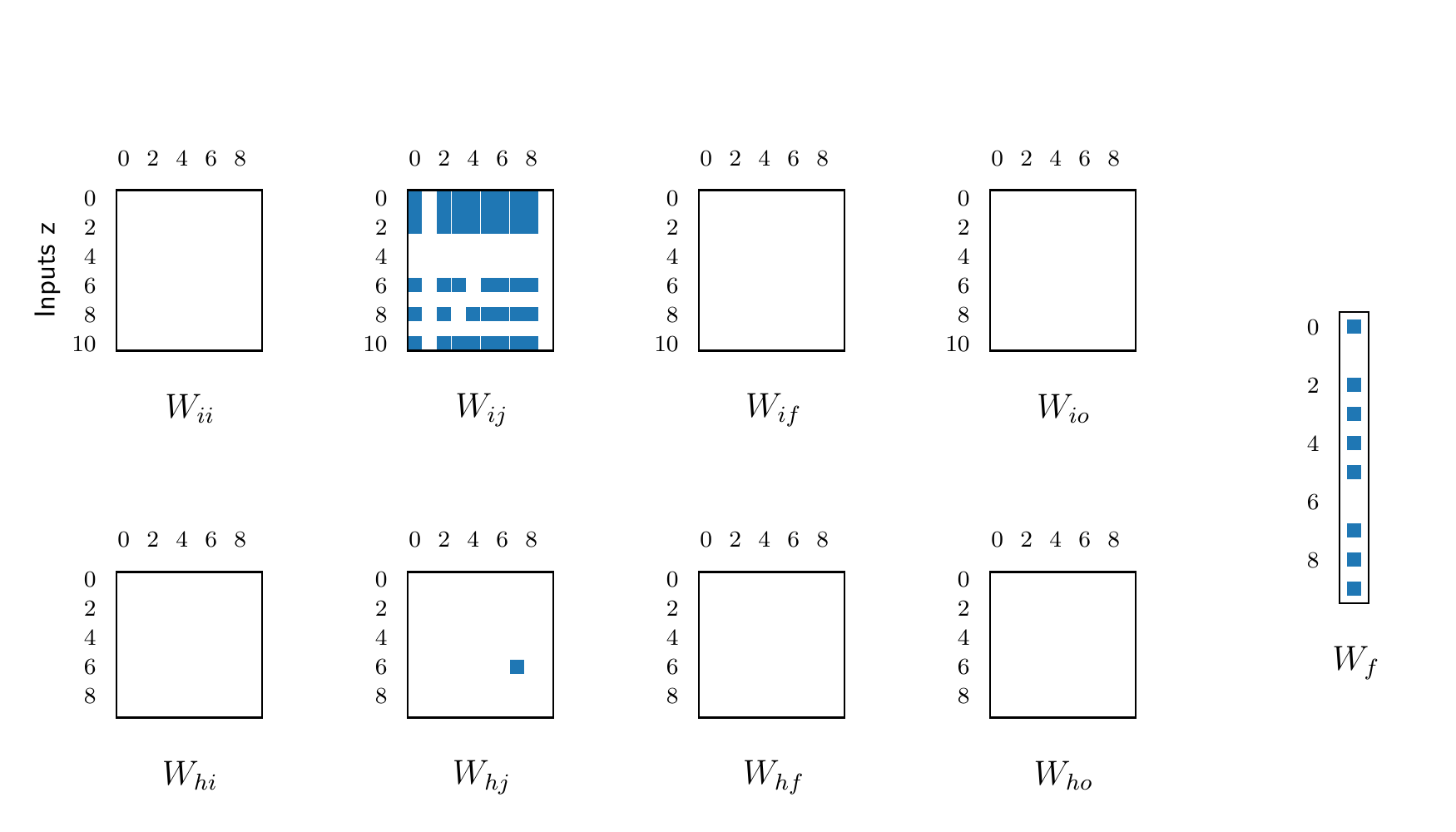}
        \caption{LSTM model}
        \label{fig:dry_lstm_model}
	\end{subfigure}
	\caption{Model sparsity plot of the identified MLP and LSTM on Hairdryer dataset. \textit{Blue indicates non-pruned connections and white indicate pruned ones.}}
\end{figure}

A fully connected MLP model with one hidden layer and 50 nodes is randomly initialised. The activation function is a linear activation without the bias term. The input and output lags chosen for the regressors are both $5$. Models are inferred through $K_{\max} = 6$ identification cycles. The best validated model is obtained in the $5$th cycle of identification with a sparsity of $88.1\%$. The model sparsity plot is shown in Fig.~\ref{fig:dry_mlp_model}. Furthermore, an RNN network is randomly initialised with one layer and 10 hidden LSTM units and no bias term. $l_u$ and $l_y$ are set to 5. The 6th and final identification cycle lead to the sparsest and best validated model with a sparsity of 93.5 \%. Fig.~\ref{fig:dry_lstm_model} shows the final model sparsity plot.

\begin{figure}[h]
	\centering
	\begin{subfigure}[h]{0.45\textwidth}
		\centering
        \includegraphics[scale=0.4]{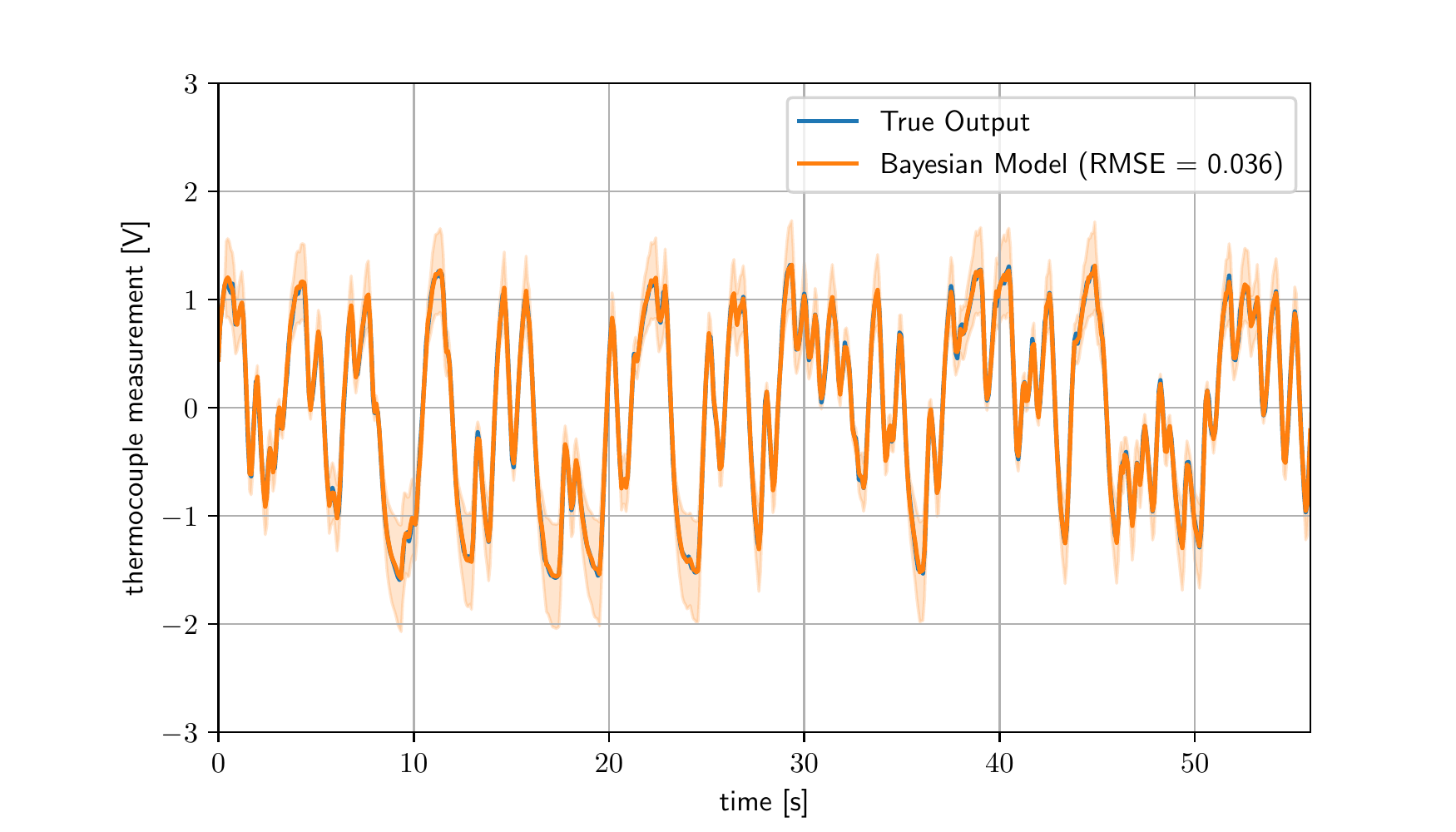}
        \caption{MLP model}
        \label{fig:dry_mlp_pred}
	\end{subfigure}
	\begin{subfigure}[h]{0.45\textwidth}
		\centering
        \includegraphics[scale=0.4]{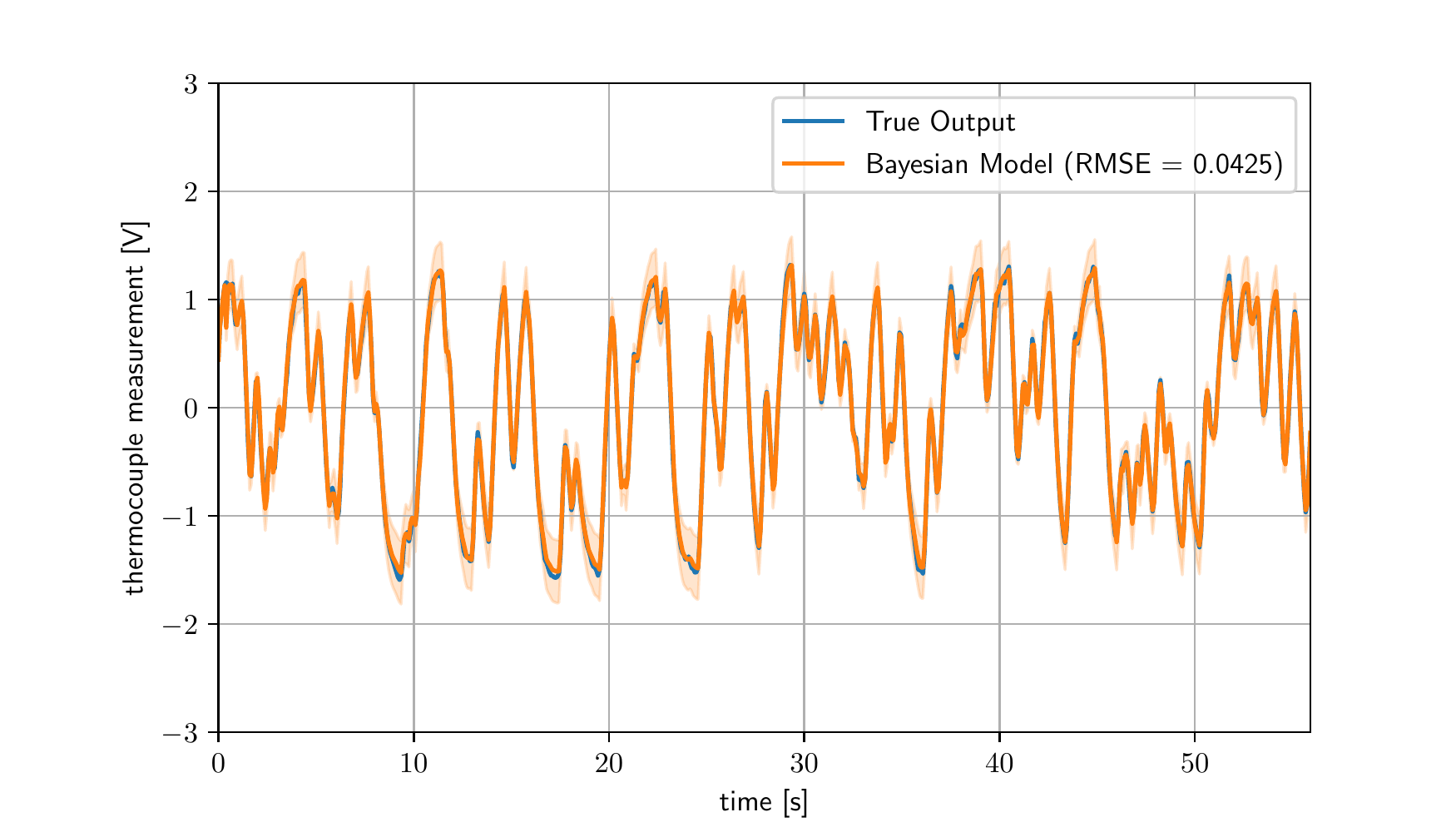}
        \caption{LSTM model}
        \label{fig:dry_lstm_pred}
	\end{subfigure}
	\caption{The identified MLP and LSTM model's output posterior mean predictions ($\pm 2\sigma$) on Hairdryer dataset.}
\end{figure}

Plots of the posterior predictive distribution's mean prediction and standard deviation obtained by sampling 10000 times from the posterior distribution of the connections' weights and by using Eq.~\eqref{eq:mean_mc} and \eqref{eq:std_mc} are shown in Fig.~\ref{fig:dry_mlp_pred} and~\ref{fig:dry_lstm_pred}. Plots of the identified models' free run simulations can be found in Fig.~\ref{fig:dry_sim}. 

\section{Heat Exchanger} \label{ap:hex}
A heat exchanger is a thermodynamic device that ensures a transfer of heat in between two fluids seperated by a wall. In this experiment, the dynamic relationship between the change in coolant temperature and the change in the product temperature is identified \cite{matlab_hex}.
The first 3000 data points are used for identification and the remaining 2000 for validation. This dataset is particularly unique among the others. The process exhibits a delay of around 1/4 of a minute \cite{matlab_hex}.  

One hidden-layer MLP with 50 nodes is initialised with a linear activation function and no bias term. The lag chosen is $l_u=l_y=$ 150 samples corresponding to the delay of $0.25$ seconds that can be observed in the first instance of the given dataset. The experiment is run for $6$ identification cycles, in which the 4th model is selected as the best validated model. The model is $99.3\%$ sparse . 

\begin{figure}[h]
	\centering
	\begin{subfigure}[h]{0.45\textwidth}
      \centering
        \includegraphics[scale=0.4]{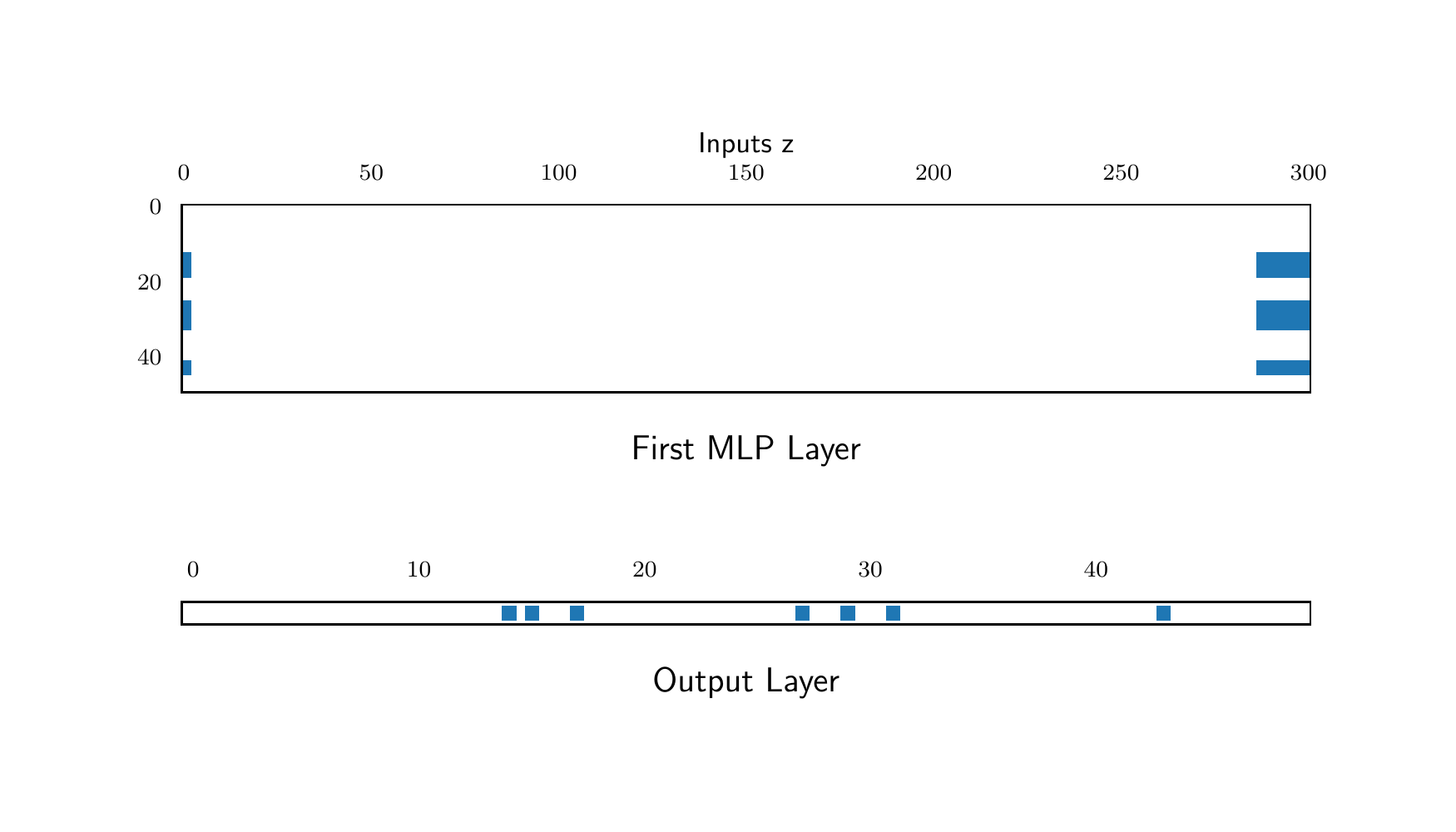}
        \vspace{-0.4cm}
        \caption{MLP model}
        \label{fig:hex_mlp_model}
	\end{subfigure}
	\begin{subfigure}[h]{0.45\textwidth}
        \centering
        \includegraphics[scale=0.4]{figures/hex_lstm_sparse_model_lag_150_lambda_3_0.pdf}
        \vspace{-0.5cm}
        \caption{LSTM model}
        \label{fig:hex_lstm_model}
	\end{subfigure}	
	\caption{Model sparsity plot of the identified MLP and LSTM on Heat Exchanger dataset.}
\end{figure}

One layer RNN network with 10 LSTM units is trained with the same lag used previously ($l_u=l_y=$ 150). The best validated model is the second out of 6 identification cycles. The accepted model's sparsity is 96.4 \% for which the sparsity plot is given in Fig.~\ref{fig:hex_lstm_model}. 

\begin{figure}[h]
	\centering
	\begin{subfigure}[h]{0.45\textwidth}
		\centering
        \includegraphics[scale=0.4]{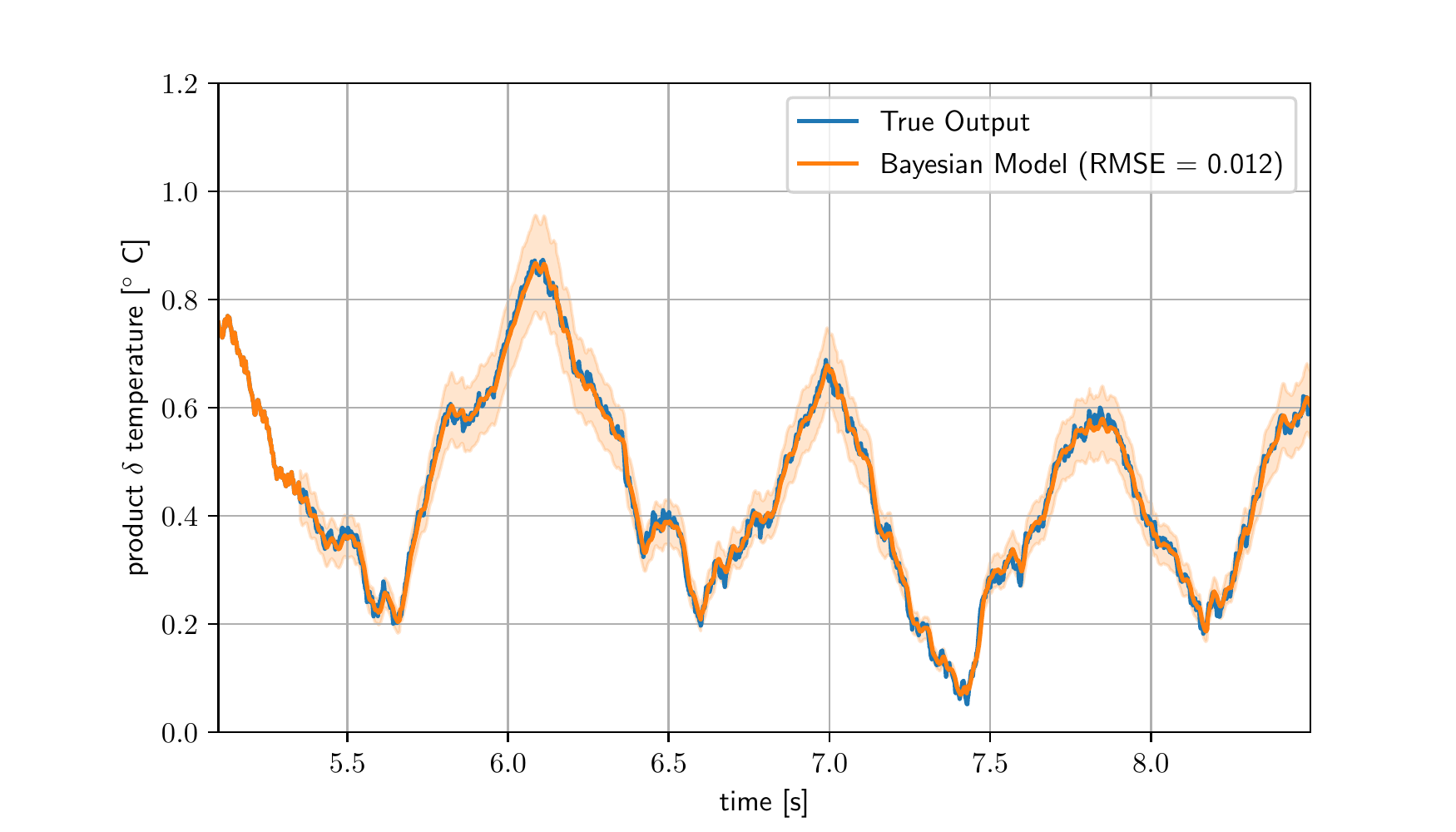}
        \caption{MLP model}
        \label{fig:hex_mlp_pred}
	\end{subfigure}
	\begin{subfigure}[h]{0.45\textwidth}
		\centering
        \includegraphics[scale=0.4]{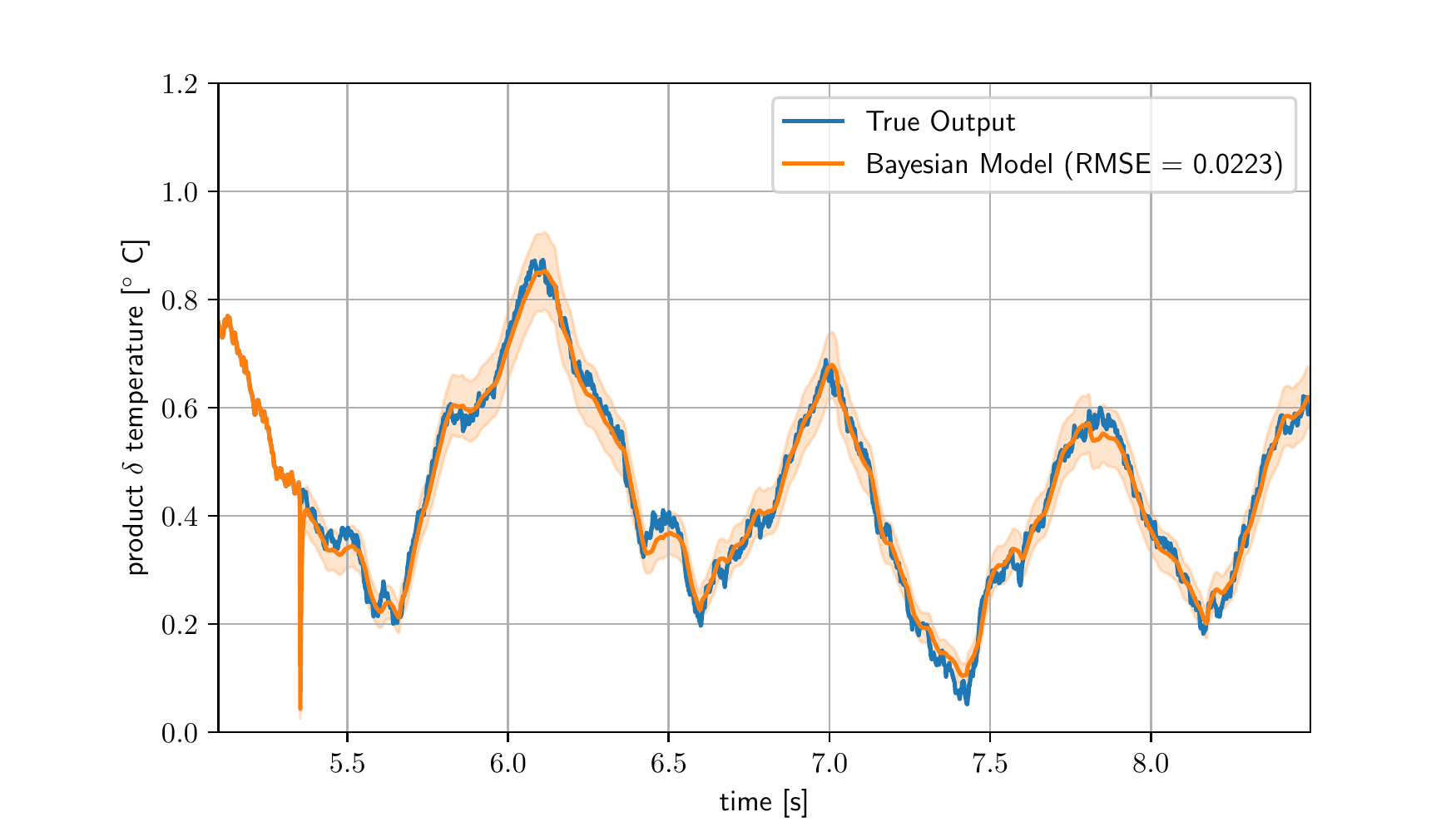}
        \caption{LSTM model}
        \label{fig:hex_lstm_pred}
	\end{subfigure}
	\caption{The identified MLP and LSTM model's output posterior mean predictions ($\pm 2\sigma$) on Heat Exchanger dataset.}
\end{figure}

The predictive mean and standard deviation of the posterior predictive distribution are shown in Fig.~\ref{fig:hex_mlp_pred}-\ref{fig:hex_lstm_pred} against the real validation signal. These are obtained using 10000 samples of the posterior distribution. Please refer to Fig.~\ref{fig:hex_sim}, for a plot of these free run simulations.

\section{Glass Tube Manufacturing Process} \label{ap:gt}
In the process of manufacturing glass tubes, the melted glass shapes around a rotating cylinder, while homogenizing. Then it is drawn on rollers to a certain length. The thickness of the obtained glass tube is measured by a laser beam outside the chamber \cite{gt_benchmark}. The objective is to identify the linear dynamic relationship between the input drawing speed and the output thickness.
The datasets are provided by the MATLAB example. These are detrended and decimated by four, to get rid of the high frequency components of the signal \cite{matlab_gt}. This results in a sampling time of 4 seconds. The data used for identification consist of the first $500$ datapoints and remaining $175$ datapoints are used for validation.

\begin{figure}[h]
	\centering
	\begin{subfigure}[h]{0.45\textwidth}
        \centering
        \includegraphics[scale=0.4]{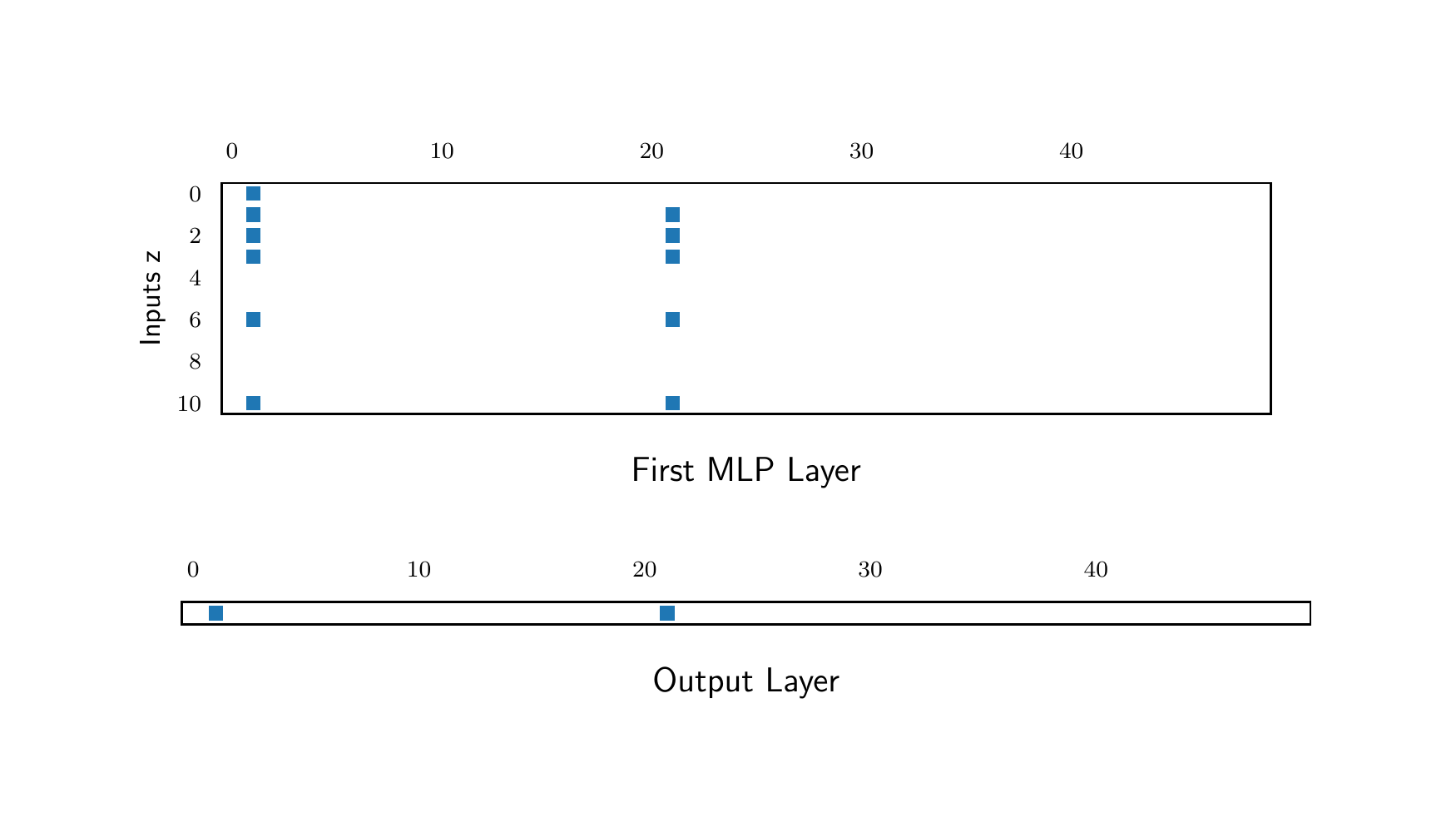}
        \vspace{-0.5cm}
        \caption{MLP model}
        \label{fig:gt_mlp_model}
	\end{subfigure}
	\begin{subfigure}[h]{0.45\textwidth}
        \centering
        \includegraphics[scale=0.4]{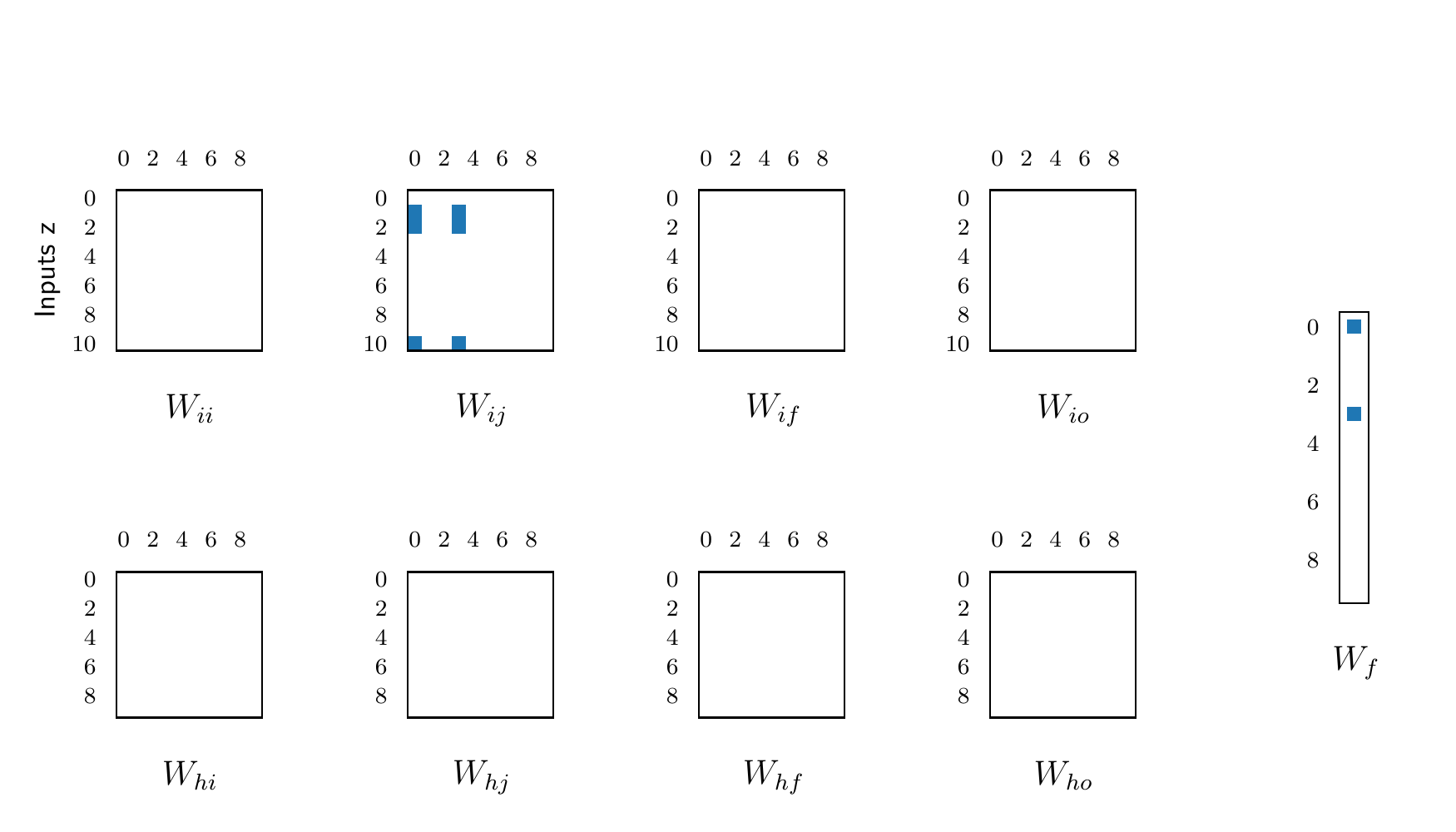}
        \caption{LSTM model}
        \label{fig:gt_lstm_model}
	\end{subfigure}	
	\caption{Model sparsity plot of the identified MLP and LSTM on Glass Tube Manufacturing dataset.}
\end{figure}

An MLP is randomly initialised with one hidden layer and 50 neurons. The input regressors are chosen such as $l_u=l_y=$ 5. The activation function used is linear without a bias term. The final obtained model is 97.8 \% sparse with a sparsity plot shown in Fig.~\ref{fig:gt_mlp_model}. This model is the third generated model out of 6 identification cycles.

With the same choice of regressors, an LSTM network is initialised with one layer of 10 LSTM units. The bias term is not used in this case. In the 6 identification cycles, the 6th generated model is the sparsest and have the best validation performance. The sparsity plot of this network is given by Fig.~\ref{fig:gt_lstm_model}. The model is 99\% sparse, and the only non-pruned parameters in the model correspond to the input to cell state operator $W_{ij}$. 
\vspace{-0.3cm}

\begin{figure}[h]
	\centering
	\begin{subfigure}[h]{0.45\textwidth}
		\centering
        \includegraphics[scale=0.4]{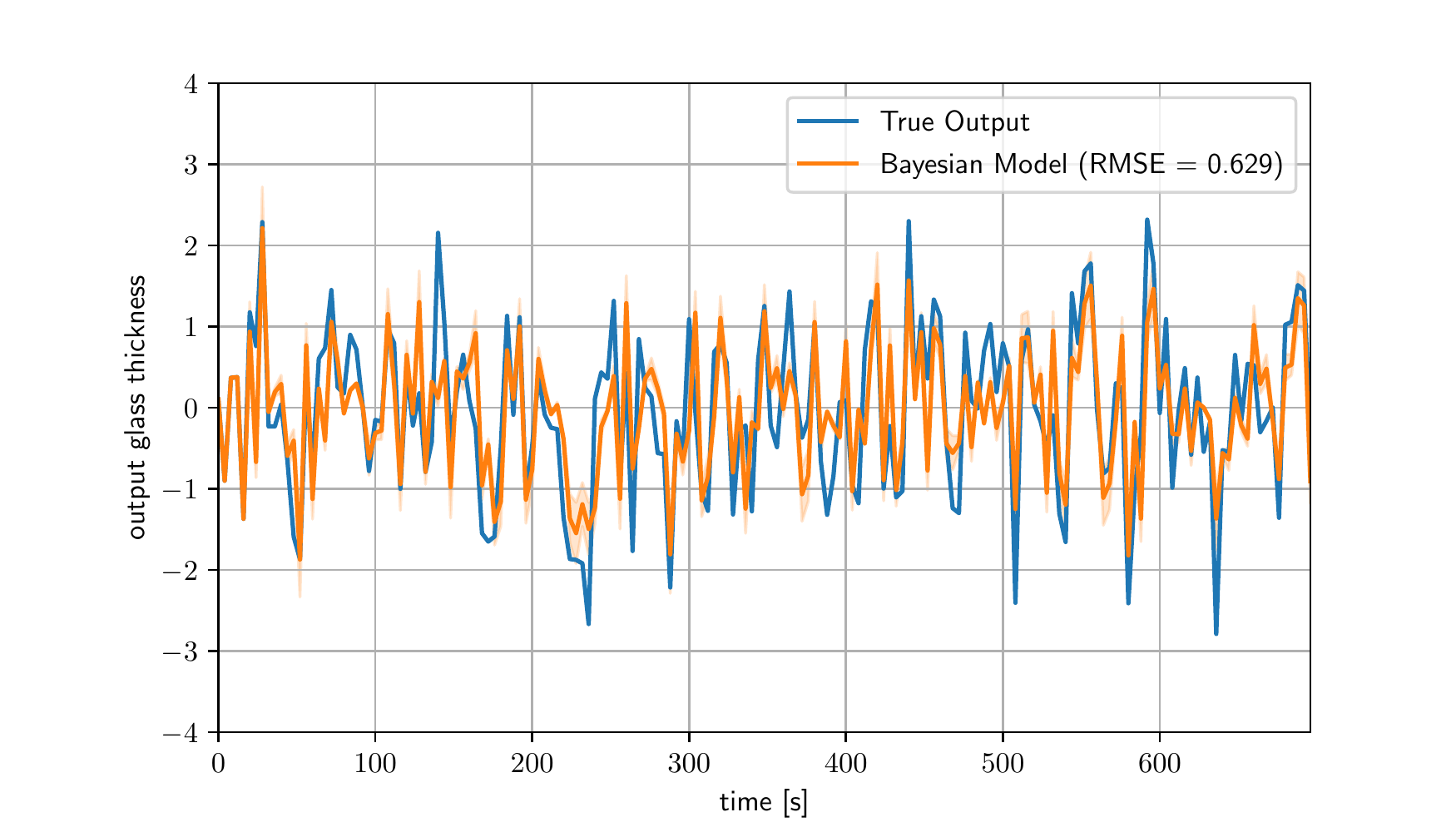}
        \caption{MLP model}
        \label{fig:gt_mlp_pred}
	\end{subfigure}
	\begin{subfigure}[h]{0.45\textwidth}
		\centering
        \includegraphics[scale=0.4]{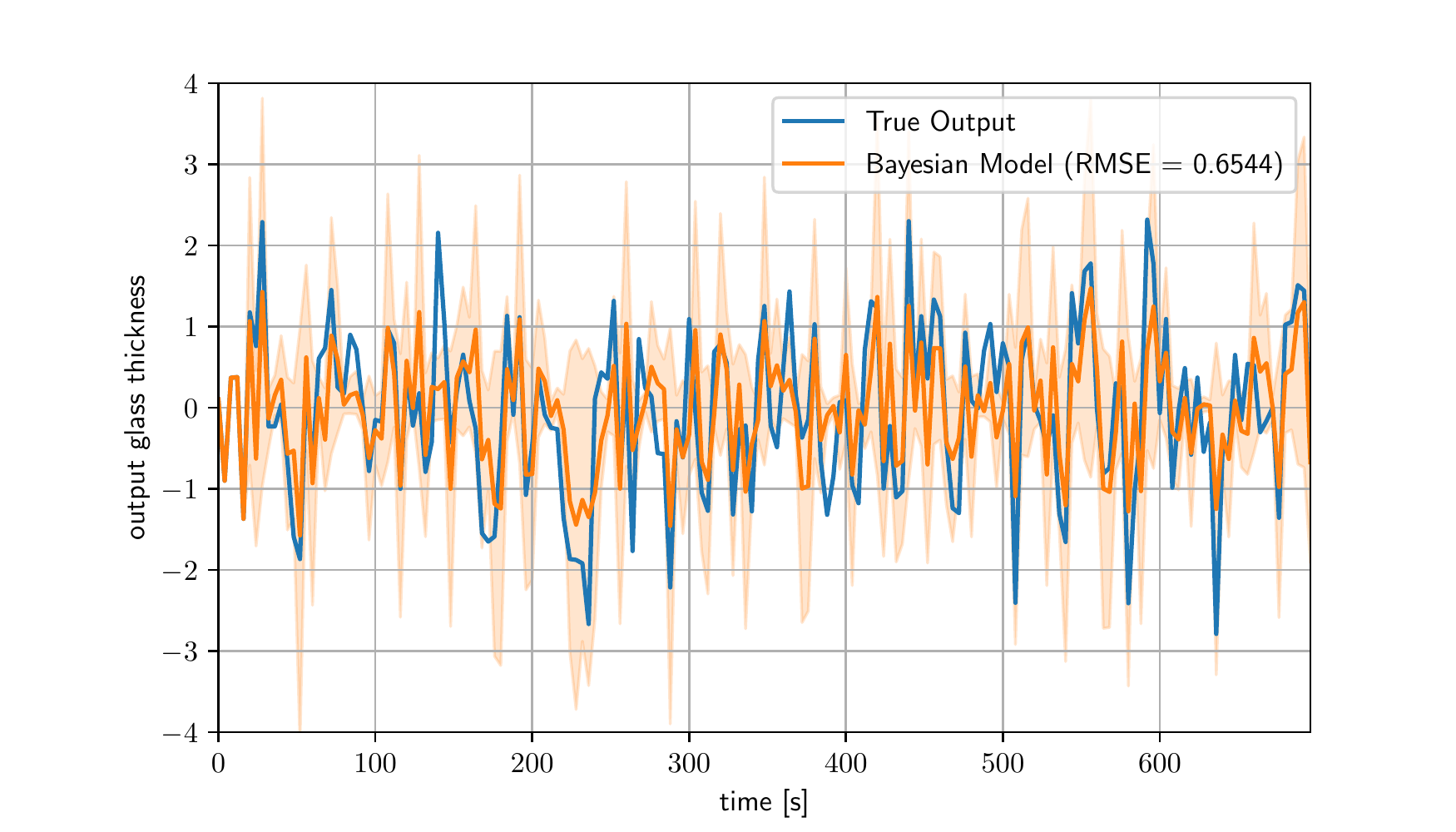}
        \caption{LSTM model}
        \label{fig:gt_lstm_pred}
	\end{subfigure}
	\caption{The identified MLP and LSTM model's output posterior mean predictions ($\pm 2\sigma$) on Glass Tube Manufacturing dataset.}
\end{figure}

The one-step ahead prediction estimates and uncertainties are obtained by Monte Carlo sampling 10000 times from the posterior and are shown in Fig.~\ref{fig:gt_mlp_pred}-\ref{fig:gt_lstm_pred} as a representation of the posterior predictive distribution. The free run simulations of the generated models in this paper are presented in Fig.~\ref{fig:gt_sim}.

\section{Cascaded Tanks} \label{ap:ct}
\begin{figure}[ht]
    \centering
    \includegraphics[scale=0.1]{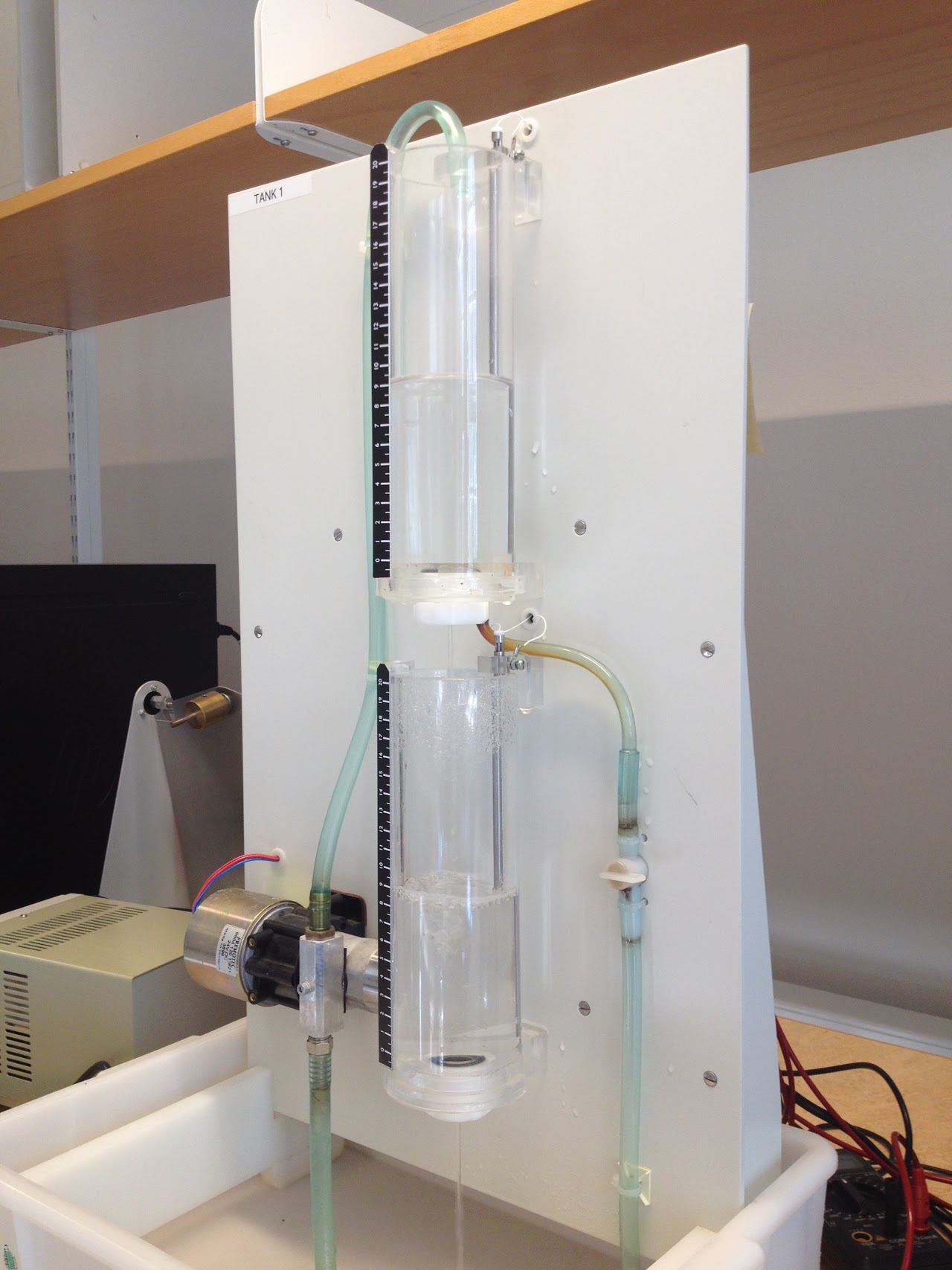}
    \caption{The cascaded tanks setup~\cite{ct_benchmark}.}
    \label{fig:cascaded_tank_setup}
\end{figure}

A pump drives water up from the reservoir to the upper tank of two vertically cascaded tanks. The upper and lower tanks are seperated by a small opening allowing water to fill the lower tank. The lower tank and the reservoir are also seperated by a small opening, from which water goes back to the reservoir. In addition to that, water can overflow from the upper tank to the lower tank and reservoir. Water can also overflow the second tank and drop into the reservoir. The small openings and overflows are sources of nonlinearity \cite{ct_benchmark}. The objective of the benchmark is the identification of the dynamic relationship between the input voltage to the pump and the output measured water level in the lower tank by a capacitive sensor \cite{ct_benchmark}.
The setup of the cascaded tanks is shown in Fig~\ref{fig:cascaded_tank_setup}. 
Two multisine input datasets and their corresponding outputs  with a sampling rate of 4 seconds are provided. The datasets contain 1024 samples and are with different initial conditions. One of the datasets is used for estimation and the other for validation. The signals provided exhibit a static bias that is dealt with in the preprocessing stage of the identification procedure by detrending.

\begin{figure}[h]
	\centering
	\begin{subfigure}[h]{0.45\textwidth}
        \centering
        \includegraphics[scale=0.2]{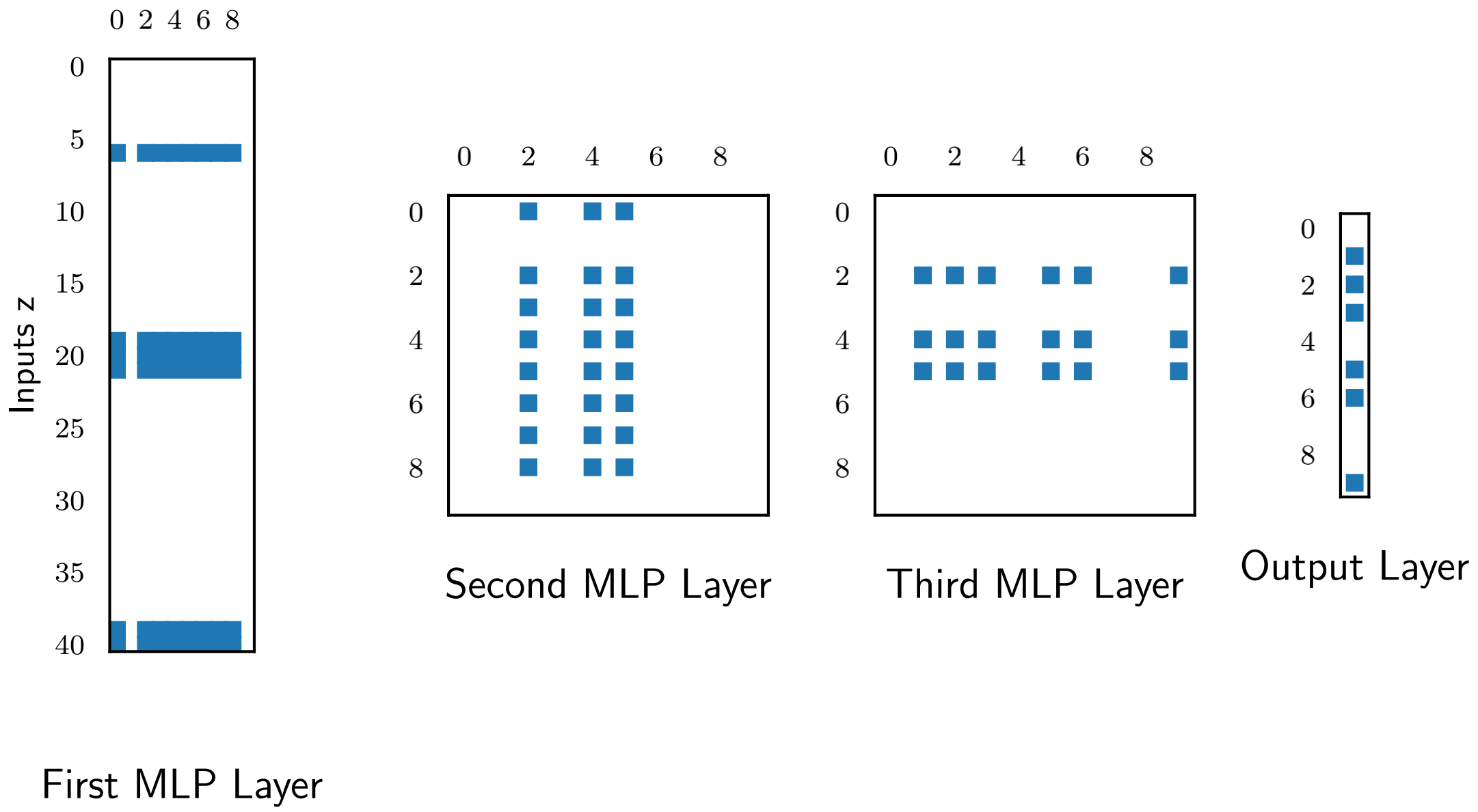}
        \vspace{-0.3cm}
        \caption{MLP model}
        \label{fig:ct_mlp_model}
	\end{subfigure}
	\begin{subfigure}[h]{0.45\textwidth}
        \centering
        \includegraphics[scale=0.2]{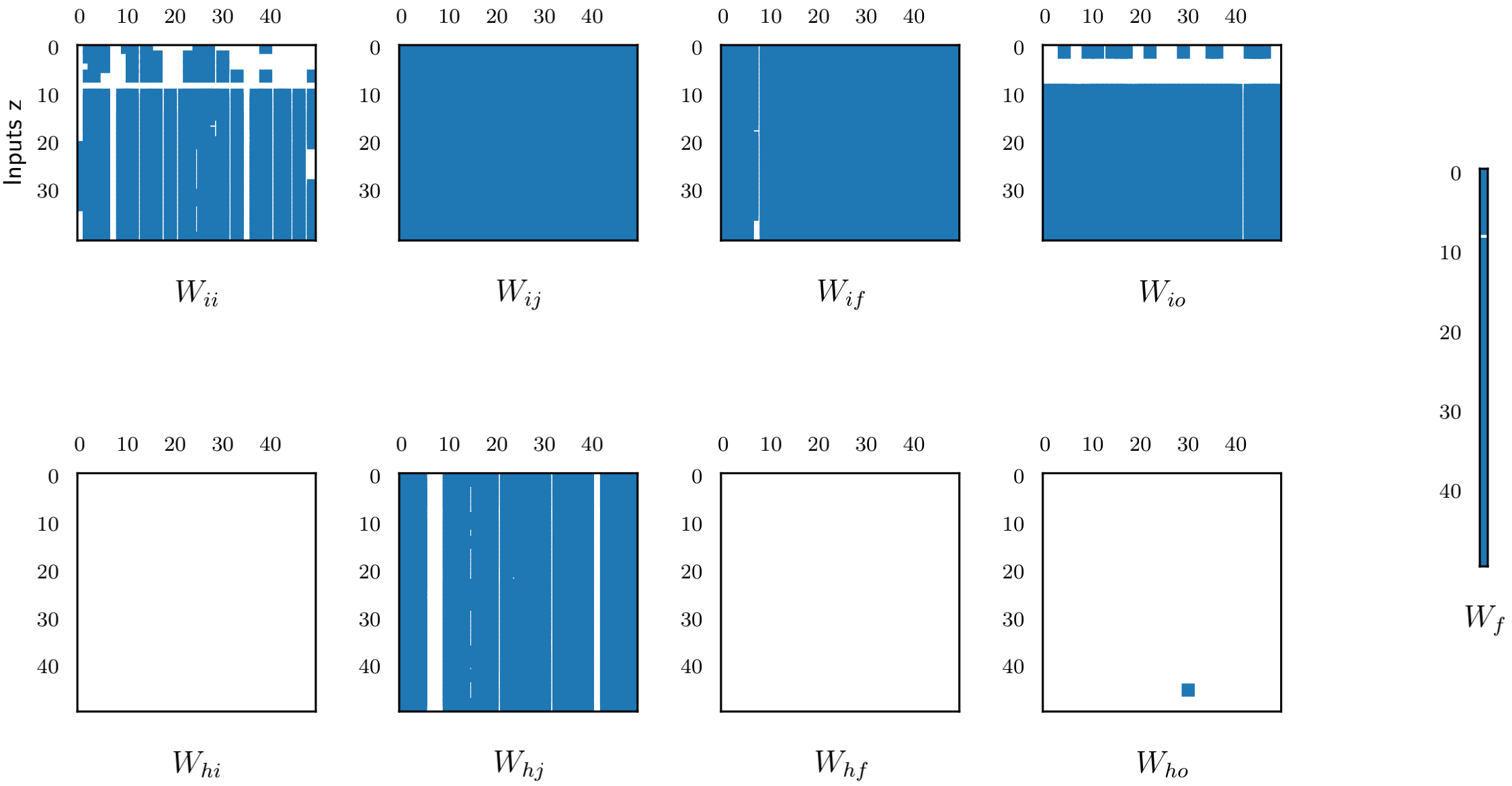}
        \caption{LSTM model}
        \label{fig:ct_lstm_model}
	\end{subfigure}	
	\caption{Model sparsity plot of the identified MLP and LSTM on Cascaded Tanks dataset.}
\end{figure}

A three hidden layers deep MLP network with 10 neurons per layer is randomly initialised. The activation function used is the Rectified Linear Unit (ReLU) activation. The input regressors are selected as $l_u=l_y=$ 20. The identification experiment is run for 10 cycles. The $9$th generated model performs the best in validation with a sparsity of 84.5\%. The model's sparsity plot is shown in Fig.~\ref{fig:ct_mlp_model}.

Moreover, a one-layer RNN with 10 LSTM units is also used as a model structure for the identification experiment. The $4$th identified model with $60.3 \%$ sparsity turns out to be the best validated model out of 10 identification cycles. The sparsity plot of the corresponding model is shown in Fig.~\ref{fig:ct_lstm_model}.
\vspace{-0.3cm}

\begin{figure}[h]
	\centering
	\begin{subfigure}[h]{0.45\textwidth}
		\centering
        \includegraphics[scale=0.4]{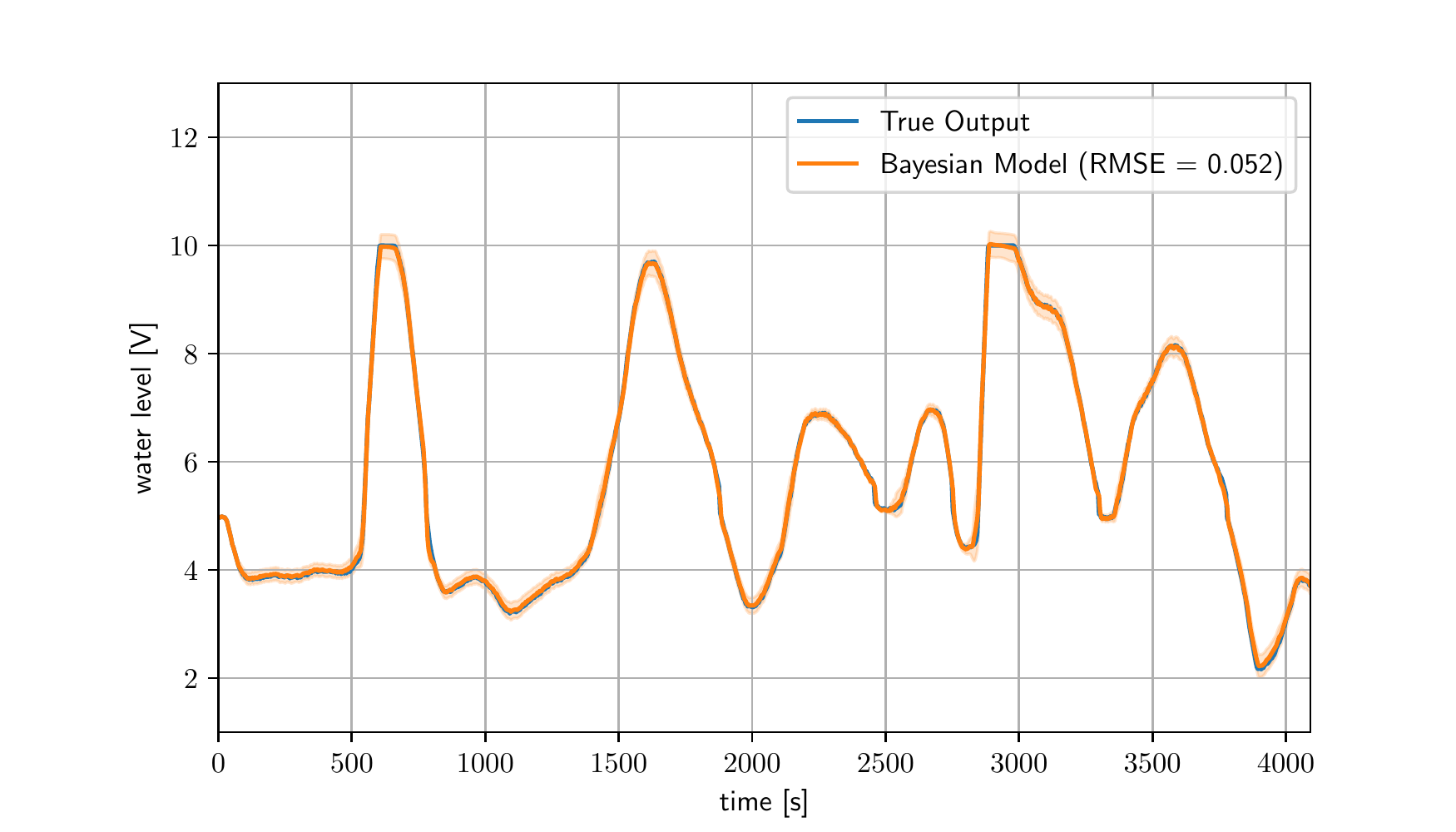}
        \caption{MLP model}
        \label{fig:ct_mlp_pred}
	\end{subfigure}
	\begin{subfigure}[h]{0.45\textwidth}
		\centering
        \includegraphics[scale=0.4]{figures/ct_lstm_pred_results_lag_20_lambda_4_0.pdf}
        \caption{LSTM model}
        \label{fig:ct_lstm_pred}
	\end{subfigure}
	\caption{The identified MLP and LSTM model's output posterior mean predictions ($\pm 2\sigma$) on Cascaded Tanks dataset.}
\end{figure}

In addition, the posterior predictive mean and standard deviation are given in Fig.~\ref{fig:ct_mlp_pred} and \ref{fig:ct_lstm_pred}. These are obtained by averaging Eqs~\ref{eq:mean_mc} and \ref{eq:std_mc} and sampling 50000 times from the infered posterior distribution of the weights. A plot of the models' free run simulations is given in Fig.~\ref{fig:ct_sim}.
\section{Coupled Electric Drives} \label{ap:ced}
\begin{figure}[ht]
    \centering
    \includegraphics[width=\linewidth,scale=0.06]{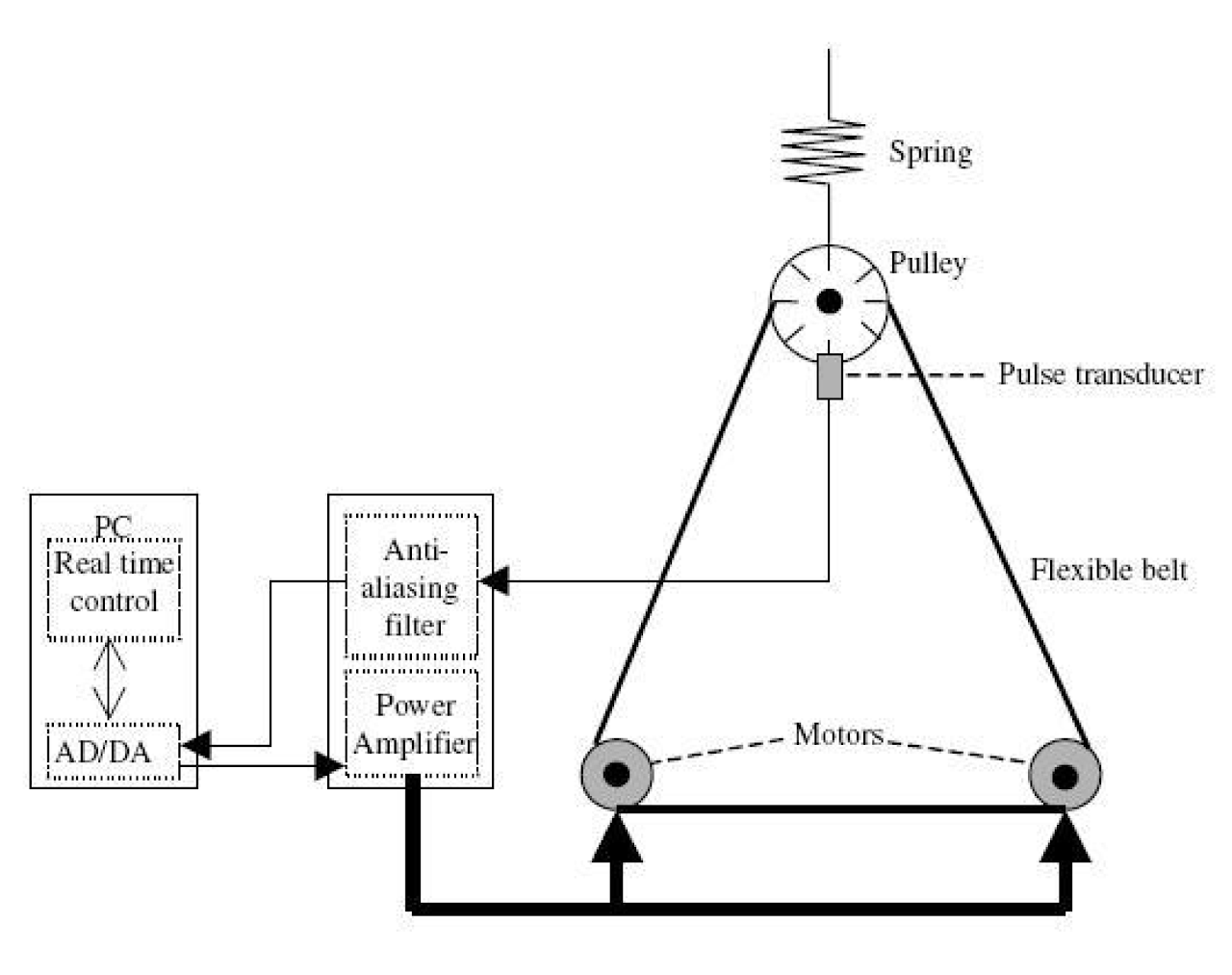}
    \caption{The coupled electric drives setup \cite{ced_benchmark}.}
    \label{fig:ce8_setup}
\end{figure}

The coupled electric drives consist of two electric motors and a pulley, connected by a flexible belt forming a triangle. The pulley is attached by a spring to a fixed frame. This results in belt tension, slippage, and pulley speed that is harder to model. In addition to that, the output pulley rotational speed is measured in ticks per second, insensitive to rotational directions.
The setup of the Coupled Electric Drives is shown in Fig.~\ref{fig:ce8_setup}.
The dynamic relationship to be identified is between the input motors voltage and the measured rotational speed of the pulley.
For this identification task, two uniformly distributed signals of $500$ samples is provided spanning $10$ seconds. With each of these datasets, the first $300$ samples are used for estimation and the remaining for validation.

\begin{figure}[h]
	\centering
	\begin{subfigure}[h]{0.45\textwidth}
        \centering
        \includegraphics[ scale=0.4]{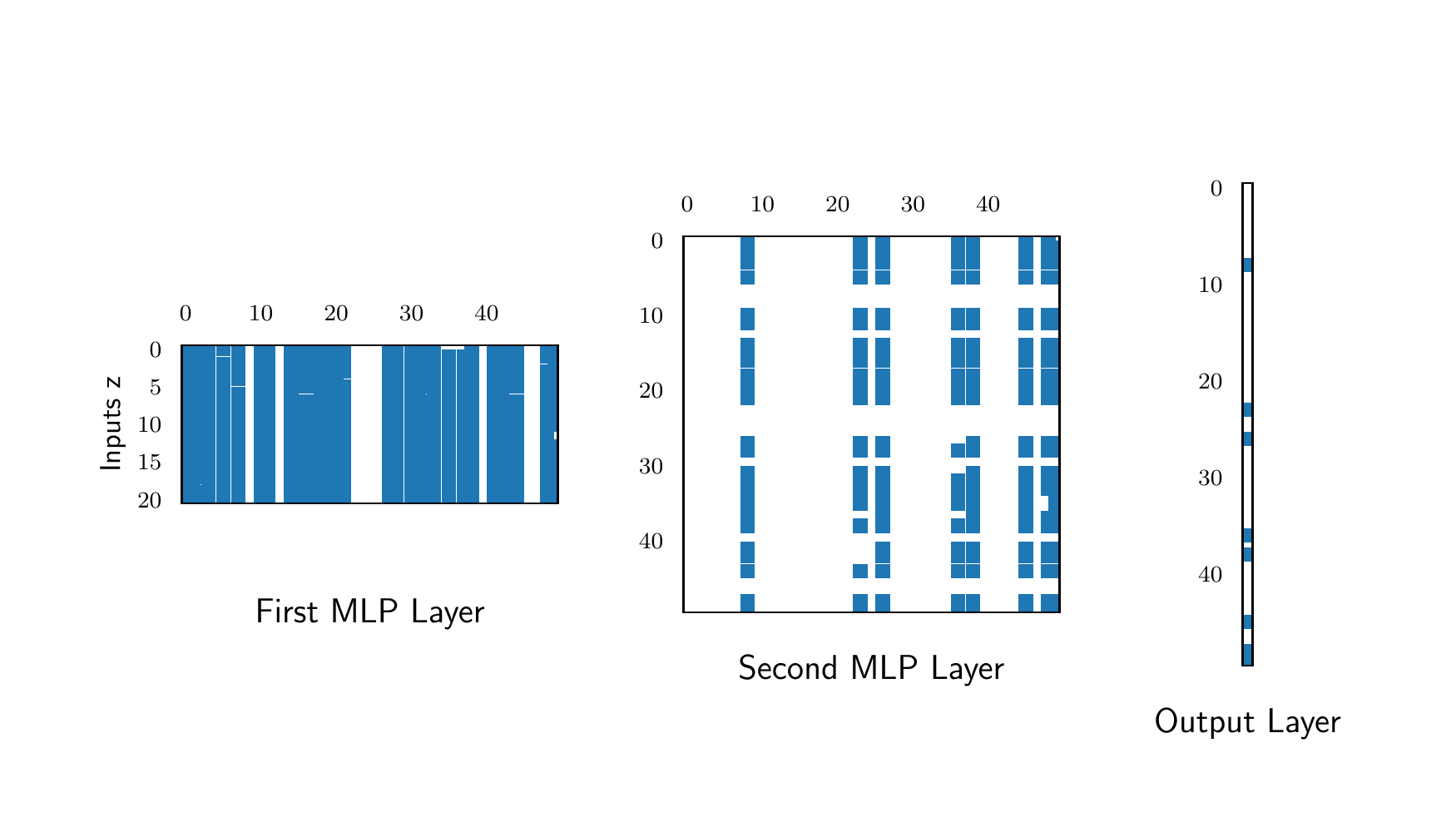}
        \vspace{-0.5cm}
        \caption{MLP model}
        \label{fig:ced_mlp_model}
	\end{subfigure}
	\begin{subfigure}[h]{0.45\textwidth}
        \centering
        \includegraphics[ scale=0.4]{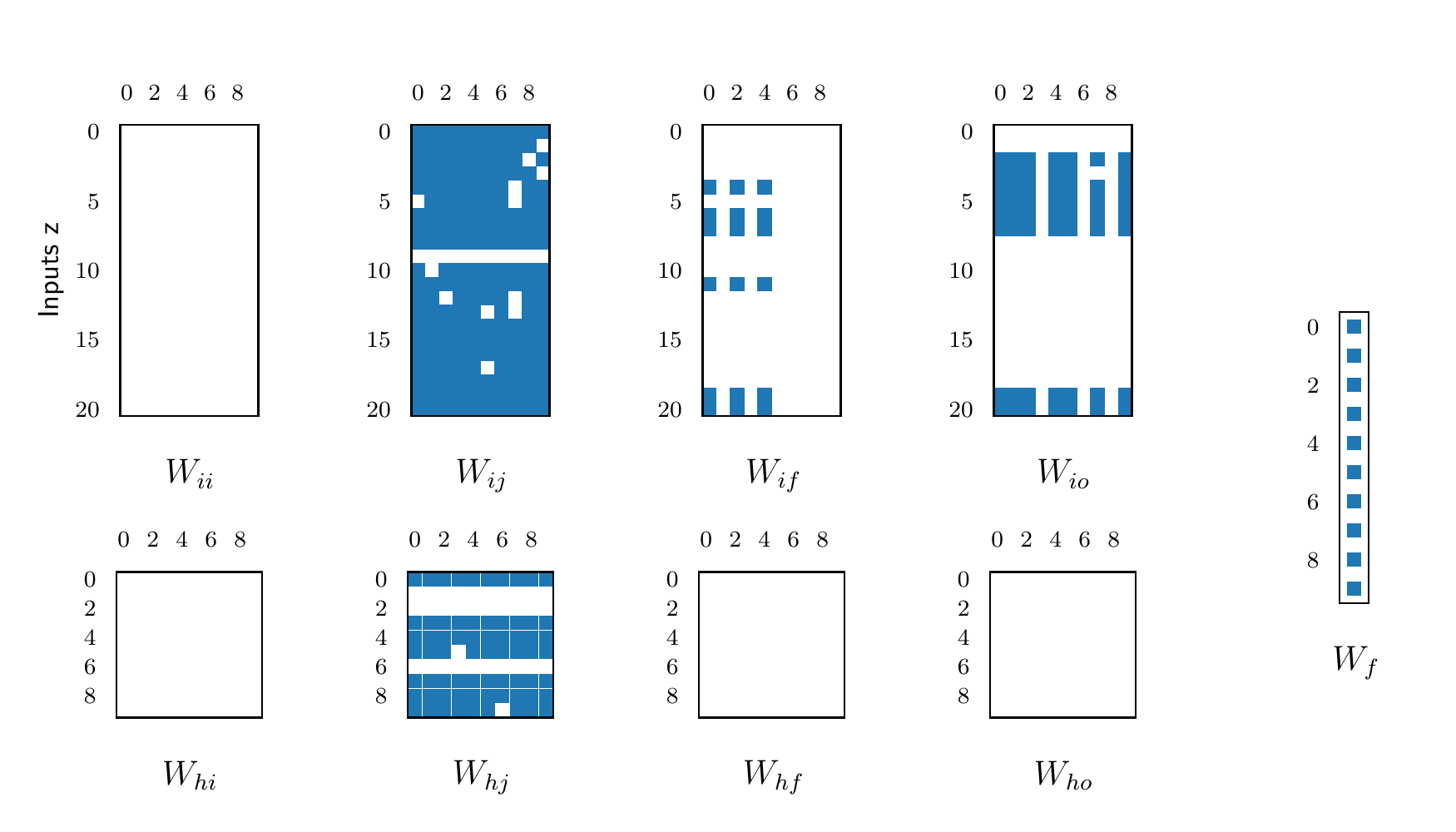}
        \caption{LSTM model}
        \label{fig:ced_lstm_model}
	\end{subfigure}
	\caption{Model sparsity plot of the identified MLP and LSTM on Coupled Electric Drives dataset. }
\end{figure}

Two hidden layers MLP with 50 neurons each and ReLu activation functions is randomly initialised and trained with the estimation data for 10 identification cycles. The model's regressors are chosen such that $l_u=l_y=$ 10. The model obtained in the 6th identification iteration is the chosen best model. This model is 78.4\% sparse for which the sparsity plot is shown in Fig.~\ref{fig:ced_mlp_model}.

The same regressors are used for the identification of RNN model structure. An RNN with one layer and 10 LSTM units is trained for 10 identification cycles. The 8th identification yields the best simulation validation results. The resulting model is 72.8\% sparse with the sparsity plot depicted in Fig.~\ref{fig:ced_lstm_model}.
\vspace{-0.3cm}

\begin{figure}[h]
	\centering
	\begin{subfigure}[h]{0.45\textwidth}
		\centering
        \includegraphics[scale=0.4]{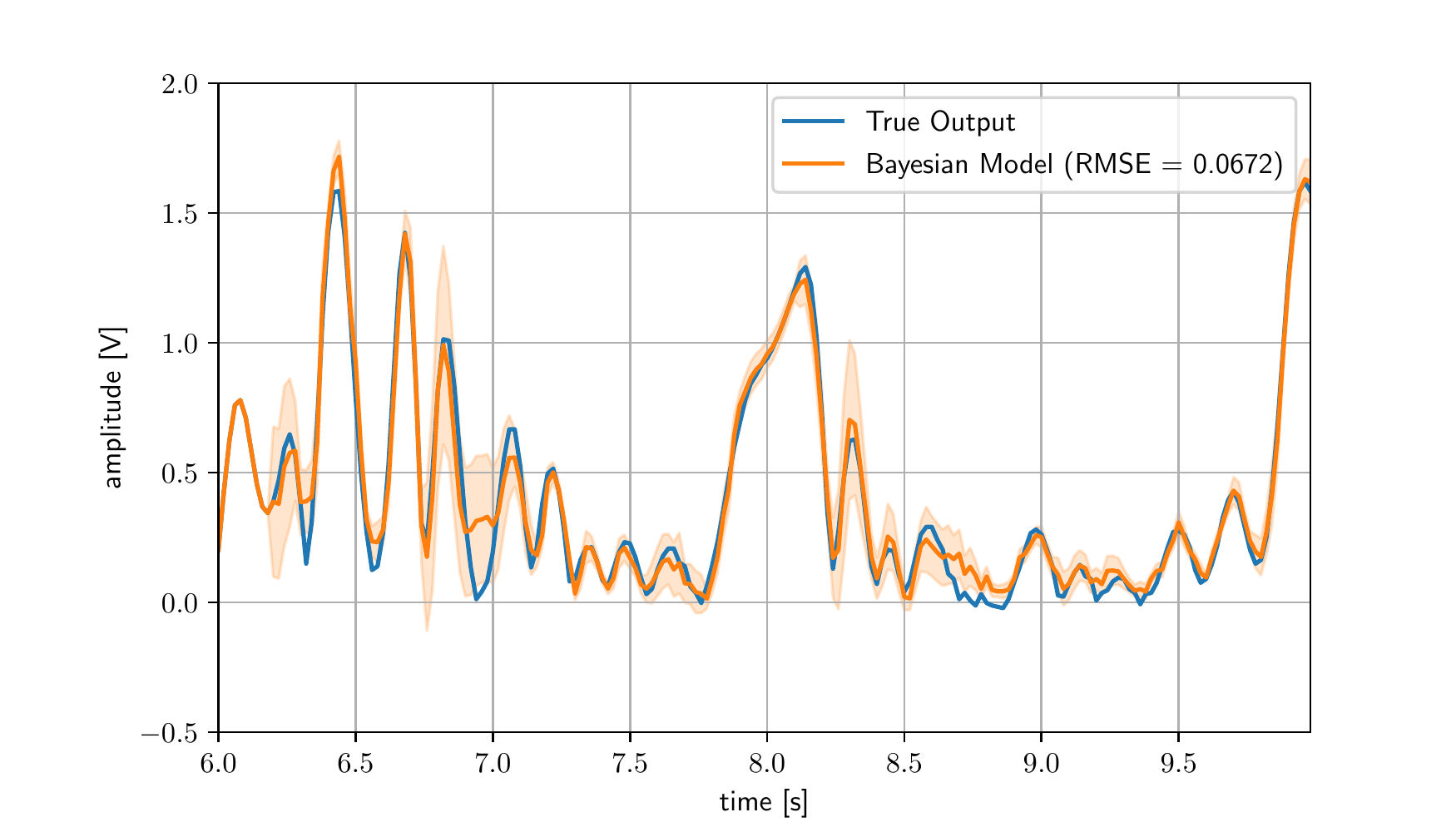}
        \caption{MLP model on first validation dataset}
        \label{fig:ced_mlp_pred1}
	\end{subfigure}
	\begin{subfigure}[h]{0.45\textwidth}
		\centering
        \includegraphics[scale=0.4]{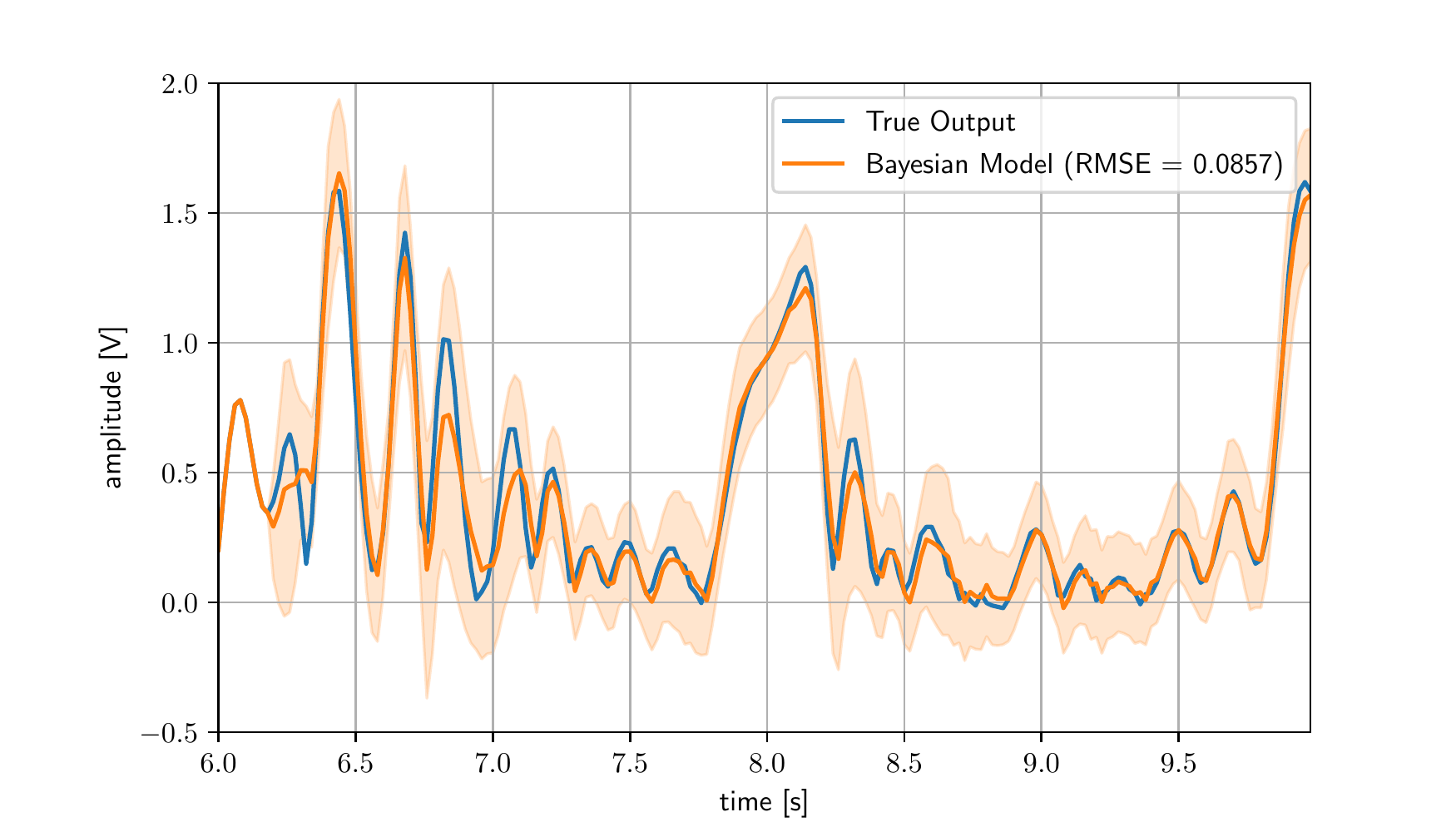}
        \caption{LSTM model on first validation dataset}
        \label{fig:ced_lstm_pred1}
	\end{subfigure}
	\begin{subfigure}[h]{0.45\textwidth}
		\centering
        \includegraphics[scale=0.4]{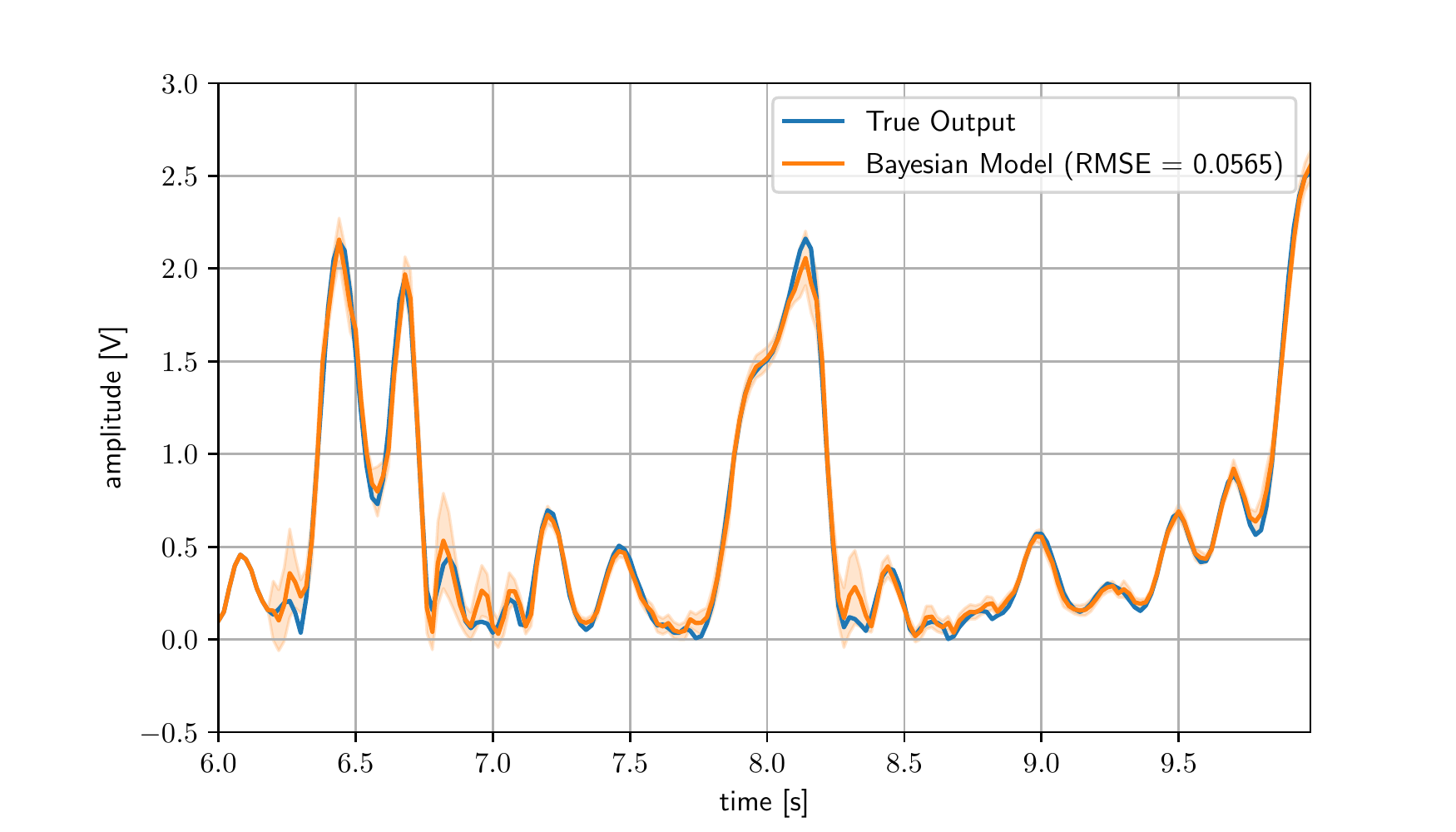}
        \caption{MLP model on second validation dataset}
        \label{fig:ced_mlp_pred2}
	\end{subfigure}
	\begin{subfigure}[h]{0.45\textwidth}
		\centering
    	 \includegraphics[scale=0.4]{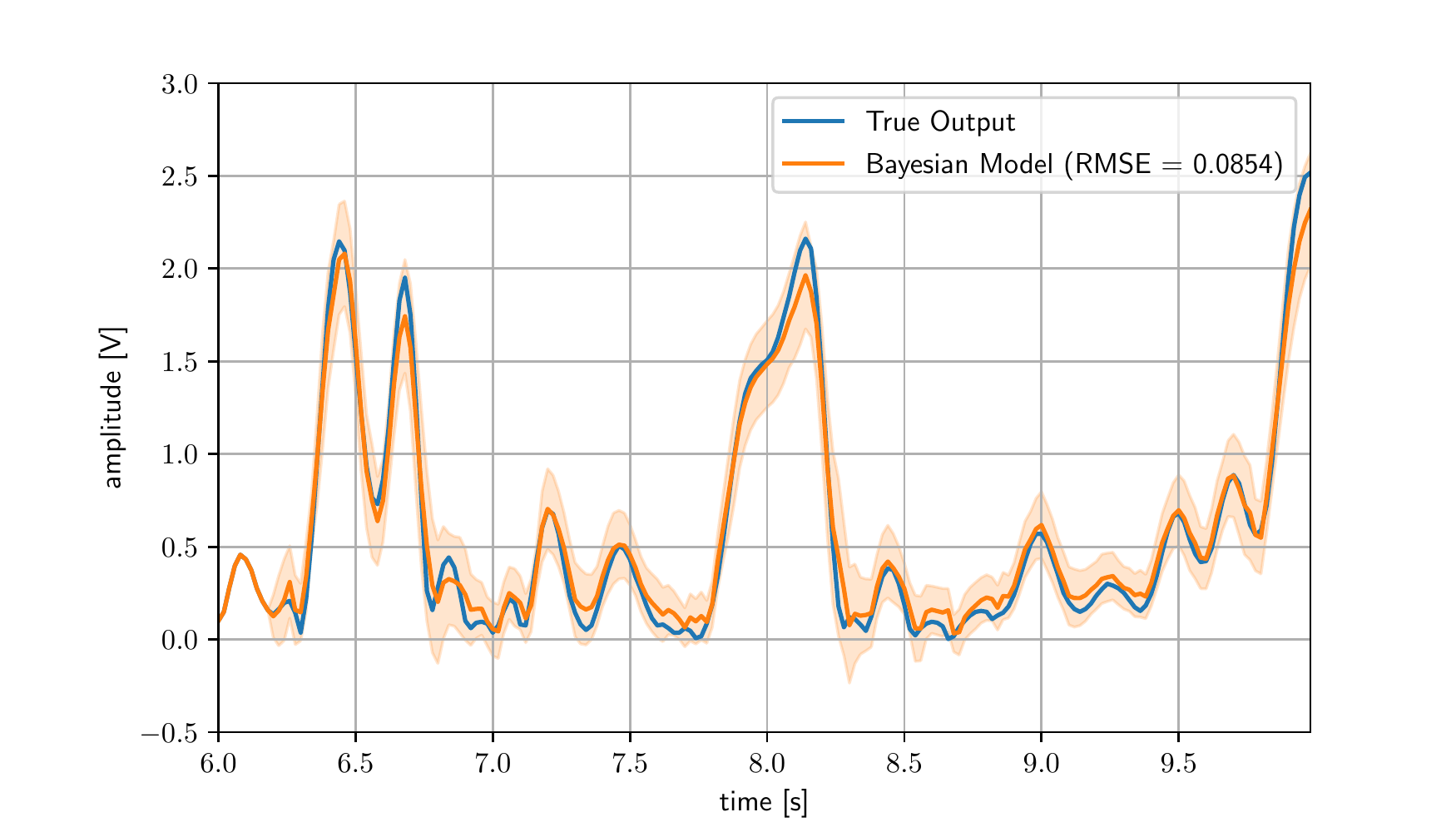}
        \caption{LSTM model on second validation dataset}
        \label{fig:ced_lstm_pred2}
	\end{subfigure}
	\caption{The identified MLP and LSTM model's output posterior mean predictions ($\pm 2\sigma$) of Coupled Electric Drives on first and second validation dataset.}
\end{figure}
By using Eqs.~\eqref{eq:mean_mc} and~\eqref{eq:std_mc}, the mean and standard deviation of the posterior predictive distributions are plotted in Fig.~\ref{fig:ced_mlp_pred1},~\ref{fig:ced_mlp_pred2}, \ref{fig:ced_lstm_pred1} and~\eqref{fig:ced_lstm_pred2} for both validation datasets. These are obtained with Eqs.~\eqref{eq:mean_mc} and \eqref{eq:std_mc} and $50000$ samples of the posterior distribution. The figures showing the resulting free run simulations are Fig.~\ref{fig:ced_sim1}-\ref{fig:ced_sim2}.

\modify{
\section{Comparison with Classical Model Types}
\label{appsec:discussion_comparison_other_methods}

In Table~\ref{tab:comparision_linear_simulation}-\ref{tab:comparision_non_linear_simulation}, the simulation performance of the proposed Bayesian approach is compared with several typical model types used in system identification, e.g.,  autoregressive with exogenous terms (ARX) model, transfer function. 
 Compared to these classical model types, the neural network model used in this paper requires less prior information and has a better generalisation ability. 
    For example, if we use the autoregressive with exogenous terms (ARX) model: 
    \begin{align}
    A(q^{-1})y(t) = B(q^{-1})u(t) + e(t)
    \label{eq:arx_model}
    \end{align}
    where \begin{math}
    A(q^{-1}) = 1 + a_1q^{-1} + a_2q^{-2} + \ldots + a_{n_a} q^{-n_a},  B(q^{-1}) = b_0 + b_1q^{-1} + b_2q^{-2} + \ldots + b_{n_b-1} q^{-n_b+1}
    \end{math}. $q^{-1}$ stands for the backward shift operator with $q^{-1}u(t) = u(t-1)$. 
    It can be found that the structure of an ARX model is decided by $n_a$ and $n_b$. According to~\cite{van2012system}, the selection of a specific model structure is very important for an identification problem. A wrong structure may lead to a bad identification result. 
    Although ARX-structure can also be regarded as a special case of neural network~\cite{sjoberg1994},
    we still have to select a proper model structure which is mainly based on the prior information of the system. 
    By contrast, the training for a neural network model is more based on the given data, which
    can be used in the situation when we don't have enough prior information.~\cite{sjoberg1994} also stated that the neural network model can scale better to high-dimensional systems, where the classical basis functions method may suffer the
    curse of dimensionality.
    ~\cite{lstm_matlab_toolbox_2021} compares the LSTM model with the typical transfer function model to estimate a linear system. The optimisation process shows that the architecture of LSTM does not change much while the complexity of the transfer function increases significantly.
    The proposed algorithm in this paper can be understood as an automatically neural network structure selection method, which can remove the unnecessary connections by employing the regularisation on model parameters.
    
    
    In general, the contribution of our simulation experiments includes two aspects: a) we can achieve good and competitive simulation performance compared with other system identification (SYSID) approaches; b) we address the non-trivial Hessian calculation problems for deep neural networks, especially for the recurrent neural networks. 
    The breakthrough of this key technology makes it possible to apply the proposed Bayesian deep learning algorithm to accelerate the training of recurrent cells. And finally, the simulation experiments can be implemented effectively. 
    Even though an approximation is introduced using diagonal elements to represent the Hessian, the proposed method turns an intractable sparse Bayesian RNN training procedure into a tractable one.}

\section{Further Discussion}
\label{sec:discussion}
\textbf{Regularisation parameter $\lambda$ in Algorithm~\ref{algo:algorithm}:} 
The regularisation parameter $\lambda$ in~\eqref{eq:loss_function} needs to be tuned many times for network training, especially when using the simulation error as the evaluation metric. 
As a well-established strategy for global optimisation, Bayesian optimisation is a promising method that can alleviate the heavy tuning burden. 
A scalable Bayesian optimisation method based on DNNs was proposed in~\cite{snoek2015scalable}.
A Bayesian optimisation framework for DNN compression was also discussed in~\cite{ma2019bayesian}. 
In the future, the application of Bayesian optimisation in SYSID to reduce the hyper-parameter tuning burden is a research topic worthy of study.   

\textbf{Identification with physical interpretability:}
As explained in Section~\ref{sec:analysis},
good and competitive simulation accuracy can be achieved across the five benchmark datasets. 
However, the identified NNs are still black-box models, which lack interpretability in physics and cannot provide an understanding of the underlying phenomenon of the system.
Recently, several methods have been proposed to identify the governing equations. To name a few, a framework to identify the governing interactions and transition logics of subsystems in cyber-physical systems was developed in~\cite{yuan2019data}.
A practical sparse Bayesian approach was proposed in~\cite{pan2016online} to perform the online selection for the Hill function of synthetic gene networks. 
In~\cite{brunton2016discovering}, the sparse identification of nonlinear dynamics (SINDy) algorithm was proposed to identify the fewest equation terms that can describe a system.
However, these approaches suffer from the nontrivial task of choosing appropriate basis functions, limiting their capacity for more general applications.
Inspired by the research of physics-informed machine learning~\cite{willard2020integrating} and symbolic regression, the problem of identifying system equations can be solved by designing a novel network structure with basic mathematical operations or encoding the prior information in the loss function. 

\textbf{Convergence of DNN training:} 
The convergence of NN training is difficult to analyse, which is influenced by many aspects, i.e., the pre-processing of training data~\cite{lawrence1998size}, the initialisation of weight matrices~\cite{zhou2021local}, proper selection of learning rate and batch size, and the complexity of NN~\cite{zhou2021local,lawrence1998size}.
A local convergence theory was developed for mildly over-parameterised two-layer NN, which shows the gradient descent can converge to zero with the initial loss below a threshold in~\cite{zhou2021local}. 
~\cite{gao2020recalling} proposed using the efficient conjugate gradient (CG) algorithm to train the RNN, which can 
accelerate the convergence procedure and help find the optimal solution.
In this paper, although there is no guarantee that the absolute global minimum can be achieved during the training process, the experimental result shows that the convergence trend is noticeable. This is also consistent with the research findings in~\cite{lawrence1998size} that state that the backpropagation process can always make it possible to meet practical stopping criteria. 
\end{appendices}

%
\end{document}